\documentclass[aps,prb,article,twocolumn,preprintnumbers,amsmath,amssymb,superscriptaddress]{revtex4}

\date{\today}
\usepackage{epsfig}
\usepackage{subfigure}
\usepackage{graphicx}
\usepackage{dcolumn}
\usepackage{bm}
\usepackage[colorlinks,linkcolor=blue,hyperindex,CJKbookmarks]{hyperref}
\usepackage{float}
\usepackage{hyperref}
\usepackage{comment}
\begin{document}

\title{Theoretical Understanding of Photon Spectroscopies in Correlated Materials In and Out of Equilibrium}
\author{Yao Wang}
\affiliation{Department of Physics, Harvard University, Cambridge, Massachusetts 02138, USA}
\author{Martin Claassen}
\affiliation{The Center for Computational Quantum Physics, The Flatiron Institute, New York, NY 10010}
\author{Chaitanya Das Pemmaraju}
\affiliation{Stanford Institute for Materials and Energy Sciences, SLAC National Laboratory and Stanford University, Menlo Park, CA 94025, USA}
\author{Chunjing Jia}
\affiliation{Stanford Institute for Materials and Energy Sciences, SLAC National Laboratory and Stanford University, Menlo Park, CA 94025, USA}
\author{Brian Moritz}
\affiliation{Stanford Institute for Materials and Energy Sciences, SLAC National Laboratory and Stanford University, Menlo Park, CA 94025, USA}
\author{Thomas P. Devereaux}
\affiliation{Stanford Institute for Materials and Energy Sciences, SLAC National Laboratory and Stanford University, Menlo Park, CA 94025, USA}
	\date{\today}
\begin{abstract}
Photon-based spectroscopies have had a significant impact on both fundamental science and applications by providing an efficient approach to investigate the microscopic physics of materials. Together with the development of synchrotron X-ray techniques, theoretical understanding of the spectroscopies themselves and the underlying physics that they reveal has progressed through advances in numerical methods and scientific computing. In this Review , we provide an overview of theories for angle-resolved photoemission spectroscopy and resonant inelastic X-ray scattering applied to quantum materials. First, we discuss methods for studying equilibrium spectroscopies, including first-principles approaches, numerical many-body methods and a few analytical advances. Second, we assess the recent development of ultrafast techniques for out-of-equilibrium spectroscopies, from characterizing equilibrium properties to generating transient or metastable states, mainly from a theoretical point of view. Finally , we identify the main challenges and provide an outlook for the future direction of the field.
\end{abstract}

\maketitle

New materials are being fabricated on the nanoscale to have surprising performance, such as robust superconductivity in poorly conducting ceramics or at a thin interface a few atoms wide between two electrical insulators -- a golden age of quantum materials. Likewise, the tools in our arsenal have been developing at a rapid pace, and we now have the capacity to measure excitations and dynamics on the fundamental time and length scales of microscopic processes with remarkable resolution -- a golden age of spectroscopy. Angle-resolved photoemission spectroscopy (ARPES), using light from synchrotrons or table-top lasers, can now pinpoint electron dispersions with detailed energy, spin and time resolution. Resonant inelastic X-ray scattering (RIXS) using X-rays from synchrotrons, as well as free-electron lasers in the near future, has revealed bosonic excitations (orbitons, magnons, phonons and other collective modes) with increasingly detailed energy and spin resolution. Moreover, the ability to time-resolve the dynamics of these excitations may soon be possible. ARPES and RIXS, as well as other electron or optical spectroscopies, are providing a wealth of new information about quantum materials.

In one femtosecond, light travels the distance of a human hair. In the same time period, electrons in solids cover a shorter distance of only a few unit cells. These are the natural time and length scales on which the collective behaviour of materials is borne, ultimately determining the functionalities of the materials that shape our world. The field of ultrafast materials science is providing a microscopic view of this world and has opened new windows into our understanding of phenomena such as superconductivity, magnetism and ferroelectricity. The interactions and processes that govern these phenomena occur over timescales from femtoseconds to milliseconds and length scales from nanometres to micrometres. The present challenge is to decipher how the collective motion of $10^{23}$ degrees of freedom gives rise to high-temperature superconductivity, rapid switching in ferromagnets and ferroelectrics, and high-capacity batteries and solar cells with high cyclability. From an experimental perspective, the challenge of covering such a wide range of time and energy scales as well as length and momentum scales has given birth to many of the spectroscopic tools discussed in this Review.

There is an urgent need for advanced theoretical and computational tools to understand and interpret photon spectroscopies, especially photon-in-photon-out scattering and time-domain pump-probe experiments. Developments in theory are moving at a rapid pace, extending many tools for equilibrium spectroscopy into the non-equilibrium domain. These advances are revealing the importance of designing tests of competing theoretical scenarios, developing new numerical techniques, implementing new algorithms and formulating a new language to describe out-of-equilibrium systems for which conventional equilibrium concepts fail. Ultimately, these developments are helping to shape the landscape for more predictive models of novel quantum phenomena and materials. Theory, modelling and interpretation of spectroscopies, especially in the time domain, are needed to extract and therefore exploit the physical and chemical information encoded in the vast volume of experimental data covering energy, momentum, spin and space-time domains across multiple scales. This task requires theories that provide a better treatment of excited-state dynamics, going beyond conventional modelling in terms of ground-state properties (that is, modelling based solely on density functional theory (DFT)).

\begin{figure}[!ht]
\begin{center}
\includegraphics[width=1.05\columnwidth]{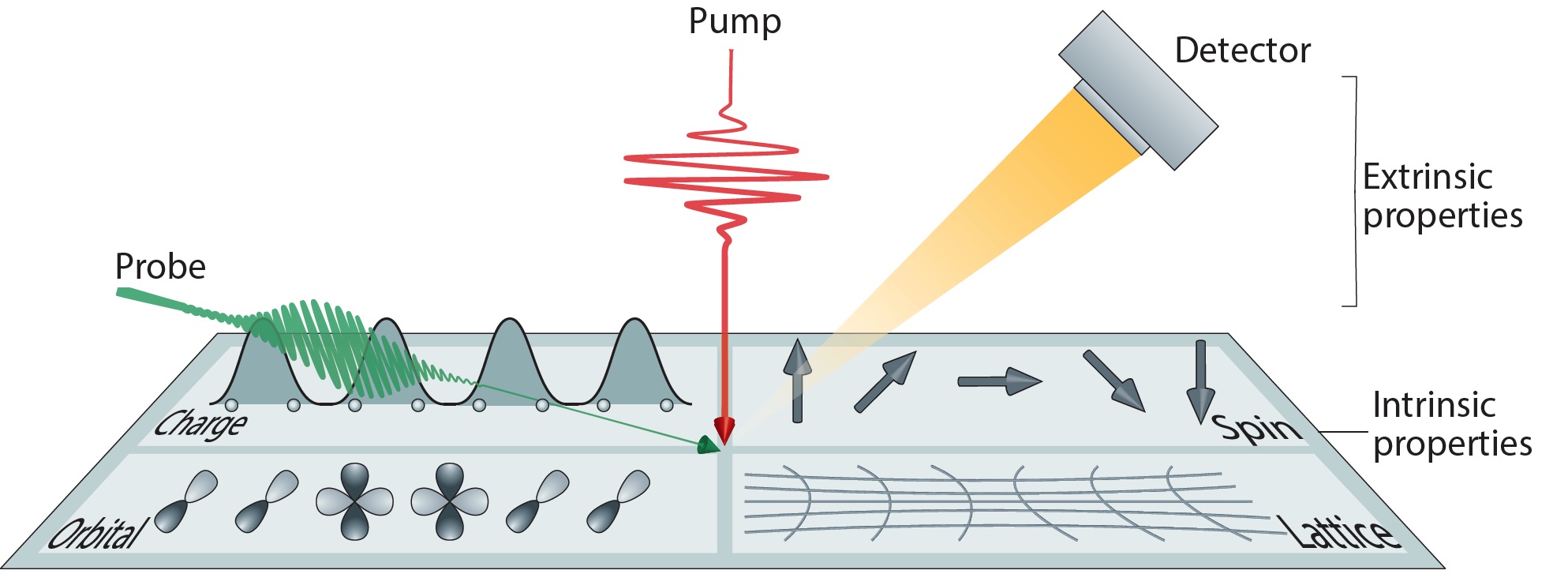}
\caption{\label{cartoon} 
\textbf{Theoretical evaluation of spectroscopies.} The schematic illustrates the two main factors (intrinsic and extrinsic) required for the theoretical evaluation of spectroscopies: first, treatment of the extrinsic measurement details that describe the light-matter interaction, and second, assessment of how the individual and collective degrees of freedom (including charge, spin, orbital and lattice) manifest in the intrinsic physical properties relevant to a specific probe.
}		
\end{center}
\end{figure}

This Review sketches the landscape of theoretical photon-based spectroscopies and outlines advances in our ability to simulate excited-state properties and spectra. New generations of codes and algorithms for the spectroscopy of quantum materials are now available, and hybrid simulations have been designed for exascale computing environments. In this Review, we focus principally on correlated materials, with the degrees of freedom treated equally, both in and out of equilibrium. We further restrict our discussion on equilibrium spectroscopies to ARPES and RIXS, which have improved both experimentally and, more importantly for the purposes of this Review, theoretically in terms of our ability to understand and simulate the spectra for quantum materials. Owing to the nascent development and implementation of out-of-equilibrium spectroscopic techniques -- both experimental and theoretical -- we also discuss the nature of the fundamental physics in the ultrafast regime.

There is an urgent need for advanced theoretical and computational tools to understand and interpret these novel photon spectroscopies, especially photon-in/photon-out scattering and time-domain pump-probe experiments.  These developments in theory are moving at a rapid pace, extending many tools for equilibrium spectroscopy into the non-equilibrium domain. These advances reveal the importance for designing tests of competing theoretical scenarios, developing new numerical techniques and implementing new algorithms, and developing a new language to describe out-of-equilibrium systems where conventional, equilibrium concepts fail.  Ultimately, these developments help to shape the landscape for more predictive models of novel quantum phenomena and materials. Theory, modeling, and interpretation of spectroscopies, especially in the time-domain, are needed to extract, and therefore exploit, the physical and chemical information encoded in the vast volume of experimental data covering energy, momentum, spin, and space/time domains across multiple scales.  This requires theories that go beyond conventional modeling in terms of ground state properties (for example, those based on density functional theory), to treat excited state dynamics.

\section{Equilibrium Spectroscopy Theory}

With the high level of control enabled by modern synchrotrons, the electronic structure of a system can be probed with fine momentum and energy resolution, providing detailed information about the states and collective orders in complex quantum materials. However, there exist significant challenges in deciphering the underlying physics from these measurements (Fig.~\ref{cartoon}). On the one hand, precise treatment of the extrinsic photon-probe processes requires characterization of the photon cross section, matrix elements and excited-state lifetimes. On the other hand, the intrinsic many-body nature of correlated quantum materials means that the relevant physics needs to be disentangled at the microscopic level to determine the influence of charge, spin, lattice and orbital degrees of freedom. Theoretical modelling of many-body systems and their response to various X-ray probes must capture both the intrinsic and extrinsic aspects in an efficient manner. In the following, we review the theoretical progress in photoemission spectroscopy, which provides single-particle information, and X-ray scattering, which provides information on collective excitations from multiple sources.

\subsection{Advances in Theories for Angle-Resolved Photoemission}

A theory for angle-resolved photoemission~\cite{damascelli2003angle,Fadley2010} with the associated degrees of freedom -- light polarization and energy as well as the momentum, energy, orbital and spin of the photoemitted electron -- involves calculation of the photocurrent under the sudden approximation~\cite{Pendry1976,Borstel1985} (Eq.\ref{eq:1}).
\begin{equation}\label{eq:1}
I(\omega) = \sum_{f,i} |\mathcal{V}_{fi}|^2 A_{i}(\omega - \varepsilon_f)
\end{equation} 
The photocurrent is generally expressed in terms of a convolution of matrix elements $\mathcal{V}_{fi}$, which describes the process of exciting an electron from an initial state i to a final state (photoelectron) f with kinetic energy $\varepsilon_f$. The intrinsic electron-removal spectral function ${A}_{i}(\omega - \varepsilon_f)$ (where $\omega$ is the photon energy) contains the single-particle information for each initial state. The matrix elements $\mathcal{V}_{fi}$ encode all the extrinsic factors in the photoemission process, such as the momentum and polarization of the incoming photon and the characteristics of the final state of the photoelectron, while also accounting for photoelectron propagation through the bulk and sample surface~\cite{Pendry1976,Borstel1985}. By contrast, the spectral function ${A}_{i}(\omega - \varepsilon_f)$ contains information about the electronic properties intrinsic to the material under investigation and is generally of primary interest.

\begin{figure}[!ht]
\begin{center}
\includegraphics[width=0.8\columnwidth]{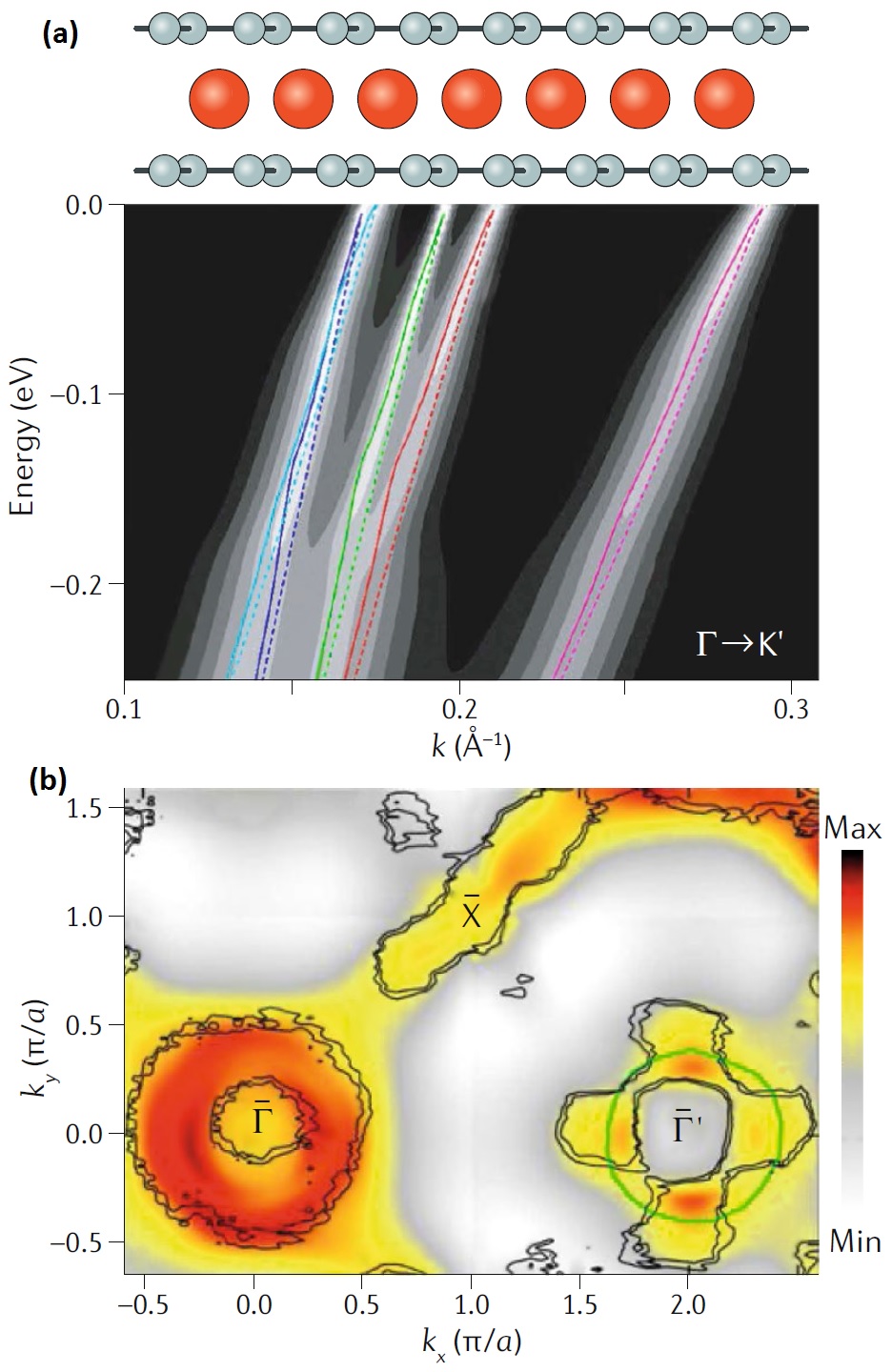}
\caption{\label{lf-evt} 
\textbf{Ab initio evaluation of the electronic structure.} (a) The top part shows the structure of C$_6$CaC$_6$ (C in grey and Ca in red). The bottom part shows the single-particle spectral function of the phonon-driven superconducting bilayer C$_6$CaC$_6$ in the normal state. The \textit{ab initio} band structure, calculated in the Migdal approximation with (solid lines) and without (dashed lines) electron-phonon coupling, is superimposed on the experimental spectral function. The spectral function exhibits clear kinks at a binding energy of 180\,meV. (b) Fermi surface cuts across several Brillouin zones for (Ba$_{0.6}$K$_{0.4}$)Fe$_2$As$_2$. The surface cuts were calculated from first principles with the local density approximation combined with dynamical mean-field theory to simulate a one-step model for angle-resolved photoemission spectroscopy at an energy of 75\,eV. The black lines correspond to the experimental data. $a$, lattice constant; $k$, momentum. Panel (a) is adapted from Ref.~\onlinecite{Margine2016}, CC-BY-4.0. Panel (b) is adapted from Ref.~\onlinecite{Derondeau2017}, CC-BY-4.0.
}		
\end{center}
\end{figure}

In most materials in which correlation effects are negligible or can be treated perturbatively, first-principles approaches can accurately describe the single-particle spectral functions relevant to ARPES. For systems in the ground state, improvements in the efficiency of algorithms\cite{Umari2010,Lambert2013,Liu2016} related to the GW (where $G$ is the single-particle Green's function and $W$ is the screened Coulomb interaction) method\cite{Hedin1965,Hedin1999} for the electronic self-energy have enabled routine simulations of quasiparticle band structures in a wide variety of materials, such as semiconductors and topological insulators\cite{Reining2017}.

Methodologies within the GW paradigm that account for both electron-electron and electron-phonon self-energy corrections have also been incorporated in band structure calculations\cite{Bernardi2014,Story2014,Ponce2016}. These corrections, in conjunction with efficient evaluation of electron-phonon couplings, enable \textit{ab initio} investigation of phonon-mediated superconductivity in the Migdal-Eliashberg framework\cite{Ponce2016,Margine2016,Sanna2018} [see Fig.~\ref{lf-evt}(a)]. Furthermore, generalizations of GW theory, in particular the GW plus cumulant formalism, have been actively researched with the aim of correctly reproducing quasiparticle renormalization and satellite features induced by electron-boson coupling\cite{Guzzo2011,Lischner2013,Story2014,Kas2014}.

For correlated systems, the local density approximation (LDA) plus dynamical mean-field theory (DMFT)\cite{Anisimov2014} (LDA + DMFT) has been widely applied to several classes of materials, such as transition metal oxides and f-electron materials. This composite approach typically uses Wannier downfolding\cite{Mostofi2014} schemes to map the lattice problem onto a correlated single-site problem embedded within a dynamical Weiss field. Notably, LDA + DMFT has been incorporated into detailed mechanistic ARPES simulations within the relativistic Korringa-Kohn-Rostoker (KKR) multiple scattering (MS) framework in order to capture self-energy effects in the initial state\cite{Minar2011}. Additionally, extensions that combine the first-principles GW method with variants of DMFT have also been developed or proposed for correlated materials\cite{Biermann2014}.
 
Starting from \textit{ab initio}-derived models of Wannier orbitals and the corresponding matrix elements, more sophisticated many-body methods can be adopted in the calculation of ARPES spectra, including exact diagonalization (ED)\cite{dagotto1994correlated}, quantum Monte Carlo (QMC)\cite{foulkes2001quantum}, the density matrix renormalization group (DMRG)\cite{schollwock2005density}, cluster perturbation theory\cite{senechal2000spectral}, the dynamical cluster approximation\cite{hettler2000dynamical} and the variational cluster approximation\cite{potthoff2003variational}. These numerical approaches typically treat the many-body effects more precisely than Hartree-Fock. For example, the strong correlation-induced `high-energy anomaly' in ARPES has been successfully characterized by ED\cite{zemljivc2008temperature}, the dynamical cluster approximation\cite{macridin2007high}, QMC\cite{moritz2009effect} and cluster perturbation theory\cite{wang2015origin}. These many-body approaches, after modifications, can be extended to multiparticle scattering, as discussed below.

In addition to the intrinsic spectral function, a reliable interpretation of ARPES spectral intensities requires matrix-element effects\cite{Bansil1999,Minar2011,Okuda2013} to be taken into account. Among the first-principles methods currently available, the KKR-MS approach\cite{Korringa1947,Kohn1954} is well suited for this purpose and has been routinely adopted for interpreting ARPES data in a wide variety of quantum materials, such as high-temperature superconductors\cite{Bansil1999,Bansil2012} and, more recently, topological insulators and semimetals\cite{Scholz,Sanchez-Barriga2014}. A recent combined experimental and theoretical ARPES study on the pnictide superconductor Ba$_{1-x}$K$_x$Fe$_2$As$_2$ illustrates the success of the KKR-MS approach\cite{Derondeau2017}; the reported ARPES simulations exhibited good agreement with experiment in terms of the photon energy and polarization dependence [see Fig.~\ref{lf-evt}(b)]. More precise treatment of light-matter interactions involving quantum electrodynamics has also been developed in the first-principles framework, providing an option for describing exotic spectral properties within an optical cavity\cite{ruggenthaler2018quantum}.
  
\subsection{Theoretical Approaches of Resonant Inelastic X-ray Scattering}

Multiparticle processes in correlated materials encode information on collective excitations, and this information can be revealed using various photon-in-photon-out X-ray scattering probes. Among these scattering approaches, RIXS is a rapidly developing and expanding technique that enables an understanding of low-energy excitations in a wide range of materials. The element specificity and electronic-state selectivity of RIXS can be tuned by changing the incident photon energy, the photon polarizations and the relative momentum transfer in the material, making it a powerful method for interrogating specific excitations of interest\cite{Kotani2001, Ament2011, Ishii2017}. In such a resonant process, the intermediate many-body state has a significant role and therefore usually cannot be treated simply with mean-field approaches or DFT.

\begin{figure*}
\includegraphics[width=2\columnwidth]{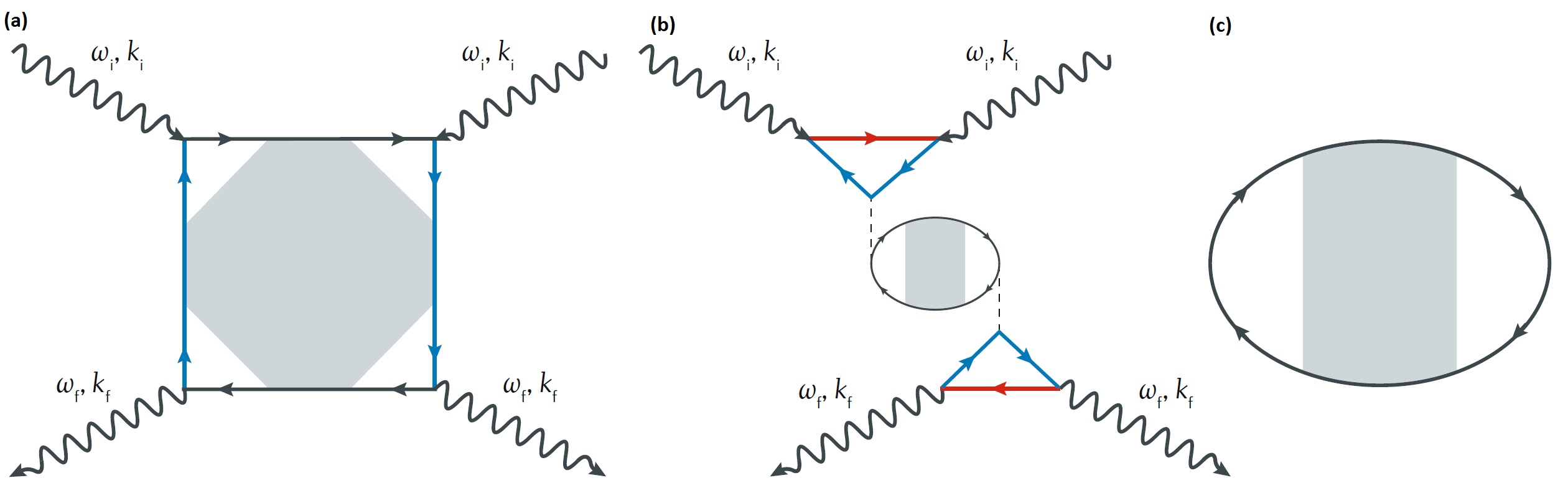}
\caption{\label{fig:fd}  
\textbf{Feynman diagrams for RIXS processes and approximations.} In these diagrams, the wavy lines represent photon propagators; the blue and black lines represent the core level and valence electron Green's functions, respectively; the red lines represent the Green's function for electronic states in bands into which core-level electrons are excited in an indirect resonant inelastic X-ray scattering (RIXS) process. (a) Full RIXS cross section for direct RIXS. The photon-in and photon-out dipole processes are indicated by the vertices; with interactions included, the cross section represents an irreducible four-point correlation. (b) Full RIXS cross section for indirect RIXS. As the core electron is excited to a high-energy state, the valence electrons contribute to the cross section through an effective two-point correlation. The dashed lines indicate the interaction between the core hole and valence electrons. (c) Charge or spin dynamical structure factor. $\omega_{\rm i}$ and $\omega_{\rm f}$, represent the energy of the incident and emitted photon, respectively; $k_{\rm i}$ and $k_{\rm f}$ represent the momentum of the incident and emitted photon, respectively.
}
\end{figure*}

RIXS has proved to be an effective tool in probing the excitations in transition metal systems, which typically contain many intertwined degrees of freedom, including charge, spin, orbitals and phonons, in the low-energy range. In the study of high-temperature superconducting materials, such as cuprates, RIXS was first used at the Cu $K$-edge to detect charge excitations\cite{Hill1998, Abbamonte1999} and then extended to the Cu $L$-edge to detect mainly spin excitations as well as orbitals and charge-transfer excitations\cite{Braicovich2009, Schlappa2009, Tacon2011, Schlappa2012, Dean2013b}. Spin excitations are accessible in the RIXS process at the Cu $L$-edge because the spin is coupled to the 2$p$ core orbital and accordingly breaks the valence spin conservation rule. In addition, RIXS has been applied at the O $K$-edge\cite{Bisogni2012a, Kim2015} and $M$-edge\cite{Ishii2017} of cuprates and used in the study of other correlated systems; for example, RIXS has been used to measure spin excitations in Fe-based superconductors\cite{Zhou2013}, to explore spin and orbital degrees of freedom in iridates\cite{Sala2014, Yin2013} and to determine the hidden order in URu$_2$Si$_2$ (Ref.~\onlinecite{Wray2015}). As RIXS provides a `fingerprint' of the electronic state of a system, it has also been used in the chemistry community for the study of transition metal complexes\cite{Groot2005, Wernet2015}.

The momentum-dependent RIXS cross section can be expressed using the Kramers-Heisenberg formula\cite{KramersHeisenberg}
\begin{equation}\label{RIXS}
I(\mathbf{q}, \Omega, \omega_i)
= \frac{1}{\pi} \mathrm{Im} \Big\langle \Psi \Big| \frac{1}{\mathcal{H} - \mathit{E}_0 -\Omega  - \mathit{i} 0^+} \Big| \Psi\Big \rangle
\end{equation}
and 
\begin{equation}\label{psi}
\Big| \Psi \Big\rangle = \sum_{j,\sigma} e^{i\mathbf{q}\cdot\mathbf{r}_j} \mathcal{D}_j^{\dagger} \frac{1}{\mathcal{H}_j^{\prime}-\mathit{E}_0-\omega_i-\mathit{i}\Gamma} \mathcal{D}_j \Big| 0 \Big\rangle,
\end{equation}
where $\mathbf{q}$ is the momentum transfer; $\omega_i$ is the incident photon energy; $\Omega$ is the energy transfer ($\Omega = \omega_{\rm i}- \omega_{\rm f}$;
where $\omega_{\rm f}$ is the energy of the emitted photon); $\Gamma$ is
the inverse core-hole lifetime; and $E_G$ and $|G\rangle$ are the ground-state energy and wavefunction, respectively. Here, $\mathcal{H}^{\prime}_j$ represents the intermediate-state Hamiltonian, which contains the interactions induced by the core hole; $\mathcal{D}_j$ is the dipole transition operator with a specific X-ray absorption edge; $\mathbf{r}$ is the electron position; $\sigma$ is the spin; and $j$ is the site index. In the direct RIXS process, an incoming photon excites a dipole transition from a core level to a valence level, whereas in the indirect RIXS process, the dipole transition occurs between a core level and a level much higher than the valence, and the Coulomb attraction from the core hole acts on the valence electrons [see Fig.~\ref{fig:fd}]. For systems with strong correlation effects, the Hilbert space dimension for many-body states exponentially increases with the system size, making full evaluation challenging or even impossible.

One way to tackle the challenge of full evaluation is to develop algorithms that embed both symmetry reduction and large-scale parallel computing techniques\cite{Jia2017} and to evaluate the Kramers-Heisenberg formula explicitly using ED. In pioneering theoretical calculations for RIXS on 2D cuprates with the single-band\cite{Tsutsui1999, Tsutsui2003} and three-band\cite{Chen2010} Hubbard model, the momentum and doping dependencies of Cu $K$-edge RIXS were calculated. The calculated momentum dependence and resonant profile showed good agreement with experiment. For example, the RIXS spectrum evaluated by ED for an undoped cuprate displays a feature at 4.7eV for zero momentum transfer, which corresponds to the charge-transfer energy in this material [see Fig.~\ref{fig:fd}(a)]. The calculations for various doping levels demonstrated that the screening effect for the intermediate states is crucial for the accurate evaluation of RIXS spectra.

\begin{figure}
\includegraphics[width=9cm, height=6cm]{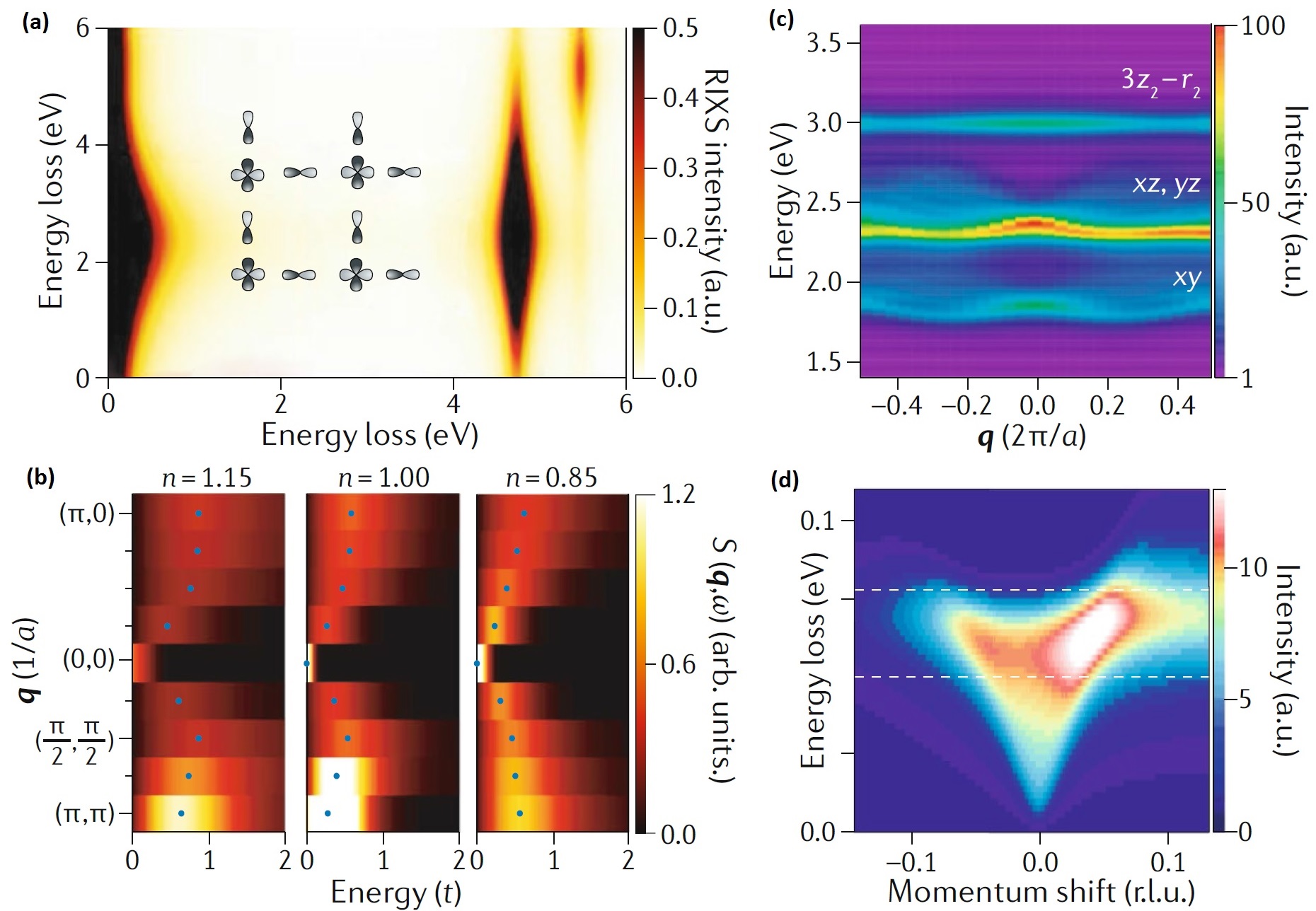}
\caption{\label{fig:results}  
\textbf{Theoretical simulations of RIXS highlighting various elementary excitations in cuprates.} (a) Charge-transfer excitations in Cu $K$-edge resonant inelastic X-ray scattering (RIXS) calculated using the 2D three-band Hubbard model. The inset shows the orbital geometry of the small 2$\times$2 cluster with four CuO$_2$ basis elements used in the calculation. (b) Paramagnon excitations in Cu $L$-edge RIXS, approximated here by the spin dynamical structure factor $S(\mathbf{q},\omega)$ (where $\mathbf{q}$ is the momentum transfer and $\omega$ is the excitation energy), simulated in the 2D single-band Hubbard model for various carrier concentrations ($n$). (c) Orbiton dispersions in a quasi-1D cuprate (SrCuO$_3$) calculated using a spin-orbiton model. (d) RIXS intensity of phonon excitations in a 1D system with a charge-density wave instability and strong electron-phonon coupling, calculated in the Migdal approximation. The momentum shift is calculated relative to 2$k_F$ (where $k_F$ is the Fermi momentum). $a$, lattice constant; r.l.u., reciprocal lattice units; $t$, hopping parameter. Panel (a) is adapted with permission from Ref.~\onlinecite{Chen2010}, American Physical Society. Panel (b) is adapted from Ref.~\onlinecite{Jia2014}, Springer Nature Limited. Panel (c) is adapted from Ref.~\onlinecite{Schlappa2012}, Springer Nature Limited. Panel (d) is adapted from Ref.~\onlinecite{chaix2017dispersive}, Springer Nature Limited.
}
\end{figure}

Theoretical studies have also established the connection between Cu $K$-edge\cite{Jia2012} and $L$-edge RIXS\cite{Jia2014, Jia2016} and the corresponding dynamical charge and spin structure factors. Such a connection provides a cheaper alternative to evaluating the more complicated four-particle RIXS diagram, which can instead be approximated by a two-particle correlation function under certain conditions, such as using different incoming and outgoing polarization combinations, specific incoming photon energies tuned to different X-ray edges and specific doping concentrations. Although the complicated full RIXS cross section can thus far be treated only diagrammatically or calculated using ED, correlation functions can also be evaluated for a much finer momentum grid with other numerical tools, such as QMC. The QMC-evaluated dynamical spin and charge structure factor $S(\textbf{q},\omega)$ captures the spin excitations across both electron and hole dopings\cite{Haverkort2010, Kourtis2012, Jia2014} [see Fig.~\ref{fig:results}(b)], including the hardening of the paramagnon energy for electron-doped cuprates and the persistence of the excitation energy for hole-doped cuprates around $(\pi,0)$, consistent with measured RIXS data\cite{Braicovich2009, Schlappa2009, Tacon2011, Schlappa2012, Dean2013b}. This consistency makes RIXS a complementary tool to inelastic neutron scattering for studying the momentum dependence of spin excitations. Compared with traditional approaches, RIXS requires much smaller sample volumes, enabling a wider range of materials to be investigated.

The endeavour to connect RIXS with correlation functions dates back to the 1990s. Fast collision approximations were first used to connect RIXS with $S(\textbf{q},\omega)$ based on the assumption that the dynamics need be considered only at the site of the core hole in the intermediate state\cite{Luo1993, Groot1998}. Later on, the ultrashort core-hole lifetime (UCL) expansion was introduced for both indirect and direct RIXS processes\cite{Ament2011, Bisogni2012a}. Under UCL, it has been shown that an indirect RIXS cross section can be reduced to the charge dynamical structure factor $N(\textbf{q},\omega)$, whereas a direct RIXS cross section in the spin-flip channel can be reduced to the spin dynamical structure factor. In higher-order expansions, it is also possible to map onto bimagnons\cite{Bisogni2012b}. Although UCL is powerful in deriving effective RIXS cross sections, caution must be taken in using it for specific cases, as convergence of the UCL expansion may break down when the intermediate-state energy manifold is not much smaller than the inverse core-hole lifetime $\Gamma$.

To test the validity of these downfolding approaches, the full RIXS cross section can be directly evaluated and compared with calculated dynamical structure factors on an equal footing. It was demonstrated that indirect RIXS is consistent with the charge dynamical structure factor when the screening effect does not have a role in the intermediate state [see Fig.~\ref{fig:fd}(b)]: by neglecting the core-hole Coulomb attraction (represented by the dashed lines), the Feynman diagram for indirect RIXS can be simplified to dynamical charge structure factors [see Fig.~\ref{fig:fd}(c)]. Moreover, it has been shown that direct RIXS in the spin-flip channel can be mapped onto the dynamical spin structure factor on a qualitative and semi-quantitative level, although the connection becomes less precise for doped models\cite{Jia2014}, while in the nonspin-flip channel, the cross section can be mapped onto a projected dynamical charge structure factor only qualitatively\cite{Jia2016}. The failure of the above attempts to simplify the description in terms of two-particle correlation functions reflects the inherent complexity of the RIXS process.

The connection between final and initial states through a complicated process involving intermediate states means that certain dipole-forbidden excitations, including orbital excitations (for example, $d$-$d$ excitations in transition metal compounds\cite{butorin1996low,ghiringhelli2004low}) and even excitations involving the lattice degree of freedom\cite{lee2013role,peng2015magnetic}, are active in RIXS. The simulated RIXS spectra from a simple spin-orbital model [see Fig.~\ref{fig:results}(c)] were shown to be consistent with experimental results demonstrating spin-orbiton coupling in the 1D cuprate Sr$_2$CuO$_3$ (Ref.~\onlinecite{Schlappa2012}), and RIXS simulations and experiments have characterized phonon modes and their interplay with the charge degrees of freedom in quantum materials at low energies\cite{chaix2017dispersive} [see Fig.~\ref{fig:results}(d)].

\begin{figure*}[ht!]
\begin{center}
\includegraphics[height=5cm, width=10cm]{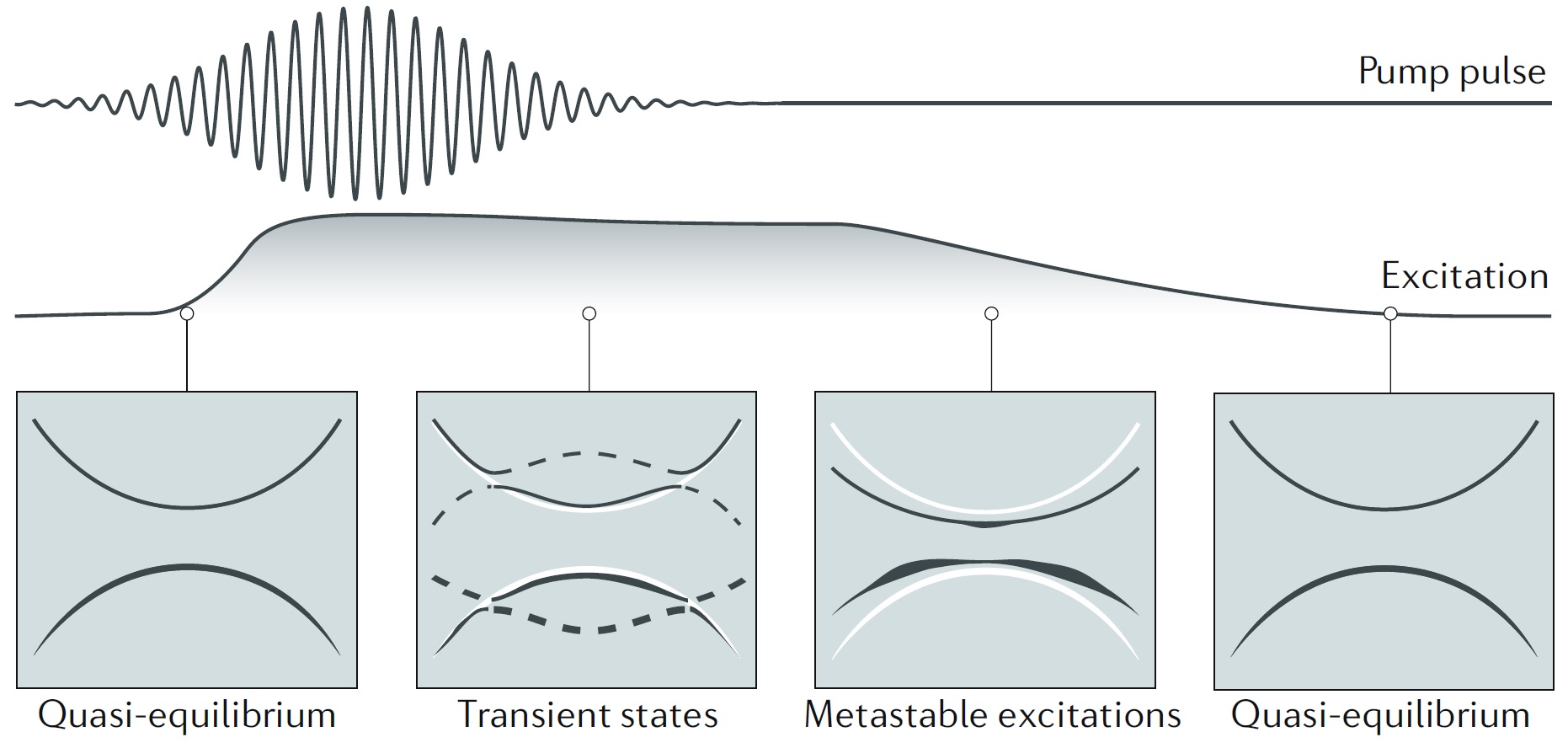}
\caption{\label{fig:noneqcartoon}
\textbf{Accessing physics out of equilibrium.} Schematic illustrating a pump pulse and the excitation profile as a function of time. A time-resolved measurement can reveal distinct non-equilibrium regimes as a function of the pump-probe delay , which can in turn provide information about both equilibrium and transient states. For example, it is possible to uncover intertwined or subleading equilibrium orders or low-energy excitations that may be easier to resolve in the time domain; create novel states of matter that do not have an equilibrium analogue; manipulate metastable states and tip the balance between competing orders. These phenomena are approximately separated in time during the pump-probe process and also differ in their deviation from equilibrium. The insets show the single-particle spectra typical for each dynamical regime during the pump. The solid black lines represent the instantaneous electron and hole distribution, the white lines denote the equilibrium bands, and the dashed lines denote the transient Floquet sidebands. The initial stages of the pump can be understood within the linear response, providing a snapshot of quasi-equilibrium behaviour. When the system is highly excited, a transient modification of both the Hamiltonian and the distribution dominates the physics. Afterwards, the system may access metastable states with pre-thermal excitations before ultimately returning to quasi-equilibrium in the final stages.
}
\end{center}
\end{figure*}
\section{Non-equilibrium spectroscopy theory}

Time adds a new dimension to the study of quantum materials. By using this extra dimension, it is possible to directly access excited states and non-equilibrium dynamics as a means to decipher underlying equilibrium properties -- that is, unoccupied states, certain elementary excitations and excited-state and quasiparticle lifetimes. The complex nature of quantum materials often makes these properties difficult to measure directly or to distinguish easily, especially at low energies, using established equilibrium techniques. Remaining close to equilibrium, within or just beyond the regime applicable to linear-response theory, while accessing this new dimension of time requires low pump fluences, which can be observed at the very beginning or end of a typical pump envelope (see `Quasi-equilibrium' in Fig.~\ref{fig:noneqcartoon}).

By pushing a quantum material out of equilibrium, novel states of matter can be stabilized, such as those that have no equilibrium analogue or may not be easily accessible by standard approaches of chemical substitution. These exotic states may emerge from the light-matter interactions through precise engineering of new terms or modification of existing terms in the many-body Hamiltonian. By modulating or manipulating the eigenstate manifold, exotic states can be stabilized through strong pump fields or pump fields that persist over a sufficient period of time to enable the formation and resolution of distinguishing characteristics and features of these states (see `Transient states' in Fig.~\ref{fig:noneqcartoon}).

The complex interplay between multiple degrees of freedom in quantum materials often gives rise to cooperating and competing phases. A single phase typically becomes dominant for a given set of equilibrium parameters, such as chemical composition, temperature or pressure. However, pump-probe time-domain techniques can be used to alter the sometimes delicate balance between intertwined orders and to manipulate the physics enough to reveal an exotic phase with subleading character that would otherwise not be expressed in equilibrium. Although the non-equilibrium virtual states that mediate such a process may be short-lived owing to the transient nature of the pump field, the exotic phase and its signatures may be metastable or at the very least detectable for some time in the immediate aftermath of the pump pulse or in its tail (see `Metastable excitations' in Fig.~\ref{fig:noneqcartoon}).

The non-equilibrium numerical techniques on which this Review focuses fall broadly into methods based on either the wavefunction and density matrix or a Green's function formalism. The former includes Krylov-subspace ED\cite{balzer2011krylov}, variational ED\cite{bonvca1999holstein}, dynamical DMRG and matrix-product state methods\cite{schollwock2011density}. Each, in some form, tracks the evolution of a time-dependent state, treated as either a vector or density matrix in Hilbert space, and measures all observables based on that state. The exponential increase in Hilbert space dimension with system size typically limits these studies to small clusters. By contrast, methods based on Green's functions describe the evolution of observables expanded on a generalized Green's function. Time-dependent Hartree-Fock, DMFT\cite{aoki2014nonequilibrium} and cluster perturbation theory\cite{balzer2011nonequilibrium} techniques belong to this class. These methods are less restricted by the Hilbert space dimension but cannot describe all many-body observables with similar accuracy.

Although there are currently only a few, the number of \textit{ab initio} simulations of single-particle spectral functions and photoemission spectra in quantum materials out of equilibrium is expected to grow in the near future owing to recent methods development\cite{Spataru2004,Park2009a,Perfetto2016b,Braun2016}. The methods and attempts at simulation include, but are not restricted to, a non-equilibrium generalization of the GW approach\cite{Spataru2004} and a first-principles lesser Green's-function approach\cite{Perfetto2016b}. Real-time, time-dependent, DFT-based approaches within the adiabatic local spin-density approximation\cite{Marques2012a} have also been adopted for simulating spin- and time-resolved ARPES\cite{Wopperer2017}.

\subsection{Characterizing hidden equilibrium properties}
Taking advantage of additional information afforded in the time domain, non-equilibrium approaches can disentangle and reveal otherwise obscured parameters of quantum materials. For example, time-domain techniques have been used widely to determine quasiparticle relaxation and infer lifetimes in the recombination of particle-hole pairs from semiconductors and superconductors\cite{yusupov2010coherent, smallwood2012tracking, dal2015snapshots, vishik2017ultrafast}. A theoretical understanding of these phenomena has come from a microscopic description that treats electron-phonon and electron-electron interactions\cite{eckstein2011thermalization,sentef2013examining, murakami2015interaction,strand2017hund}, tying the relaxation and lifetimes to the effective scattering rates due to these interactions. More recently, a combination of two time-domain spectroscopic techniques has made strides in elucidating the electron-phonon coupling strength in FeSe (Ref.~\onlinecite{gerber2017femtosecond}); the extracted deformation potential is consistent with theoretical predictions from an equilibrium DFT+DMFT approach\cite{mandal2014strong}. Deciphering material-specific properties using these time-domain methods can help to test and correct intuitive models and theoretical descriptions.

\begin{figure}[!ht]
\begin{center}
\includegraphics[height=14.2cm, width=7.5cm]{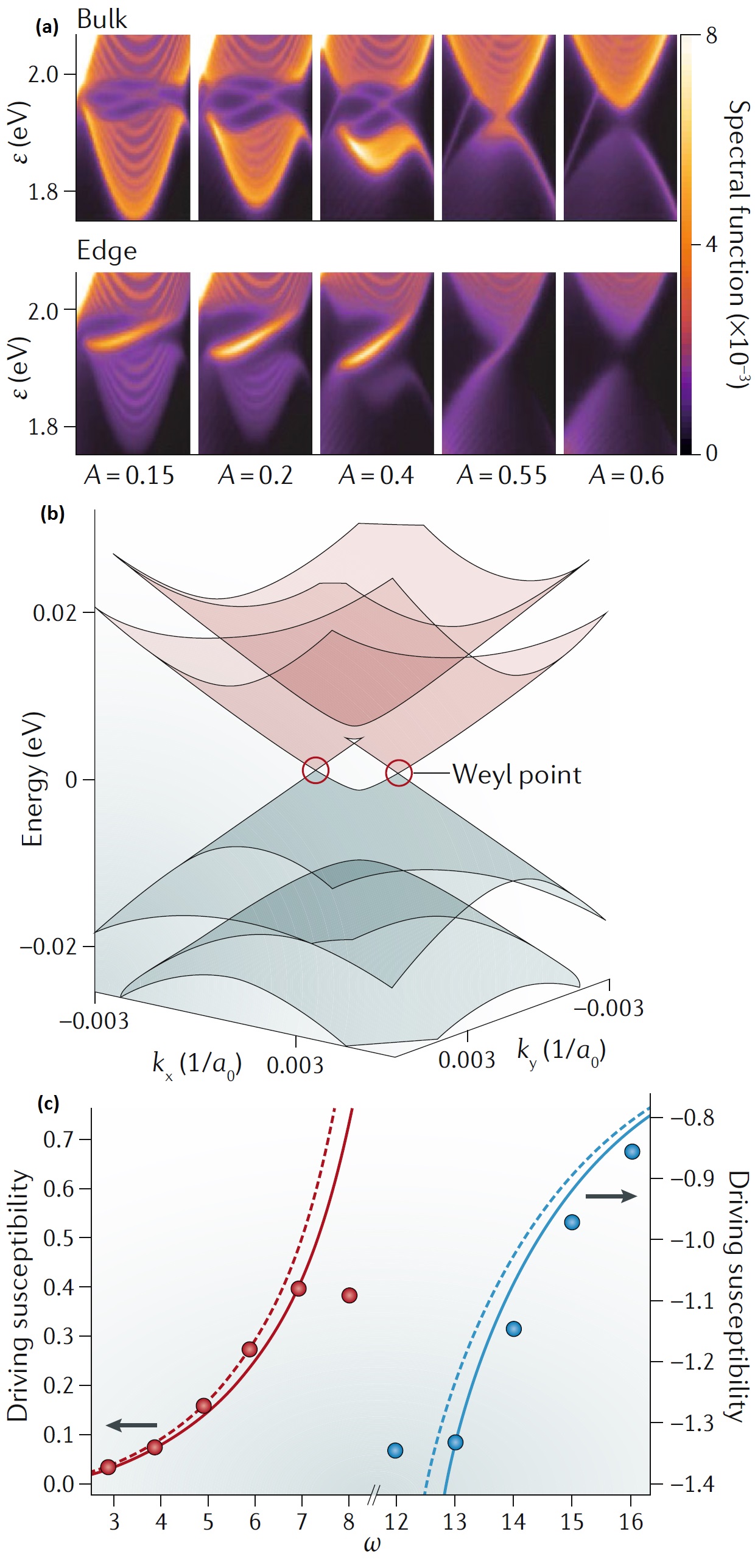}
\caption{\label{fig:floquet}
\textbf{Pumped-induced Floquet physics.} (a) Floquet physics in a semiconductor. A red-detuned pump field induces a hybridization gap at the bottom of the conduction band in bulk WS$_2$, whereas a conducting state forms in this gap at the edge of a finite-sized sample in a ribbon geometry. A is the strength of the pump field in natural units of the simulation, and $\varepsilon$ is the electron energy. (b) Floquet physics in a semimetal. When driven by a periodic pump field, Floquet-Weyl points form 3D cones in the Brillouin zone of Na$_3$Bi. The band structure was calculated using density functional theory and is shown here in the $k_x$-$k_y$ plane. (c) Floquet physics in a correlated Mott insulator. The driving susceptibility $\Delta J_{\rm ex}/(J_{\rm ex}A^2)$, where $J_{\rm ex}$ is the exchange interaction, for frequencies $\omega$ evaluated in the Hubbard model above (blue) and below the Mott gap (red), obtained from dynamical mean-field theory (circles), from the numerical Floquet spectrum of a two-site cluster (solid lines) and from perturbation theory (dashed lines). $a_0$, lattice constant. Panel (a) is adapted from Ref.~\onlinecite{claassen2016all}, CC-BY-4.0. Panel (b) is adapted from Ref.~\onlinecite{hubener2017creating}, CC-BY-4.0. Panel c is adapted from Ref.~\onlinecite{mentink2015ultrafast}, CC-BY-4.0
}
\end{center}
\end{figure}

Basic physical properties in condensed-matter systems are usually directly tied to the electronic structure. With unique developments and improvements in ARPES\cite{damascelli2003angle}, the single-particle spectral functions can be measured with fine momentum and energy resolutions. However, this information is available only for the occupied single-particle states with similar information about the unoccupied band structure not as easily accessible owing to the much poorer resolution and cross section for inverse photoemission. Non-equilibrium spectroscopic approaches, such as two-photon photoemission, seek to circumvent this problem. A fraction of electrons can be excited to unoccupied states above the Fermi energy, and subsequent photoexcitation liberates these electrons. This technique of two-photon photoemission has been used to elucidate the unoccupied states of topological insulators (from Be$_2$Se$_3$ and Be$_2$Te$_3$ families)\cite{sobota2012ultrafast,niesner2012unoccupied, sobota2013direct, sobota2014distinguishing, sobota2014ultrafast}, which can be well captured by DFT methods. The spin textures that accompany these states can be mapped owing to experimental advances in resolving the electronic spin for ARPES, and the observed spin-orbital locking can be captured in DFT simulations\cite{jozwiak2016spin}. Compared with weakly correlated topological insulators, the characterization of unoccupied states in correlated systems is more challenging\cite{sonoda2004unoccupied}. However, in a recent study, the unoccupied states of cuprates were successfully characterized through the comparison with many-body numerical calculations\cite{yang2017revealing}.

Beyond single-particle states, non-equilibrium approaches have been applied to the characterization of collective excitations. Differences in the relaxation time structure of pump-probe dynamics allow for characterization of charge-density waves, spin fluctuations and phonon degrees of freedom\cite{hellmann2012time,conte2012disentangling,yang2015thickness}, providing a tool to disentangle intertwined degrees of freedom in a correlated system. With fine control of the ultrafast pump, the `Higgs' mode, or amplitude mode, in superconductors has been identified in NbN (Refs.~\onlinecite{matsunaga2013higgs,matsunaga2014light,sherman2015higgs}) and validated by comparison with mean-field theory\cite{bittner2015leggett,krull2016coupling} and by microscopic calculation of photoemission spectroscopy in the superconducting state\cite{nosarzewski2017amplitude,kemper2015direct}. Additional theoretical work has demonstrated how ultrafast approaches can not only separate equilibrium modes for cleaner detection, but can also be used to characterize the intertwined nature of the degrees of freedom near a quantum phase transition where fermionic and bosonic excitations become entangled\cite{wang2016using}.

\subsection{Creating novel transient states of matter}

The most intriguing aspect of non-equilibrium approaches might be the potential to exploit the change in electronic or magnetic dynamics during the pump pulse to achieve exotic states of matter that have no equilibrium analogue. Consider a wide pump pulse such that the combined Hamiltonian of the system and this non-equilibrium driving field approximately obey discrete time translation symmetry over a wide time window. In this regime, the transient dynamics can be mapped onto an effective static eigenproblem by virtue of Floquet's theorem. The resulting ladder of virtual states represents a steady-state solution to the problem given a continuous driving field with a well-defined frequency, and this steady-state approximation will apply over a range of times near the centre of relatively wide pump pulses in time-resolved pump-probe schemes.

Experimentally, in materials, the photon-dressed sidebands and dynamical symmetry breaking associated with this Floquet physics were first observed directly in the single-particle spectrum of the surface state of a topological insulator using time-resolved ARPES (trARPES)\cite{wang2013observation,mahmood2016selective}. In general, the effective Floquet-Bloch band structures of a driven Dirac fermion\cite{oka2009photovoltaic} can entail a change in band topology\cite{inoue2010photoinduced, kitagawa2011transport, lindner2011floquet}; this was first proposed for graphene\cite{sentef2015theory} and then extended to monolayer transition metal dichalcogenides\cite{claassen2016all} [see Fig.~\ref{fig:floquet}(a)]. More recently, this type of dynamical Floquet engineering has been adopted numerically to create a Weyl semimetal by implementing a version of Floquet time-dependent DFT\cite{hubener2017creating} [see Fig.~\ref{fig:floquet}(b)].

\begin{figure*}[ht!]
\begin{center}
\includegraphics[width=18cm]{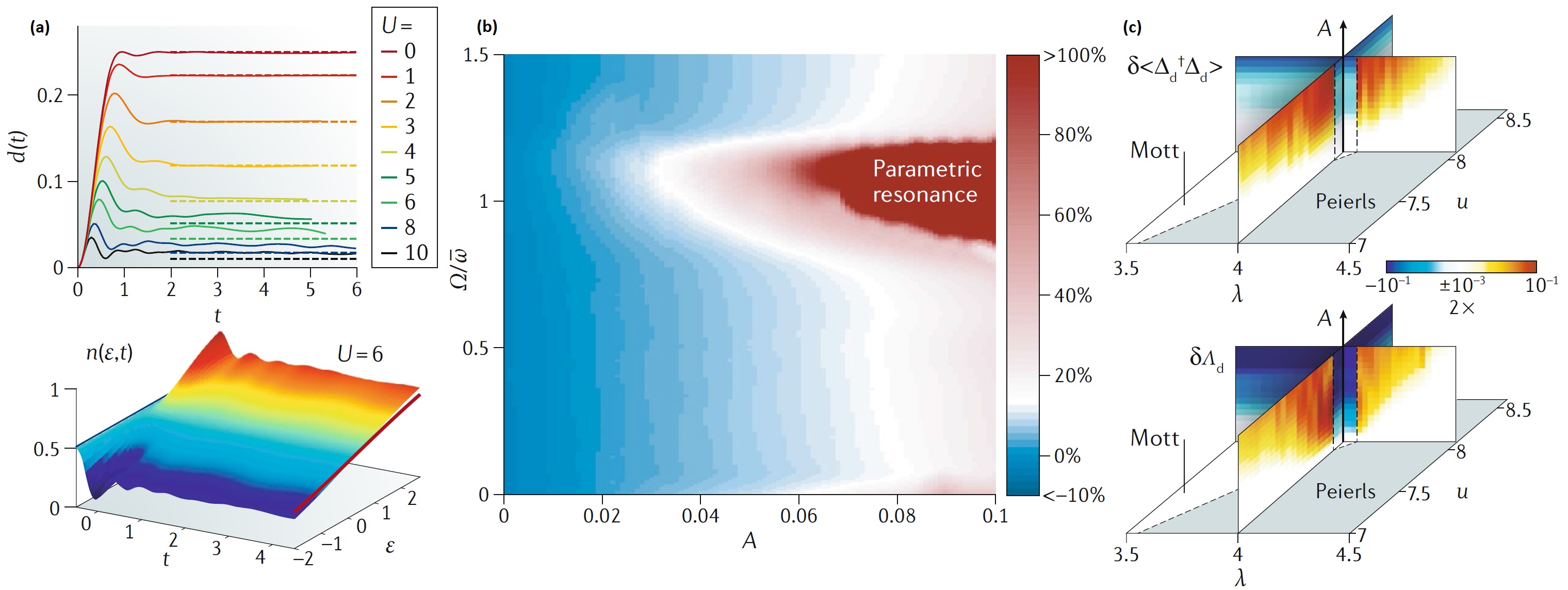}
\caption{\label{fig:controlOrders}
\textbf{Non-equilibrium excitations and phase change.} (a) Electronic excitations induced by a quantum quench of a N\'{e}el system. The time ($t$) evolution of the electron double occupancy $d(t)$ for different values of the Hubbard interaction $U$ (top), and the evolution of the electron distribution as a function of energy ($\varepsilon$), $n(\varepsilon,t)$, for $U$=6 (bottom). (b) Superconductivity induced by parametrically driven phonons. The plot shows the relative change in the superconducting transition temperature with respect to equilibrium in an electron-phonon system as a function of the pump frequency $\Omega$ and the driving amplitude $A$. The data are evaluated for linearly dispersing phonons with mean frequency $\bar{\omega}$, relative spread $\omega/\bar{\omega}$=0.2 and negative quartic couplings between Raman and infrared-active modes. (c) Unconventional superconductivity and spin fluctuations induced by a pulse pump. Change in the d-wave pairing correlation ( $\langle \Delta_d^\dagger \Delta_d\rangle$ , top) and projected spin fluctuations ($\Lambda_d$, bottom) evaluated for various pump strengths $A$ and model parameters (dimensionless electron-phonon coupling strength $\lambda$ and electron-electron coupling strength $u$) near the boundary between the Mott and Peierls phases. The calculation is based on a Hubbard-Holstein model, with the pump field coupled directly to the electrons using a Peierls substitution. Panel (a) is adapted from Ref.~\onlinecite{balzer2015nonthermal}, CC-BY-3.0. Panel (b) is adapted with permission from Ref.~\onlinecite{knap2016dynamical}, American Physical Society. Panel (c) is adapted with permission from Ref.~\onlinecite{wang2017light}, American Physical Society.
}
\end{center}
\end{figure*}

Although a great deal of effort has been devoted to Floquet studies in systems with negligible or weak interactions, the possibility of manipulating transient dynamics in strongly interacting systems is an enticing prospect. To achieve this, it is essential to not only introduce the pump frequency as a new energy scale for determining the one-particle and two-particle response functions, but also to `reshape' the underlying interacting Hamiltonian to stabilize phases of matter that might be inaccessible at equilibrium. The central challenge is to understand and control both the effective transient dynamics that are determined by the pump plateau and the effective distribution that is set by the transient envelope and relaxation processes. Theoretically, arguments based on the eigenstate thermalization hypothesis entail that driven systems continuously absorb energy and heat to infinite temperature\cite{DAlessioPRX2014,LazaridesPRE2014} unless the system is integrable or many-body localized\cite{DAlessioAnnPhys2013,PontePRL2015,LazaridesPRL2015}. However, it has been shown that long-lived `Floquet pre-thermal' regimes with effective engineered local Hamiltonians can persist for a suitable separation of energy scales between materials degrees of freedom and the external drive\cite{CanoviPRE2016,kuwahara2016floquet,MoriPRL2016,bukov2015prethermal,AbaninPRL2015,HoARXIV2016}. The simplest example is the single-band Hubbard model, for which perturbation theory leads to a renormalized spin exchange $J_{\rm ex}(A)/J_{\rm ex}(A\!=\!0)=\sum_{m=-\infty}^{+\infty}\mathcal{J}_{|m|}(A)^2/(1+m\omega/U)$ for an off-resonant pump field with frequency $\omega$ and pump strength $A$, where $U$ is the value of the Hubbard interaction, $\mathcal{J}_{m}$ is the Bessel function of the first kind and $m$ indexes the rungs of the Floquet ladder\cite{mentink2015ultrafast,ItinPRL2015} [see Fig.~\ref{fig:floquet}(c)]. Beyond modifying local spin exchange, circularly polarized pumping of frustrated Mott insulators can dynamically break time-reversal symmetry and induce a transient chiral spin liquid in a frustrated Mott insulator\cite{claassen2017dynamical}, thus changing not merely the band structure but also the topological order of the system.

In addition to manipulating quantum magnets, Floquet engineering of effective Hubbard models with a resonant drive or at finite doping can similarly exhibit pre-thermalized regimes, which are described by exotic correlated pair hopping models for photoinduced doublon-holon pairs\cite{bukov2016schrieffer, eckardt2017colloquium} or photo-enhanced Cooper pairing\cite{coulthard2016enhancement}, respectively. These systems can similarly exhibit long pre-thermalized regimes even for resonant excitation\cite{kuwahara2016floquet, bukov2015prethermal, bukov2016heating}. Microscopic modelling and full time-domain simulations have revealed that the transient states can mediate post-pump metastable excitations in a coherent manner\cite{wang2017producing}. Thus, such artificially designed transient states could provide a platform for the study of novel physics not accessible through standard materials synthesis and equilibrium controls, such as temperature or pressure.

\subsection{Controlling orders and metastable states}

By taking advantage of transient dynamics out of equilibrium, it becomes possible to induce transitions between competing or metastable phases that otherwise would be subleading orders in equilibrium. The conceptually simplest application uses a strong pump field to induce a change in the electronic distribution, photodoping or heating of the system to a higher effective temperature. Experimentally, Mott\cite{perfetti2008femtosecond, liu2012terahertz}, charge-density wave\cite{schmitt2008transient,stojchevska2014ultrafast,zhang2016cooperative} and other bandgaps\cite{wegkamp2014instantaneous} have been melted by an ultrafast pump. Photoexcitation and thermalization can be explained using numerical methods such as non-equilibrium DMFT\cite{aoki2014nonequilibrium}. For example, a Mott insulator has been shown to melt\cite{perfetti2008femtosecond} through a series of metastable metallic states induced either by pumping or by a quantum quench of the Hamiltonian parameters (the ratio $t/U$)\cite{werner2012nonthermal}. Although pre-thermalized states can be accessed after perturbing a Hubbard model\cite{moeckel2008interaction}, strong correlations present in the model preclude integrability, leading to a smooth thermalization towards an effective photodoping condition\cite{balzer2015nonthermal} [see Fig.~\ref{fig:controlOrders}(a)].

In comparison to the rather dramatic effects of melting an insulator or existing order, a more challenging task has been to stabilize an out-of-equilibrium order in a metal or weakly correlated material. Experimentally, the superconducting transition temperature has been observed to increase with coherent pumping in K$_3$C$_{60}$, a fulleride compound\cite{mankowsky2014nonlinear}. Subsequently, numerous theoretical efforts have focused on first explaining the observed effect and then predicting other systems that may display a similar sort of non-equilibrium superconductivity. There are two main explanations for the effect: dynamical cooling, which suppresses thermal fluctuations, increasing the transition temperature\cite{denny2015proposed, hoeppner2015redistribution, nava2018cooling}; and an increase in effective electron-phonon interactions as a result of photoinduced phonon deformation or squeezing\cite{werner2015field, babadi2017theory, kennes2017transient, murakami2017nonequilibrium, sentef2017light}, which amplifies the superconducting order parameter at the level of Bardeen-Cooper-Schrieffer and Migdal-Eliashberg theory\cite{komnik2016bcs, sentef2016theory, knap2016dynamical} [see Fig.~\ref{fig:controlOrders}(b)]. A first-principles study of the A$_3$C$_{60}$ family further reveals that photoinduced deformation of the $T_{1u}$ phonon mode can cause an interaction imbalance\cite{kim2016enhancing}, favouring superconductivity\cite{mazza2017nonequilibrium}.

Ultrafast control of order can have additional practical meaning in materials with multiple competing phases, such as the high-temperature superconducting cuprates. The transient dynamics in such systems can lead to phenomena such as ultrafast switching between metallic (superconducting) and insulating phases. Following experiments that demonstrated transient light-enhanced superconductivity in insulating, charge-ordered La$_{1.8-x}$Eu$_{0.2}$Sr$_x$CuO$_4$ (Ref.~\onlinecite{fausti2011light}), it was expected that pumped correlated materials could display a rich variety of phases resulting from a change in the balance between dominant and subleading instabilities. It was shown that superconductivity can be enhanced by a pump through competition with bond-density wave\cite{raines2015enhancement,patel2016light} and charge-density wave\cite{sentef2017theory} orders. More recently, the discussion has progressed to the concept of inducing unconventional $d$-wave superconductivity by driving a material with competing charge-density wave and spin-density wave ordering tendencies near a quantum phase transition\cite{wang2017light} [see Fig.~\ref{fig:controlOrders}(c)]. Photoinduced localization effects have been suggested as a possible pathway to further enhance correlations and the competition between orders\cite{ido2017correlation}.

Beyond materials with charge-density wave, spin-density wave and superconducting ordering tendencies, the idea of manipulating competing orders in correlated materials has been extended to excitonic insulators\cite{mor2017ultrafast, murakami2017photoinduced}. The transient nature of these photoinduced metastable phase transitions implies that the traditional paradigm of symmetry breaking and long-range order does not necessarily apply in simple toy models. Thus, current theoretical work focuses on instabilities extracted from correlation functions or susceptibilities, re-expressed out of equilibrium. What is lacking currently is precise treatments of non-equilibrium phases with a unified definition connected directly to potential experimental observables.

\section{Summary and Outlook}

With the help of advances in scientific computing, the theoretical understanding of photon-based spectroscopies has advanced significantly in the past two decades. From an equilibrium perspective, this understanding has enabled the separate treatment of extrinsic measurement and intrinsic electronic properties. The combination of first-principles and many-body approaches has been used to successfully decipher the properties of numerous materials, from semiconductors and topological insulators to complex transition metal oxides and unconventional superconductors. Out-of-equilibrium, pump-probe techniques hold great advantages in characterizing hidden properties of materials, creating novel states of matter and controlling phase transitions. With appropriate modifications, equilibrium numerical approaches can be advanced to describe pump-probe experiments. These methods can quantitatively explain transient phenomena, such as superconductivity, and help to predict exotic non-equilibrium phases that are accessible by a fine control of pump conditions.

The main challenge for current theories lies in the oversimplification of materials descriptions owing to computational limitations. The mean-field-based methods, including the variants of Hartree-Fock, random-phase approximation, DMFT and other variational approaches, are computationally efficient and provide a direct connection between the physical picture and experimental observables. However, oversimplification of many-body Hamiltonians means that they can provide only a biased, posterior perspective on a specific problem. The many-body approaches overcome some of these issues, providing an exact description of the correlated physics induced by various many-body interactions through a complicated numerical evaluation. However, as a compromise, these approaches are limited by the mathematical complexity of the problem. For example, ED is restricted to small clusters, DMRG is restricted to low dimensions and short-range entanglement, and QMC is restricted to high temperatures. These issues limit their applicability in resolving fine details of experimental measurements in realistic materials. First-principles approaches extend mean-field methods by including correlation effects through exchange-correlation functionals, pseudopotentials and force fields. Based on material-specific, atomic ingredients, these first-principles approaches offer a description of both intrinsic material properties and extrinsic details of the process for each spectroscopy. However, such a first-principles approach obscures underlying physical intuition, and current exchange-correlation functional treatments underestimate many-body correlations and are somewhat dependent on known material-specific data sets.

These issues are even more severe out of equilibrium. Green's function methods based on the Keldysh formulation, valid only in the perturbative regime, require retention of the full two-time dependence of correlation functions even if only equal-time quantities are desired, placing a constraint on the maximum achievable simulation time. Many-body methods fare even worse: real-time QMC suffers from a severe phase problem even for models that are free of sign problems in equilibrium, limiting its applicability to ultrashort time behaviour only, and DMRG and tensor network methods are restricted by the exponential increase of the bond dimension with time. First-principles approaches such as time-dependent DFT, although computationally scalable, have a limited domain of applicability owing to the shortcomings of currently available exchange-correlation functionals. The simulation of spectroscopies proves especially difficult for strongly correlated methods, as more complicated multi-time correlation functions such as RIXS or Raman cross sections are currently computationally inaccessible even for small systems out of equilibrium. Finally, oversimplification of the materials description is an even more severe issue in a non-equilibrium setting. For example, proper modelling of pump-probe experiments should require a microscopic description of the light-matter interaction beyond a Peierls substitution in effective low-energy models, as well as correct treatment of the lattice and multi-orbital effects.

Despite the challenges faced by each of these approaches, there has nevertheless been remarkable progress in theoretical and numerical methods for advanced spectroscopies, occurring in tandem with novel experimental advances in both table-top and large-scale facility investigations of quantum matter. As the golden age of spectroscopy progresses, we can be sure that advances in both theory and experiment will bring us closer to understanding the behaviour of materials on their intrinsic time and length scales, with the hope of unravelling the phenomena of emergence through predictive tools for novel quantum phenomena and materials.

\section*{Acknowledgement}
C.D.P., C.J., B.M. and T.P.D. acknowledge support from the US Department of Energy, Office of Science, Office of Basic Energy Sciences, Division of Materials Sciences and Engineering, under Contract No. DE-AC02-76SF00515. Y.W. is supported by a Postdoctoral Fellowship in Quantum Science from the Harvard-Max Planck Institute of Quantum Optics. M.C. is supported by the Flatiron Institute, a division of the Simons Foundation.

\bibliography{refList}

\begin{thebibliography}{176}
\expandafter\ifx\csname natexlab\endcsname\relax\def\natexlab#1{#1}\fi
\expandafter\ifx\csname bibnamefont\endcsname\relax
  \def\bibnamefont#1{#1}\fi
\expandafter\ifx\csname bibfnamefont\endcsname\relax
  \def\bibfnamefont#1{#1}\fi
\expandafter\ifx\csname citenamefont\endcsname\relax
  \def\citenamefont#1{#1}\fi
\expandafter\ifx\csname url\endcsname\relax
  \def\url#1{\texttt{#1}}\fi
\expandafter\ifx\csname urlprefix\endcsname\relax\def\urlprefix{URL }\fi
\providecommand{\bibinfo}[2]{#2}
\providecommand{\eprint}[2][]{\url{#2}}

\bibitem[{\citenamefont{Damascelli et~al.}(2003)\citenamefont{Damascelli,
  Hussain, and Shen}}]{damascelli2003angle}
\bibinfo{author}{\bibfnamefont{A.}~\bibnamefont{Damascelli}},
  \bibinfo{author}{\bibfnamefont{Z.}~\bibnamefont{Hussain}}, \bibnamefont{and}
  \bibinfo{author}{\bibfnamefont{Z.-X.} \bibnamefont{Shen}},
  \bibinfo{journal}{Rev. Mod. Phys.} \textbf{\bibinfo{volume}{75}},
  \bibinfo{pages}{473} (\bibinfo{year}{2003}).

\bibitem[{\citenamefont{Fadley}(2010)}]{Fadley2010}
\bibinfo{author}{\bibfnamefont{C.}~\bibnamefont{Fadley}}, \bibinfo{journal}{J
  Electron Spectrosc.} \textbf{\bibinfo{volume}{178-179}}, \bibinfo{pages}{2}
  (\bibinfo{year}{2010}).

\bibitem[{\citenamefont{Pendry}(1976)}]{Pendry1976}
\bibinfo{author}{\bibfnamefont{J.~B.} \bibnamefont{Pendry}},
  \bibinfo{journal}{Surf. Sci.} \textbf{\bibinfo{volume}{57}},
  \bibinfo{pages}{679} (\bibinfo{year}{1976}).

\bibitem[{\citenamefont{Borstel}(1985)}]{Borstel1985}
\bibinfo{author}{\bibfnamefont{G.}~\bibnamefont{Borstel}},
  \bibinfo{journal}{Appl. Phys. A} \textbf{\bibinfo{volume}{38}},
  \bibinfo{pages}{193} (\bibinfo{year}{1985}).

\bibitem[{\citenamefont{Margine et~al.}(2016)\citenamefont{Margine, Lambert,
  and Giustino}}]{Margine2016}
\bibinfo{author}{\bibfnamefont{E.~R.} \bibnamefont{Margine}},
  \bibinfo{author}{\bibfnamefont{H.}~\bibnamefont{Lambert}}, \bibnamefont{and}
  \bibinfo{author}{\bibfnamefont{F.}~\bibnamefont{Giustino}},
  \bibinfo{journal}{Sci. Rep.} \textbf{\bibinfo{volume}{6}},
  \bibinfo{pages}{21414} (\bibinfo{year}{2016}).

\bibitem[{\citenamefont{Derondeau et~al.}(2017)\citenamefont{Derondeau, Bisti,
  Kobayashi, Braun, Ebert, Rogalev, Shi, Schmitt, Ma, Ding
  et~al.}}]{Derondeau2017}
\bibinfo{author}{\bibfnamefont{G.}~\bibnamefont{Derondeau}},
  \bibinfo{author}{\bibfnamefont{F.}~\bibnamefont{Bisti}},
  \bibinfo{author}{\bibfnamefont{M.}~\bibnamefont{Kobayashi}},
  \bibinfo{author}{\bibfnamefont{J.}~\bibnamefont{Braun}},
  \bibinfo{author}{\bibfnamefont{H.}~\bibnamefont{Ebert}},
  \bibinfo{author}{\bibfnamefont{V.~A.} \bibnamefont{Rogalev}},
  \bibinfo{author}{\bibfnamefont{M.}~\bibnamefont{Shi}},
  \bibinfo{author}{\bibfnamefont{T.}~\bibnamefont{Schmitt}},
  \bibinfo{author}{\bibfnamefont{J.}~\bibnamefont{Ma}},
  \bibinfo{author}{\bibfnamefont{H.}~\bibnamefont{Ding}}, \bibnamefont{et~al.},
  \bibinfo{journal}{Sci. Rep.} \textbf{\bibinfo{volume}{7}},
  \bibinfo{pages}{8787} (\bibinfo{year}{2017}).

\bibitem[{\citenamefont{Umari et~al.}(2010)\citenamefont{Umari, Stenuit, and
  Baroni}}]{Umari2010}
\bibinfo{author}{\bibfnamefont{P.}~\bibnamefont{Umari}},
  \bibinfo{author}{\bibfnamefont{G.}~\bibnamefont{Stenuit}}, \bibnamefont{and}
  \bibinfo{author}{\bibfnamefont{S.}~\bibnamefont{Baroni}},
  \bibinfo{journal}{Phys. Rev. B} \textbf{\bibinfo{volume}{81}},
  \bibinfo{pages}{115104} (\bibinfo{year}{2010}).

\bibitem[{\citenamefont{Lambert and Giustino}(2013)}]{Lambert2013}
\bibinfo{author}{\bibfnamefont{H.}~\bibnamefont{Lambert}} \bibnamefont{and}
  \bibinfo{author}{\bibfnamefont{F.}~\bibnamefont{Giustino}},
  \bibinfo{journal}{Phys. Rev. B} \textbf{\bibinfo{volume}{88}},
  \bibinfo{pages}{075117} (\bibinfo{year}{2013}).

\bibitem[{\citenamefont{Liu et~al.}(2016)\citenamefont{Liu, Kaltak,
  Klime{\v{s}}, and Kresse}}]{Liu2016}
\bibinfo{author}{\bibfnamefont{P.}~\bibnamefont{Liu}},
  \bibinfo{author}{\bibfnamefont{M.}~\bibnamefont{Kaltak}},
  \bibinfo{author}{\bibfnamefont{J.}~\bibnamefont{Klime{\v{s}}}},
  \bibnamefont{and} \bibinfo{author}{\bibfnamefont{G.}~\bibnamefont{Kresse}},
  \bibinfo{journal}{Phys. Rev. B} \textbf{\bibinfo{volume}{94}},
  \bibinfo{pages}{165109} (\bibinfo{year}{2016}), \eprint{1607.02859}.

\bibitem[{\citenamefont{Hedin}(1965)}]{Hedin1965}
\bibinfo{author}{\bibfnamefont{L.}~\bibnamefont{Hedin}},
  \bibinfo{journal}{Phys. Rev.} \textbf{\bibinfo{volume}{139}},
  \bibinfo{pages}{A796} (\bibinfo{year}{1965}).

\bibitem[{\citenamefont{Hedin}(1999)}]{Hedin1999}
\bibinfo{author}{\bibfnamefont{L.}~\bibnamefont{Hedin}}, \bibinfo{journal}{J.
  Phys. Condens. Matter} \textbf{\bibinfo{volume}{489}}, \bibinfo{pages}{R489}
  (\bibinfo{year}{1999}).

\bibitem[{\citenamefont{Reining}(2017)}]{Reining2017}
\bibinfo{author}{\bibfnamefont{L.}~\bibnamefont{Reining}},
  \bibinfo{journal}{WIRES Comput. Mol. Sci.} \textbf{\bibinfo{volume}{8}},
  \bibinfo{pages}{e1344} (\bibinfo{year}{2017}).

\bibitem[{\citenamefont{Bernardi et~al.}(2014)\citenamefont{Bernardi,
  Vigil-Fowler, Lischner, Neaton, and Louie}}]{Bernardi2014}
\bibinfo{author}{\bibfnamefont{M.}~\bibnamefont{Bernardi}},
  \bibinfo{author}{\bibfnamefont{D.}~\bibnamefont{Vigil-Fowler}},
  \bibinfo{author}{\bibfnamefont{J.}~\bibnamefont{Lischner}},
  \bibinfo{author}{\bibfnamefont{J.~B.} \bibnamefont{Neaton}},
  \bibnamefont{and} \bibinfo{author}{\bibfnamefont{S.~G.} \bibnamefont{Louie}},
  \bibinfo{journal}{Phys. Rev. Lett.} \textbf{\bibinfo{volume}{112}},
  \bibinfo{pages}{257402} (\bibinfo{year}{2014}).

\bibitem[{\citenamefont{Story et~al.}(2014)\citenamefont{Story, Kas, Vila,
  Verstraete, and Rehr}}]{Story2014}
\bibinfo{author}{\bibfnamefont{S.~M.} \bibnamefont{Story}},
  \bibinfo{author}{\bibfnamefont{J.~J.} \bibnamefont{Kas}},
  \bibinfo{author}{\bibfnamefont{F.~D.} \bibnamefont{Vila}},
  \bibinfo{author}{\bibfnamefont{M.~J.} \bibnamefont{Verstraete}},
  \bibnamefont{and} \bibinfo{author}{\bibfnamefont{J.~J.} \bibnamefont{Rehr}},
  \bibinfo{journal}{Phys. Rev. B} \textbf{\bibinfo{volume}{90}},
  \bibinfo{pages}{195135} (\bibinfo{year}{2014}).

\bibitem[{\citenamefont{Ponc{\'{e}} et~al.}(2016)\citenamefont{Ponc{\'{e}},
  Margine, Verdi, and Giustino}}]{Ponce2016}
\bibinfo{author}{\bibfnamefont{S.}~\bibnamefont{Ponc{\'{e}}}},
  \bibinfo{author}{\bibfnamefont{E.~R.} \bibnamefont{Margine}},
  \bibinfo{author}{\bibfnamefont{C.}~\bibnamefont{Verdi}}, \bibnamefont{and}
  \bibinfo{author}{\bibfnamefont{F.}~\bibnamefont{Giustino}},
  \bibinfo{journal}{Comput. Phys. Commun.} \textbf{\bibinfo{volume}{209}},
  \bibinfo{pages}{116} (\bibinfo{year}{2016}).

\bibitem[{\citenamefont{Sanna et~al.}(2018)\citenamefont{Sanna, Flores-Livas,
  Davydov, Profeta, Dewhurst, Sharma, and Gross}}]{Sanna2018}
\bibinfo{author}{\bibfnamefont{A.}~\bibnamefont{Sanna}},
  \bibinfo{author}{\bibfnamefont{J.~A.} \bibnamefont{Flores-Livas}},
  \bibinfo{author}{\bibfnamefont{A.}~\bibnamefont{Davydov}},
  \bibinfo{author}{\bibfnamefont{G.}~\bibnamefont{Profeta}},
  \bibinfo{author}{\bibfnamefont{K.}~\bibnamefont{Dewhurst}},
  \bibinfo{author}{\bibfnamefont{S.}~\bibnamefont{Sharma}}, \bibnamefont{and}
  \bibinfo{author}{\bibfnamefont{E.~K.~U.} \bibnamefont{Gross}},
  \bibinfo{journal}{J Phys. Soc. Jpn.} \textbf{\bibinfo{volume}{87}},
  \bibinfo{pages}{041012} (\bibinfo{year}{2018}).

\bibitem[{\citenamefont{Guzzo et~al.}(2011)\citenamefont{Guzzo, Lani, Sottile,
  Romaniello, Gatti, Kas, Rehr, Silly, Sirotti, and Reining}}]{Guzzo2011}
\bibinfo{author}{\bibfnamefont{M.}~\bibnamefont{Guzzo}},
  \bibinfo{author}{\bibfnamefont{G.}~\bibnamefont{Lani}},
  \bibinfo{author}{\bibfnamefont{F.}~\bibnamefont{Sottile}},
  \bibinfo{author}{\bibfnamefont{P.}~\bibnamefont{Romaniello}},
  \bibinfo{author}{\bibfnamefont{M.}~\bibnamefont{Gatti}},
  \bibinfo{author}{\bibfnamefont{J.~J.} \bibnamefont{Kas}},
  \bibinfo{author}{\bibfnamefont{J.~J.} \bibnamefont{Rehr}},
  \bibinfo{author}{\bibfnamefont{M.~G.} \bibnamefont{Silly}},
  \bibinfo{author}{\bibfnamefont{F.}~\bibnamefont{Sirotti}}, \bibnamefont{and}
  \bibinfo{author}{\bibfnamefont{L.}~\bibnamefont{Reining}},
  \bibinfo{journal}{Phys. Rev. Lett.} \textbf{\bibinfo{volume}{107}},
  \bibinfo{pages}{166401} (\bibinfo{year}{2011}), \eprint{1107.2207}.

\bibitem[{\citenamefont{Lischner et~al.}(2013)\citenamefont{Lischner,
  Vigil-Fowler, and Louie}}]{Lischner2013}
\bibinfo{author}{\bibfnamefont{J.}~\bibnamefont{Lischner}},
  \bibinfo{author}{\bibfnamefont{D.}~\bibnamefont{Vigil-Fowler}},
  \bibnamefont{and} \bibinfo{author}{\bibfnamefont{S.~G.} \bibnamefont{Louie}},
  \bibinfo{journal}{Phys. Rev. Lett.} \textbf{\bibinfo{volume}{110}},
  \bibinfo{pages}{146801} (\bibinfo{year}{2013}).

\bibitem[{\citenamefont{Kas et~al.}(2014)\citenamefont{Kas, Rehr, and
  Reining}}]{Kas2014}
\bibinfo{author}{\bibfnamefont{J.~J.} \bibnamefont{Kas}},
  \bibinfo{author}{\bibfnamefont{J.~J.} \bibnamefont{Rehr}}, \bibnamefont{and}
  \bibinfo{author}{\bibfnamefont{L.}~\bibnamefont{Reining}},
  \bibinfo{journal}{Phys. Rev. B} \textbf{\bibinfo{volume}{90}},
  \bibinfo{pages}{085112} (\bibinfo{year}{2014}).

\bibitem[{\citenamefont{Anisimov and Lukoyanov}(2014)}]{Anisimov2014}
\bibinfo{author}{\bibfnamefont{V.~I.} \bibnamefont{Anisimov}} \bibnamefont{and}
  \bibinfo{author}{\bibfnamefont{A.~V.} \bibnamefont{Lukoyanov}},
  \bibinfo{journal}{Acta Crystallogr} \textbf{\bibinfo{volume}{70}},
  \bibinfo{pages}{137} (\bibinfo{year}{2014}).

\bibitem[{\citenamefont{Mostofi et~al.}(2014)\citenamefont{Mostofi, Yates,
  Pizzi, Lee, Souza, Vanderbilt, and Marzari}}]{Mostofi2014}
\bibinfo{author}{\bibfnamefont{A.~A.} \bibnamefont{Mostofi}},
  \bibinfo{author}{\bibfnamefont{J.~R.} \bibnamefont{Yates}},
  \bibinfo{author}{\bibfnamefont{G.}~\bibnamefont{Pizzi}},
  \bibinfo{author}{\bibfnamefont{Y.~S.} \bibnamefont{Lee}},
  \bibinfo{author}{\bibfnamefont{I.}~\bibnamefont{Souza}},
  \bibinfo{author}{\bibfnamefont{D.}~\bibnamefont{Vanderbilt}},
  \bibnamefont{and} \bibinfo{author}{\bibfnamefont{N.}~\bibnamefont{Marzari}},
  \bibinfo{journal}{Comput. Phys. Commun.} \textbf{\bibinfo{volume}{185}},
  \bibinfo{pages}{2309} (\bibinfo{year}{2014}).

\bibitem[{\citenamefont{Min{\'{a}}r et~al.}(2011)\citenamefont{Min{\'{a}}r,
  Braun, Mankovsky, and Ebert}}]{Minar2011}
\bibinfo{author}{\bibfnamefont{J.}~\bibnamefont{Min{\'{a}}r}},
  \bibinfo{author}{\bibfnamefont{J.}~\bibnamefont{Braun}},
  \bibinfo{author}{\bibfnamefont{S.}~\bibnamefont{Mankovsky}},
  \bibnamefont{and} \bibinfo{author}{\bibfnamefont{H.}~\bibnamefont{Ebert}},
  \bibinfo{journal}{J Electron Spectrosc.} \textbf{\bibinfo{volume}{184}},
  \bibinfo{pages}{91} (\bibinfo{year}{2011}).

\bibitem[{\citenamefont{Biermann}(2014)}]{Biermann2014}
\bibinfo{author}{\bibfnamefont{S.}~\bibnamefont{Biermann}},
  \bibinfo{journal}{J. Phys. Condens. Matter} \textbf{\bibinfo{volume}{26}},
  \bibinfo{pages}{173202} (\bibinfo{year}{2014}).

\bibitem[{\citenamefont{Dagotto}(1994)}]{dagotto1994correlated}
\bibinfo{author}{\bibfnamefont{E.}~\bibnamefont{Dagotto}},
  \bibinfo{journal}{Rev. Mod. Phys.} \textbf{\bibinfo{volume}{66}},
  \bibinfo{pages}{763} (\bibinfo{year}{1994}).

\bibitem[{\citenamefont{Foulkes et~al.}(2001)\citenamefont{Foulkes, Mitas,
  Needs, and Rajagopal}}]{foulkes2001quantum}
\bibinfo{author}{\bibfnamefont{W.}~\bibnamefont{Foulkes}},
  \bibinfo{author}{\bibfnamefont{L.}~\bibnamefont{Mitas}},
  \bibinfo{author}{\bibfnamefont{R.}~\bibnamefont{Needs}}, \bibnamefont{and}
  \bibinfo{author}{\bibfnamefont{G.}~\bibnamefont{Rajagopal}},
  \bibinfo{journal}{Rev. Mod. Phys.} \textbf{\bibinfo{volume}{73}},
  \bibinfo{pages}{33} (\bibinfo{year}{2001}).

\bibitem[{\citenamefont{Schollw{\"o}ck}(2005)}]{schollwock2005density}
\bibinfo{author}{\bibfnamefont{U.}~\bibnamefont{Schollw{\"o}ck}},
  \bibinfo{journal}{Rev. Mod. Phys.} \textbf{\bibinfo{volume}{77}},
  \bibinfo{pages}{259} (\bibinfo{year}{2005}).

\bibitem[{\citenamefont{S{\'e}n{\'e}chal
  et~al.}(2000)\citenamefont{S{\'e}n{\'e}chal, Perez, and
  Pioro-Ladriere}}]{senechal2000spectral}
\bibinfo{author}{\bibfnamefont{D.}~\bibnamefont{S{\'e}n{\'e}chal}},
  \bibinfo{author}{\bibfnamefont{D.}~\bibnamefont{Perez}}, \bibnamefont{and}
  \bibinfo{author}{\bibfnamefont{M.}~\bibnamefont{Pioro-Ladriere}},
  \bibinfo{journal}{Phys. Rev. Lett.} \textbf{\bibinfo{volume}{84}},
  \bibinfo{pages}{522} (\bibinfo{year}{2000}).

\bibitem[{\citenamefont{Hettler et~al.}(2000)\citenamefont{Hettler, Mukherjee,
  Jarrell, and Krishnamurthy}}]{hettler2000dynamical}
\bibinfo{author}{\bibfnamefont{M.}~\bibnamefont{Hettler}},
  \bibinfo{author}{\bibfnamefont{M.}~\bibnamefont{Mukherjee}},
  \bibinfo{author}{\bibfnamefont{M.}~\bibnamefont{Jarrell}}, \bibnamefont{and}
  \bibinfo{author}{\bibfnamefont{H.}~\bibnamefont{Krishnamurthy}},
  \bibinfo{journal}{Phys. Rev. B} \textbf{\bibinfo{volume}{61}},
  \bibinfo{pages}{12739} (\bibinfo{year}{2000}).

\bibitem[{\citenamefont{Potthoff et~al.}(2003)\citenamefont{Potthoff, Aichhorn,
  and Dahnken}}]{potthoff2003variational}
\bibinfo{author}{\bibfnamefont{M.}~\bibnamefont{Potthoff}},
  \bibinfo{author}{\bibfnamefont{M.}~\bibnamefont{Aichhorn}}, \bibnamefont{and}
  \bibinfo{author}{\bibfnamefont{C.}~\bibnamefont{Dahnken}},
  \bibinfo{journal}{Phys. Rev. Lett.} \textbf{\bibinfo{volume}{91}},
  \bibinfo{pages}{206402} (\bibinfo{year}{2003}).

\bibitem[{\citenamefont{Zemlji{\v{c}} et~al.}(2008)\citenamefont{Zemlji{\v{c}},
  Prelov{\v{s}}ek, and Tohyama}}]{zemljivc2008temperature}
\bibinfo{author}{\bibfnamefont{M.-M.} \bibnamefont{Zemlji{\v{c}}}},
  \bibinfo{author}{\bibfnamefont{P.}~\bibnamefont{Prelov{\v{s}}ek}},
  \bibnamefont{and} \bibinfo{author}{\bibfnamefont{T.}~\bibnamefont{Tohyama}},
  \bibinfo{journal}{Phys. Rev. Lett.} \textbf{\bibinfo{volume}{100}},
  \bibinfo{pages}{036402} (\bibinfo{year}{2008}).

\bibitem[{\citenamefont{Macridin et~al.}(2007)\citenamefont{Macridin, Jarrell,
  Maier, and Scalapino}}]{macridin2007high}
\bibinfo{author}{\bibfnamefont{A.}~\bibnamefont{Macridin}},
  \bibinfo{author}{\bibfnamefont{M.}~\bibnamefont{Jarrell}},
  \bibinfo{author}{\bibfnamefont{T.}~\bibnamefont{Maier}}, \bibnamefont{and}
  \bibinfo{author}{\bibfnamefont{D.}~\bibnamefont{Scalapino}},
  \bibinfo{journal}{Phys. Rev. Lett.} \textbf{\bibinfo{volume}{99}},
  \bibinfo{pages}{237001} (\bibinfo{year}{2007}).

\bibitem[{\citenamefont{Moritz et~al.}(2009)\citenamefont{Moritz, Schmitt,
  Meevasana, Johnston, Motoyama, Greven, Lu, Kim, Scalettar, Shen
  et~al.}}]{moritz2009effect}
\bibinfo{author}{\bibfnamefont{B.}~\bibnamefont{Moritz}},
  \bibinfo{author}{\bibfnamefont{F.}~\bibnamefont{Schmitt}},
  \bibinfo{author}{\bibfnamefont{W.}~\bibnamefont{Meevasana}},
  \bibinfo{author}{\bibfnamefont{S.}~\bibnamefont{Johnston}},
  \bibinfo{author}{\bibfnamefont{E.}~\bibnamefont{Motoyama}},
  \bibinfo{author}{\bibfnamefont{M.}~\bibnamefont{Greven}},
  \bibinfo{author}{\bibfnamefont{D.}~\bibnamefont{Lu}},
  \bibinfo{author}{\bibfnamefont{C.}~\bibnamefont{Kim}},
  \bibinfo{author}{\bibfnamefont{R.}~\bibnamefont{Scalettar}},
  \bibinfo{author}{\bibfnamefont{Z.}~\bibnamefont{Shen}}, \bibnamefont{et~al.},
  \bibinfo{journal}{New J Phys.} \textbf{\bibinfo{volume}{11}},
  \bibinfo{pages}{093020} (\bibinfo{year}{2009}).

\bibitem[{\citenamefont{Wang et~al.}(2015)\citenamefont{Wang, Wohlfeld, Moritz,
  Jia, van Veenendaal, Wu, Chen, and Devereaux}}]{wang2015origin}
\bibinfo{author}{\bibfnamefont{Y.}~\bibnamefont{Wang}},
  \bibinfo{author}{\bibfnamefont{K.}~\bibnamefont{Wohlfeld}},
  \bibinfo{author}{\bibfnamefont{B.}~\bibnamefont{Moritz}},
  \bibinfo{author}{\bibfnamefont{C.}~\bibnamefont{Jia}},
  \bibinfo{author}{\bibfnamefont{M.}~\bibnamefont{van Veenendaal}},
  \bibinfo{author}{\bibfnamefont{K.}~\bibnamefont{Wu}},
  \bibinfo{author}{\bibfnamefont{C.-C.} \bibnamefont{Chen}}, \bibnamefont{and}
  \bibinfo{author}{\bibfnamefont{T.~P.} \bibnamefont{Devereaux}},
  \bibinfo{journal}{Phys. Rev. B} \textbf{\bibinfo{volume}{92}},
  \bibinfo{pages}{075119} (\bibinfo{year}{2015}).

\bibitem[{\citenamefont{Bansil and Lindroos}(1999)}]{Bansil1999}
\bibinfo{author}{\bibfnamefont{A.}~\bibnamefont{Bansil}} \bibnamefont{and}
  \bibinfo{author}{\bibfnamefont{M.}~\bibnamefont{Lindroos}},
  \bibinfo{journal}{Phys. Rev. Lett.} \textbf{\bibinfo{volume}{83}},
  \bibinfo{pages}{5154} (\bibinfo{year}{1999}).

\bibitem[{\citenamefont{Okuda and Kimura}(2013)}]{Okuda2013}
\bibinfo{author}{\bibfnamefont{T.}~\bibnamefont{Okuda}} \bibnamefont{and}
  \bibinfo{author}{\bibfnamefont{A.}~\bibnamefont{Kimura}}, \bibinfo{journal}{J
  Phys. Soc. Jpn.} \textbf{\bibinfo{volume}{82}}, \bibinfo{pages}{021002}
  (\bibinfo{year}{2013}).

\bibitem[{\citenamefont{Korringa}(1947)}]{Korringa1947}
\bibinfo{author}{\bibfnamefont{J.}~\bibnamefont{Korringa}},
  \bibinfo{journal}{Physica} \textbf{\bibinfo{volume}{13}},
  \bibinfo{pages}{392} (\bibinfo{year}{1947}).

\bibitem[{\citenamefont{Kohn and Rostoker}(1954)}]{Kohn1954}
\bibinfo{author}{\bibfnamefont{W.}~\bibnamefont{Kohn}} \bibnamefont{and}
  \bibinfo{author}{\bibfnamefont{N.}~\bibnamefont{Rostoker}},
  \bibinfo{journal}{Phys. Rev.} \textbf{\bibinfo{volume}{94}},
  \bibinfo{pages}{1111} (\bibinfo{year}{1954}).

\bibitem[{\citenamefont{Bansil et~al.}(2012)\citenamefont{Bansil, Barbiellini,
  Basak, Das, Lin, Lindroos, Nieminen, Suominen, JuiWang, and
  Markiewicz}}]{Bansil2012}
\bibinfo{author}{\bibfnamefont{A.}~\bibnamefont{Bansil}},
  \bibinfo{author}{\bibfnamefont{B.}~\bibnamefont{Barbiellini}},
  \bibinfo{author}{\bibfnamefont{S.}~\bibnamefont{Basak}},
  \bibinfo{author}{\bibfnamefont{T.}~\bibnamefont{Das}},
  \bibinfo{author}{\bibfnamefont{H.}~\bibnamefont{Lin}},
  \bibinfo{author}{\bibfnamefont{M.}~\bibnamefont{Lindroos}},
  \bibinfo{author}{\bibfnamefont{J.}~\bibnamefont{Nieminen}},
  \bibinfo{author}{\bibfnamefont{I.}~\bibnamefont{Suominen}},
  \bibinfo{author}{\bibfnamefont{Y.}~\bibnamefont{JuiWang}}, \bibnamefont{and}
  \bibinfo{author}{\bibfnamefont{R.~S.} \bibnamefont{Markiewicz}},
  \bibinfo{journal}{Journal of Superconductivity and Novel Magnetism}
  \textbf{\bibinfo{volume}{25}}, \bibinfo{pages}{2135} (\bibinfo{year}{2012}).

\bibitem[{\citenamefont{Scholz et~al.}(2013)\citenamefont{Scholz,
  S{\'{a}}nchez-Barriga, Braun, Marchenko, Varykhalov, Lindroos, Wang, Lin,
  Bansil, Min{\'{a}}r et~al.}}]{Scholz}
\bibinfo{author}{\bibfnamefont{M.~R.} \bibnamefont{Scholz}},
  \bibinfo{author}{\bibfnamefont{J.}~\bibnamefont{S{\'{a}}nchez-Barriga}},
  \bibinfo{author}{\bibfnamefont{J.}~\bibnamefont{Braun}},
  \bibinfo{author}{\bibfnamefont{D.}~\bibnamefont{Marchenko}},
  \bibinfo{author}{\bibfnamefont{A.}~\bibnamefont{Varykhalov}},
  \bibinfo{author}{\bibfnamefont{M.}~\bibnamefont{Lindroos}},
  \bibinfo{author}{\bibfnamefont{Y.~J.} \bibnamefont{Wang}},
  \bibinfo{author}{\bibfnamefont{H.}~\bibnamefont{Lin}},
  \bibinfo{author}{\bibfnamefont{A.}~\bibnamefont{Bansil}},
  \bibinfo{author}{\bibfnamefont{J.}~\bibnamefont{Min{\'{a}}r}},
  \bibnamefont{et~al.}, \bibinfo{journal}{Phys. Rev. Lett.}
  \textbf{\bibinfo{volume}{110}} (\bibinfo{year}{2013}), \eprint{1303.6161}.

\bibitem[{\citenamefont{S{\'{a}}nchez-Barriga
  et~al.}(2014)\citenamefont{S{\'{a}}nchez-Barriga, Varykhalov, Braun, Xu,
  Alidoust, Kornilov, Min{\'{a}}r, Hummer, Springholz, Bauer
  et~al.}}]{Sanchez-Barriga2014}
\bibinfo{author}{\bibfnamefont{J.}~\bibnamefont{S{\'{a}}nchez-Barriga}},
  \bibinfo{author}{\bibfnamefont{A.}~\bibnamefont{Varykhalov}},
  \bibinfo{author}{\bibfnamefont{J.}~\bibnamefont{Braun}},
  \bibinfo{author}{\bibfnamefont{S.~Y.} \bibnamefont{Xu}},
  \bibinfo{author}{\bibfnamefont{N.}~\bibnamefont{Alidoust}},
  \bibinfo{author}{\bibfnamefont{O.}~\bibnamefont{Kornilov}},
  \bibinfo{author}{\bibfnamefont{J.}~\bibnamefont{Min{\'{a}}r}},
  \bibinfo{author}{\bibfnamefont{K.}~\bibnamefont{Hummer}},
  \bibinfo{author}{\bibfnamefont{G.}~\bibnamefont{Springholz}},
  \bibinfo{author}{\bibfnamefont{G.}~\bibnamefont{Bauer}},
  \bibnamefont{et~al.}, \bibinfo{journal}{Phys. Rev. X}
  \textbf{\bibinfo{volume}{4}}, \bibinfo{pages}{011046} (\bibinfo{year}{2014}).

\bibitem[{\citenamefont{Ruggenthaler et~al.}(2018)\citenamefont{Ruggenthaler,
  Tancogne-Dejean, Flick, Appel, and Rubio}}]{ruggenthaler2018quantum}
\bibinfo{author}{\bibfnamefont{M.}~\bibnamefont{Ruggenthaler}},
  \bibinfo{author}{\bibfnamefont{N.}~\bibnamefont{Tancogne-Dejean}},
  \bibinfo{author}{\bibfnamefont{J.}~\bibnamefont{Flick}},
  \bibinfo{author}{\bibfnamefont{H.}~\bibnamefont{Appel}}, \bibnamefont{and}
  \bibinfo{author}{\bibfnamefont{A.}~\bibnamefont{Rubio}},
  \bibinfo{journal}{Nat. Rev. Chem.} \textbf{\bibinfo{volume}{2}},
  \bibinfo{pages}{0118} (\bibinfo{year}{2018}).

\bibitem[{\citenamefont{Kotani and Shin}(2001)}]{Kotani2001}
\bibinfo{author}{\bibfnamefont{A.}~\bibnamefont{Kotani}} \bibnamefont{and}
  \bibinfo{author}{\bibfnamefont{S.}~\bibnamefont{Shin}},
  \bibinfo{journal}{Rev. Mod. Phys.} \textbf{\bibinfo{volume}{73}},
  \bibinfo{pages}{203} (\bibinfo{year}{2001}).

\bibitem[{\citenamefont{Ament et~al.}(2011)\citenamefont{Ament, van Veenendaal,
  Devereaux, Hill, and van~den Brink}}]{Ament2011}
\bibinfo{author}{\bibfnamefont{L.~J.~P.} \bibnamefont{Ament}},
  \bibinfo{author}{\bibfnamefont{M.}~\bibnamefont{van Veenendaal}},
  \bibinfo{author}{\bibfnamefont{T.~P.} \bibnamefont{Devereaux}},
  \bibinfo{author}{\bibfnamefont{J.~P.} \bibnamefont{Hill}}, \bibnamefont{and}
  \bibinfo{author}{\bibfnamefont{J.}~\bibnamefont{van~den Brink}},
  \bibinfo{journal}{Rev. Mod. Phys.} \textbf{\bibinfo{volume}{83}},
  \bibinfo{pages}{705} (\bibinfo{year}{2011}).

\bibitem[{\citenamefont{Ishii}(2017)}]{Ishii2017}
\bibinfo{author}{\bibfnamefont{K.}~\bibnamefont{Ishii}},
  \emph{\bibinfo{title}{Springer Tracts in Modern Physics}}, vol.
  \bibinfo{volume}{269} (\bibinfo{publisher}{Springer, Berlin, Heidelberg},
  \bibinfo{year}{2017}).

\bibitem[{\citenamefont{Hill et~al.}(1998)\citenamefont{Hill, Kao, Caliebe,
  Matsubara, Kotani, Peng, and Greene}}]{Hill1998}
\bibinfo{author}{\bibfnamefont{J.~P.} \bibnamefont{Hill}},
  \bibinfo{author}{\bibfnamefont{C.-C.} \bibnamefont{Kao}},
  \bibinfo{author}{\bibfnamefont{W.~A.~L.} \bibnamefont{Caliebe}},
  \bibinfo{author}{\bibfnamefont{M.}~\bibnamefont{Matsubara}},
  \bibinfo{author}{\bibfnamefont{A.}~\bibnamefont{Kotani}},
  \bibinfo{author}{\bibfnamefont{J.~L.} \bibnamefont{Peng}}, \bibnamefont{and}
  \bibinfo{author}{\bibfnamefont{R.~L.} \bibnamefont{Greene}},
  \bibinfo{journal}{Phys. Rev. Lett.} \textbf{\bibinfo{volume}{80}},
  \bibinfo{pages}{4967} (\bibinfo{year}{1998}).

\bibitem[{\citenamefont{Abbamonte et~al.}(1999)\citenamefont{Abbamonte, Burns,
  Isaacs, Platzman, Miller, Cheong, and Klein}}]{Abbamonte1999}
\bibinfo{author}{\bibfnamefont{P.}~\bibnamefont{Abbamonte}},
  \bibinfo{author}{\bibfnamefont{C.~A.} \bibnamefont{Burns}},
  \bibinfo{author}{\bibfnamefont{E.~D.} \bibnamefont{Isaacs}},
  \bibinfo{author}{\bibfnamefont{P.~M.} \bibnamefont{Platzman}},
  \bibinfo{author}{\bibfnamefont{L.~L.} \bibnamefont{Miller}},
  \bibinfo{author}{\bibfnamefont{S.~W.} \bibnamefont{Cheong}},
  \bibnamefont{and} \bibinfo{author}{\bibfnamefont{M.~V.} \bibnamefont{Klein}},
  \bibinfo{journal}{Phys. Rev. Lett.} \textbf{\bibinfo{volume}{83}},
  \bibinfo{pages}{860} (\bibinfo{year}{1999}).

\bibitem[{\citenamefont{Braicovich et~al.}(2009)\citenamefont{Braicovich,
  Ament, Bisogni, Forte, Aruta, Balestrino, Brookes, {De Luca}, Medaglia,
  Granozio et~al.}}]{Braicovich2009}
\bibinfo{author}{\bibfnamefont{L.}~\bibnamefont{Braicovich}},
  \bibinfo{author}{\bibfnamefont{L.~J.~P.} \bibnamefont{Ament}},
  \bibinfo{author}{\bibfnamefont{V.}~\bibnamefont{Bisogni}},
  \bibinfo{author}{\bibfnamefont{F.}~\bibnamefont{Forte}},
  \bibinfo{author}{\bibfnamefont{C.}~\bibnamefont{Aruta}},
  \bibinfo{author}{\bibfnamefont{G.}~\bibnamefont{Balestrino}},
  \bibinfo{author}{\bibfnamefont{N.~B.} \bibnamefont{Brookes}},
  \bibinfo{author}{\bibfnamefont{G.~M.} \bibnamefont{{De Luca}}},
  \bibinfo{author}{\bibfnamefont{P.~G.} \bibnamefont{Medaglia}},
  \bibinfo{author}{\bibfnamefont{F.~M.} \bibnamefont{Granozio}},
  \bibnamefont{et~al.}, \bibinfo{journal}{Phys. Rev. Lett.}
  \textbf{\bibinfo{volume}{102}}, \bibinfo{pages}{167401}
  (\bibinfo{year}{2009}).

\bibitem[{\citenamefont{Schlappa et~al.}(2009)\citenamefont{Schlappa, Schmitt,
  Vernay, Strocov, Ilakovac, Thielemann, R\o{}nnow, Vanishri, Piazzalunga, Wang
  et~al.}}]{Schlappa2009}
\bibinfo{author}{\bibfnamefont{J.}~\bibnamefont{Schlappa}},
  \bibinfo{author}{\bibfnamefont{T.}~\bibnamefont{Schmitt}},
  \bibinfo{author}{\bibfnamefont{F.}~\bibnamefont{Vernay}},
  \bibinfo{author}{\bibfnamefont{V.~N.} \bibnamefont{Strocov}},
  \bibinfo{author}{\bibfnamefont{V.}~\bibnamefont{Ilakovac}},
  \bibinfo{author}{\bibfnamefont{B.}~\bibnamefont{Thielemann}},
  \bibinfo{author}{\bibfnamefont{H.~M.} \bibnamefont{R\o{}nnow}},
  \bibinfo{author}{\bibfnamefont{S.}~\bibnamefont{Vanishri}},
  \bibinfo{author}{\bibfnamefont{A.}~\bibnamefont{Piazzalunga}},
  \bibinfo{author}{\bibfnamefont{X.}~\bibnamefont{Wang}}, \bibnamefont{et~al.},
  \bibinfo{journal}{Phys. Rev. Lett.} \textbf{\bibinfo{volume}{103}},
  \bibinfo{pages}{047401} (\bibinfo{year}{2009}).

\bibitem[{\citenamefont{{Le Tacon} et~al.}({2011})\citenamefont{{Le Tacon},
  Ghiringhelli, Chaloupka, Sala, Hinkov, Haverkort, Minola, Bakr, Zhou,
  Blanco-Canosa et~al.}}]{Tacon2011}
\bibinfo{author}{\bibfnamefont{M.}~\bibnamefont{{Le Tacon}}},
  \bibinfo{author}{\bibfnamefont{G.}~\bibnamefont{Ghiringhelli}},
  \bibinfo{author}{\bibfnamefont{J.}~\bibnamefont{Chaloupka}},
  \bibinfo{author}{\bibfnamefont{M.~M.} \bibnamefont{Sala}},
  \bibinfo{author}{\bibfnamefont{V.}~\bibnamefont{Hinkov}},
  \bibinfo{author}{\bibfnamefont{M.~W.} \bibnamefont{Haverkort}},
  \bibinfo{author}{\bibfnamefont{M.}~\bibnamefont{Minola}},
  \bibinfo{author}{\bibfnamefont{M.}~\bibnamefont{Bakr}},
  \bibinfo{author}{\bibfnamefont{K.~J.} \bibnamefont{Zhou}},
  \bibinfo{author}{\bibfnamefont{S.}~\bibnamefont{Blanco-Canosa}},
  \bibnamefont{et~al.}, \bibinfo{journal}{{Nat. Phys.}}
  \textbf{\bibinfo{volume}{{7}}}, \bibinfo{pages}{725}
  (\bibinfo{year}{{2011}}).

\bibitem[{\citenamefont{Schlappa et~al.}(2012)\citenamefont{Schlappa, Wohlfeld,
  Zhou, Mourigal, Haverkort, Strocov, Hozoi, Monney, Nishimoto, Singh
  et~al.}}]{Schlappa2012}
\bibinfo{author}{\bibfnamefont{J.}~\bibnamefont{Schlappa}},
  \bibinfo{author}{\bibfnamefont{K.}~\bibnamefont{Wohlfeld}},
  \bibinfo{author}{\bibfnamefont{K.~J.} \bibnamefont{Zhou}},
  \bibinfo{author}{\bibfnamefont{M.}~\bibnamefont{Mourigal}},
  \bibinfo{author}{\bibfnamefont{M.~W.} \bibnamefont{Haverkort}},
  \bibinfo{author}{\bibfnamefont{V.~N.} \bibnamefont{Strocov}},
  \bibinfo{author}{\bibfnamefont{L.}~\bibnamefont{Hozoi}},
  \bibinfo{author}{\bibfnamefont{C.}~\bibnamefont{Monney}},
  \bibinfo{author}{\bibfnamefont{S.}~\bibnamefont{Nishimoto}},
  \bibinfo{author}{\bibfnamefont{S.}~\bibnamefont{Singh}},
  \bibnamefont{et~al.}, \bibinfo{journal}{Nature}
  \textbf{\bibinfo{volume}{485}}, \bibinfo{pages}{82} (\bibinfo{year}{2012}).

\bibitem[{\citenamefont{Dean et~al.}(2013)\citenamefont{Dean, Dellea,
  Springell, Yakhou-Harris, Kummer, Brookes, Liu, Sun, Strle, Schmitt
  et~al.}}]{Dean2013b}
\bibinfo{author}{\bibfnamefont{M.~P.~M.} \bibnamefont{Dean}},
  \bibinfo{author}{\bibfnamefont{G.}~\bibnamefont{Dellea}},
  \bibinfo{author}{\bibfnamefont{R.~S.} \bibnamefont{Springell}},
  \bibinfo{author}{\bibfnamefont{F.}~\bibnamefont{Yakhou-Harris}},
  \bibinfo{author}{\bibfnamefont{K.}~\bibnamefont{Kummer}},
  \bibinfo{author}{\bibfnamefont{N.~B.} \bibnamefont{Brookes}},
  \bibinfo{author}{\bibfnamefont{X.}~\bibnamefont{Liu}},
  \bibinfo{author}{\bibfnamefont{Y.-J.} \bibnamefont{Sun}},
  \bibinfo{author}{\bibfnamefont{J.}~\bibnamefont{Strle}},
  \bibinfo{author}{\bibfnamefont{T.}~\bibnamefont{Schmitt}},
  \bibnamefont{et~al.}, \bibinfo{journal}{Nat. Mater.}
  \textbf{\bibinfo{volume}{12}}, \bibinfo{pages}{1019} (\bibinfo{year}{2013}).

\bibitem[{\citenamefont{Bisogni
  et~al.}(2012{\natexlab{a}})\citenamefont{Bisogni, Simonelli, Ament, Forte,
  Moretti~Sala, Minola, Huotari, van~den Brink, Ghiringhelli, Brookes
  et~al.}}]{Bisogni2012a}
\bibinfo{author}{\bibfnamefont{V.}~\bibnamefont{Bisogni}},
  \bibinfo{author}{\bibfnamefont{L.}~\bibnamefont{Simonelli}},
  \bibinfo{author}{\bibfnamefont{L.~J.~P.} \bibnamefont{Ament}},
  \bibinfo{author}{\bibfnamefont{F.}~\bibnamefont{Forte}},
  \bibinfo{author}{\bibfnamefont{M.}~\bibnamefont{Moretti~Sala}},
  \bibinfo{author}{\bibfnamefont{M.}~\bibnamefont{Minola}},
  \bibinfo{author}{\bibfnamefont{S.}~\bibnamefont{Huotari}},
  \bibinfo{author}{\bibfnamefont{J.}~\bibnamefont{van~den Brink}},
  \bibinfo{author}{\bibfnamefont{G.}~\bibnamefont{Ghiringhelli}},
  \bibinfo{author}{\bibfnamefont{N.~B.} \bibnamefont{Brookes}},
  \bibnamefont{et~al.}, \bibinfo{journal}{Phys. Rev. B}
  \textbf{\bibinfo{volume}{85}}, \bibinfo{pages}{214527}
  (\bibinfo{year}{2012}{\natexlab{a}}).

\bibitem[{\citenamefont{Kim and van~den Brink}(2015)}]{Kim2015}
\bibinfo{author}{\bibfnamefont{B.~H.} \bibnamefont{Kim}} \bibnamefont{and}
  \bibinfo{author}{\bibfnamefont{J.}~\bibnamefont{van~den Brink}},
  \bibinfo{journal}{Phys. Rev. B} \textbf{\bibinfo{volume}{92}},
  \bibinfo{pages}{081105} (\bibinfo{year}{2015}).

\bibitem[{\citenamefont{Zhou et~al.}(2013)\citenamefont{Zhou, Huang, Monney,
  Dai, Strocov, Wang, Chen, Zhang, Dai, Patthey et~al.}}]{Zhou2013}
\bibinfo{author}{\bibfnamefont{K.-J.} \bibnamefont{Zhou}},
  \bibinfo{author}{\bibfnamefont{Y.-B.} \bibnamefont{Huang}},
  \bibinfo{author}{\bibfnamefont{C.}~\bibnamefont{Monney}},
  \bibinfo{author}{\bibfnamefont{X.}~\bibnamefont{Dai}},
  \bibinfo{author}{\bibfnamefont{V.~N.} \bibnamefont{Strocov}},
  \bibinfo{author}{\bibfnamefont{N.-L.} \bibnamefont{Wang}},
  \bibinfo{author}{\bibfnamefont{Z.-G.} \bibnamefont{Chen}},
  \bibinfo{author}{\bibfnamefont{C.}~\bibnamefont{Zhang}},
  \bibinfo{author}{\bibfnamefont{P.}~\bibnamefont{Dai}},
  \bibinfo{author}{\bibfnamefont{L.}~\bibnamefont{Patthey}},
  \bibnamefont{et~al.}, \bibinfo{journal}{Nat. Commun.}
  \textbf{\bibinfo{volume}{4}}, \bibinfo{pages}{1470} (\bibinfo{year}{2013}).

\bibitem[{\citenamefont{Sala et~al.}(2014)\citenamefont{Sala, Ohguishi,
  Al-Zein, Hirata, Monaco, and Krisch}}]{Sala2014}
\bibinfo{author}{\bibfnamefont{M.~M.} \bibnamefont{Sala}},
  \bibinfo{author}{\bibfnamefont{K.}~\bibnamefont{Ohguishi}},
  \bibinfo{author}{\bibfnamefont{A.}~\bibnamefont{Al-Zein}},
  \bibinfo{author}{\bibfnamefont{Y.}~\bibnamefont{Hirata}},
  \bibinfo{author}{\bibfnamefont{G.}~\bibnamefont{Monaco}}, \bibnamefont{and}
  \bibinfo{author}{\bibfnamefont{M.}~\bibnamefont{Krisch}},
  \bibinfo{journal}{Phys. Rev. Lett.} \textbf{\bibinfo{volume}{112}},
  \bibinfo{pages}{176402} (\bibinfo{year}{2014}).

\bibitem[{\citenamefont{Yin et~al.}(2013)\citenamefont{Yin, Liu, Tsvelik, Dean,
  Upton, Kim, Casa, Said, Gog, Qi et~al.}}]{Yin2013}
\bibinfo{author}{\bibfnamefont{W.-G.} \bibnamefont{Yin}},
  \bibinfo{author}{\bibfnamefont{X.}~\bibnamefont{Liu}},
  \bibinfo{author}{\bibfnamefont{A.~M.} \bibnamefont{Tsvelik}},
  \bibinfo{author}{\bibfnamefont{M.~P.~M.} \bibnamefont{Dean}},
  \bibinfo{author}{\bibfnamefont{M.~H.} \bibnamefont{Upton}},
  \bibinfo{author}{\bibfnamefont{J.}~\bibnamefont{Kim}},
  \bibinfo{author}{\bibfnamefont{D.}~\bibnamefont{Casa}},
  \bibinfo{author}{\bibfnamefont{A.}~\bibnamefont{Said}},
  \bibinfo{author}{\bibfnamefont{T.}~\bibnamefont{Gog}},
  \bibinfo{author}{\bibfnamefont{T.~F.} \bibnamefont{Qi}},
  \bibnamefont{et~al.}, \bibinfo{journal}{Phys. Rev. Lett.}
  \textbf{\bibinfo{volume}{111}}, \bibinfo{pages}{057202}
  (\bibinfo{year}{2013}).

\bibitem[{\citenamefont{Wray et~al.}(2015)\citenamefont{Wray, Denlinger, Huang,
  He, Butch, Maple, Hussain, and Chuang}}]{Wray2015}
\bibinfo{author}{\bibfnamefont{L.~A.} \bibnamefont{Wray}},
  \bibinfo{author}{\bibfnamefont{J.}~\bibnamefont{Denlinger}},
  \bibinfo{author}{\bibfnamefont{S.-W.} \bibnamefont{Huang}},
  \bibinfo{author}{\bibfnamefont{H.}~\bibnamefont{He}},
  \bibinfo{author}{\bibfnamefont{N.~P.} \bibnamefont{Butch}},
  \bibinfo{author}{\bibfnamefont{M.~B.} \bibnamefont{Maple}},
  \bibinfo{author}{\bibfnamefont{Z.}~\bibnamefont{Hussain}}, \bibnamefont{and}
  \bibinfo{author}{\bibfnamefont{Y.-D.} \bibnamefont{Chuang}},
  \bibinfo{journal}{Phys. Rev. Lett.} \textbf{\bibinfo{volume}{114}},
  \bibinfo{pages}{236401} (\bibinfo{year}{2015}).

\bibitem[{\citenamefont{de~Groot et~al.}(2005)\citenamefont{de~Groot, Glatzel,
  Bergmann, van Aken, Barrea, Klemme, Hävecker, Knop-Gericke, Heijboer, and
  Weckhuysen}}]{Groot2005}
\bibinfo{author}{\bibfnamefont{F.~M.~F.} \bibnamefont{de~Groot}},
  \bibinfo{author}{\bibfnamefont{P.}~\bibnamefont{Glatzel}},
  \bibinfo{author}{\bibfnamefont{U.}~\bibnamefont{Bergmann}},
  \bibinfo{author}{\bibfnamefont{P.~A.} \bibnamefont{van Aken}},
  \bibinfo{author}{\bibfnamefont{R.~A.} \bibnamefont{Barrea}},
  \bibinfo{author}{\bibfnamefont{S.}~\bibnamefont{Klemme}},
  \bibinfo{author}{\bibfnamefont{M.}~\bibnamefont{Hävecker}},
  \bibinfo{author}{\bibfnamefont{A.}~\bibnamefont{Knop-Gericke}},
  \bibinfo{author}{\bibfnamefont{W.~M.} \bibnamefont{Heijboer}},
  \bibnamefont{and} \bibinfo{author}{\bibfnamefont{B.~M.}
  \bibnamefont{Weckhuysen}}, \bibinfo{journal}{The J Phys. Chem. B}
  \textbf{\bibinfo{volume}{109}}, \bibinfo{pages}{20751}
  (\bibinfo{year}{2005}).

\bibitem[{\citenamefont{Wernet et~al.}(2015)\citenamefont{Wernet, Kunnus,
  Josefsson, Rajkovic, Quevedo, Beye, Schreck, Gr{\"u}bel, Scholz, Nordlund
  et~al.}}]{Wernet2015}
\bibinfo{author}{\bibfnamefont{P.}~\bibnamefont{Wernet}},
  \bibinfo{author}{\bibfnamefont{K.}~\bibnamefont{Kunnus}},
  \bibinfo{author}{\bibfnamefont{I.}~\bibnamefont{Josefsson}},
  \bibinfo{author}{\bibfnamefont{I.}~\bibnamefont{Rajkovic}},
  \bibinfo{author}{\bibfnamefont{W.}~\bibnamefont{Quevedo}},
  \bibinfo{author}{\bibfnamefont{M.}~\bibnamefont{Beye}},
  \bibinfo{author}{\bibfnamefont{S.}~\bibnamefont{Schreck}},
  \bibinfo{author}{\bibfnamefont{S.}~\bibnamefont{Gr{\"u}bel}},
  \bibinfo{author}{\bibfnamefont{M.}~\bibnamefont{Scholz}},
  \bibinfo{author}{\bibfnamefont{D.}~\bibnamefont{Nordlund}},
  \bibnamefont{et~al.}, \bibinfo{journal}{Nature}
  \textbf{\bibinfo{volume}{520}}, \bibinfo{pages}{78} (\bibinfo{year}{2015}).

\bibitem[{\citenamefont{Kramers and Heisenberg}(1925)}]{KramersHeisenberg}
\bibinfo{author}{\bibfnamefont{H.~A.} \bibnamefont{Kramers}} \bibnamefont{and}
  \bibinfo{author}{\bibfnamefont{W.}~\bibnamefont{Heisenberg}},
  \bibinfo{journal}{Z. Phys.} \textbf{\bibinfo{volume}{31}},
  \bibinfo{pages}{681} (\bibinfo{year}{1925}).

\bibitem[{\citenamefont{Jia et~al.}(2018)\citenamefont{Jia, Wang, Mendl,
  Moritz, and Devereaux}}]{Jia2017}
\bibinfo{author}{\bibfnamefont{C.}~\bibnamefont{Jia}},
  \bibinfo{author}{\bibfnamefont{Y.}~\bibnamefont{Wang}},
  \bibinfo{author}{\bibfnamefont{C.}~\bibnamefont{Mendl}},
  \bibinfo{author}{\bibfnamefont{B.}~\bibnamefont{Moritz}}, \bibnamefont{and}
  \bibinfo{author}{\bibfnamefont{T.}~\bibnamefont{Devereaux}},
  \bibinfo{journal}{Comput. Phys. Commun.} \textbf{\bibinfo{volume}{224}},
  \bibinfo{pages}{81 } (\bibinfo{year}{2018}).

\bibitem[{\citenamefont{Tsutsui et~al.}(1999)\citenamefont{Tsutsui, Tohyama,
  and Maekawa}}]{Tsutsui1999}
\bibinfo{author}{\bibfnamefont{K.}~\bibnamefont{Tsutsui}},
  \bibinfo{author}{\bibfnamefont{T.}~\bibnamefont{Tohyama}}, \bibnamefont{and}
  \bibinfo{author}{\bibfnamefont{S.}~\bibnamefont{Maekawa}},
  \bibinfo{journal}{Phys. Rev. Lett.} \textbf{\bibinfo{volume}{83}},
  \bibinfo{pages}{3705} (\bibinfo{year}{1999}).

\bibitem[{\citenamefont{Tsutsui et~al.}(2003)\citenamefont{Tsutsui, Tohyama,
  and Maekawa}}]{Tsutsui2003}
\bibinfo{author}{\bibfnamefont{K.}~\bibnamefont{Tsutsui}},
  \bibinfo{author}{\bibfnamefont{T.}~\bibnamefont{Tohyama}}, \bibnamefont{and}
  \bibinfo{author}{\bibfnamefont{S.}~\bibnamefont{Maekawa}},
  \bibinfo{journal}{Phys. Rev. Lett.} \textbf{\bibinfo{volume}{91}},
  \bibinfo{pages}{117001} (\bibinfo{year}{2003}).

\bibitem[{\citenamefont{Chen et~al.}(2010)\citenamefont{Chen, Moritz, Vernay,
  Hancock, Johnston, Jia, Chabot-Couture, Greven, Elfimov, Sawatzky
  et~al.}}]{Chen2010}
\bibinfo{author}{\bibfnamefont{C.-C.} \bibnamefont{Chen}},
  \bibinfo{author}{\bibfnamefont{B.}~\bibnamefont{Moritz}},
  \bibinfo{author}{\bibfnamefont{F.}~\bibnamefont{Vernay}},
  \bibinfo{author}{\bibfnamefont{J.~N.} \bibnamefont{Hancock}},
  \bibinfo{author}{\bibfnamefont{S.}~\bibnamefont{Johnston}},
  \bibinfo{author}{\bibfnamefont{C.~J.} \bibnamefont{Jia}},
  \bibinfo{author}{\bibfnamefont{G.}~\bibnamefont{Chabot-Couture}},
  \bibinfo{author}{\bibfnamefont{M.}~\bibnamefont{Greven}},
  \bibinfo{author}{\bibfnamefont{I.}~\bibnamefont{Elfimov}},
  \bibinfo{author}{\bibfnamefont{G.~A.} \bibnamefont{Sawatzky}},
  \bibnamefont{et~al.}, \bibinfo{journal}{Phys. Rev. Lett.}
  \textbf{\bibinfo{volume}{105}}, \bibinfo{pages}{177401}
  (\bibinfo{year}{2010}).

\bibitem[{\citenamefont{{Jia} et~al.}(2014)\citenamefont{{Jia}, {Nowadnick},
  {Wohlfeld}, {Kung}, {Chen}, {Johnston}, {Tohyama}, {Moritz}, and
  {Devereaux}}}]{Jia2014}
\bibinfo{author}{\bibfnamefont{C.~J.} \bibnamefont{{Jia}}},
  \bibinfo{author}{\bibfnamefont{E.~A.} \bibnamefont{{Nowadnick}}},
  \bibinfo{author}{\bibfnamefont{K.}~\bibnamefont{{Wohlfeld}}},
  \bibinfo{author}{\bibfnamefont{Y.~F.} \bibnamefont{{Kung}}},
  \bibinfo{author}{\bibfnamefont{C.-C.} \bibnamefont{{Chen}}},
  \bibinfo{author}{\bibfnamefont{S.}~\bibnamefont{{Johnston}}},
  \bibinfo{author}{\bibfnamefont{T.}~\bibnamefont{{Tohyama}}},
  \bibinfo{author}{\bibfnamefont{B.}~\bibnamefont{{Moritz}}}, \bibnamefont{and}
  \bibinfo{author}{\bibfnamefont{T.~P.} \bibnamefont{{Devereaux}}},
  \bibinfo{journal}{Nat. Commun.} \textbf{\bibinfo{volume}{5}},
  \bibinfo{pages}{3314} (\bibinfo{year}{2014}), \eprint{1308.3717}.

\bibitem[{\citenamefont{Chaix et~al.}(2017)\citenamefont{Chaix, Ghiringhelli,
  Peng, Hashimoto, Moritz, Kummer, Brookes, He, Chen, Ishida
  et~al.}}]{chaix2017dispersive}
\bibinfo{author}{\bibfnamefont{L.}~\bibnamefont{Chaix}},
  \bibinfo{author}{\bibfnamefont{G.}~\bibnamefont{Ghiringhelli}},
  \bibinfo{author}{\bibfnamefont{Y.}~\bibnamefont{Peng}},
  \bibinfo{author}{\bibfnamefont{M.}~\bibnamefont{Hashimoto}},
  \bibinfo{author}{\bibfnamefont{B.}~\bibnamefont{Moritz}},
  \bibinfo{author}{\bibfnamefont{K.}~\bibnamefont{Kummer}},
  \bibinfo{author}{\bibfnamefont{N.}~\bibnamefont{Brookes}},
  \bibinfo{author}{\bibfnamefont{Y.}~\bibnamefont{He}},
  \bibinfo{author}{\bibfnamefont{S.}~\bibnamefont{Chen}},
  \bibinfo{author}{\bibfnamefont{S.}~\bibnamefont{Ishida}},
  \bibnamefont{et~al.}, \bibinfo{journal}{Nat. Phys.}
  \textbf{\bibinfo{volume}{13}}, \bibinfo{pages}{952} (\bibinfo{year}{2017}).

\bibitem[{\citenamefont{Jia et~al.}(2012)\citenamefont{Jia, Chen, Sorini,
  Moritz, and Devereaux}}]{Jia2012}
\bibinfo{author}{\bibfnamefont{C.~J.} \bibnamefont{Jia}},
  \bibinfo{author}{\bibfnamefont{C.-C.} \bibnamefont{Chen}},
  \bibinfo{author}{\bibfnamefont{A.~P.} \bibnamefont{Sorini}},
  \bibinfo{author}{\bibfnamefont{B.}~\bibnamefont{Moritz}}, \bibnamefont{and}
  \bibinfo{author}{\bibfnamefont{T.~P.} \bibnamefont{Devereaux}},
  \bibinfo{journal}{New J Phys.} \textbf{\bibinfo{volume}{14}},
  \bibinfo{pages}{113038} (\bibinfo{year}{2012}).

\bibitem[{\citenamefont{Jia et~al.}(2016)\citenamefont{Jia, Wohlfeld, Wang,
  Moritz, and Devereaux}}]{Jia2016}
\bibinfo{author}{\bibfnamefont{C.}~\bibnamefont{Jia}},
  \bibinfo{author}{\bibfnamefont{K.}~\bibnamefont{Wohlfeld}},
  \bibinfo{author}{\bibfnamefont{Y.}~\bibnamefont{Wang}},
  \bibinfo{author}{\bibfnamefont{B.}~\bibnamefont{Moritz}}, \bibnamefont{and}
  \bibinfo{author}{\bibfnamefont{T.~P.} \bibnamefont{Devereaux}},
  \bibinfo{journal}{Phys. Rev. X} \textbf{\bibinfo{volume}{6}},
  \bibinfo{pages}{021020} (\bibinfo{year}{2016}).

\bibitem[{\citenamefont{Haverkort}(2010)}]{Haverkort2010}
\bibinfo{author}{\bibfnamefont{M.~W.} \bibnamefont{Haverkort}},
  \bibinfo{journal}{Phys. Rev. Lett.} \textbf{\bibinfo{volume}{105}},
  \bibinfo{pages}{167404} (\bibinfo{year}{2010}).

\bibitem[{\citenamefont{Kourtis et~al.}(2012)\citenamefont{Kourtis, van~den
  Brink, and Daghofer}}]{Kourtis2012}
\bibinfo{author}{\bibfnamefont{S.}~\bibnamefont{Kourtis}},
  \bibinfo{author}{\bibfnamefont{J.}~\bibnamefont{van~den Brink}},
  \bibnamefont{and} \bibinfo{author}{\bibfnamefont{M.}~\bibnamefont{Daghofer}},
  \bibinfo{journal}{Phys. Rev. B} \textbf{\bibinfo{volume}{85}},
  \bibinfo{pages}{064423} (\bibinfo{year}{2012}).

\bibitem[{\citenamefont{Luo et~al.}(1993)\citenamefont{Luo, Trammell, and
  Hannon}}]{Luo1993}
\bibinfo{author}{\bibfnamefont{J.}~\bibnamefont{Luo}},
  \bibinfo{author}{\bibfnamefont{G.~T.} \bibnamefont{Trammell}},
  \bibnamefont{and} \bibinfo{author}{\bibfnamefont{J.~P.}
  \bibnamefont{Hannon}}, \bibinfo{journal}{Phys. Rev. Lett.}
  \textbf{\bibinfo{volume}{71}}, \bibinfo{pages}{287} (\bibinfo{year}{1993}).

\bibitem[{\citenamefont{de~Groot et~al.}(1998)\citenamefont{de~Groot, Kuiper,
  and Sawatzky}}]{Groot1998}
\bibinfo{author}{\bibfnamefont{F.~M.~F.} \bibnamefont{de~Groot}},
  \bibinfo{author}{\bibfnamefont{P.}~\bibnamefont{Kuiper}}, \bibnamefont{and}
  \bibinfo{author}{\bibfnamefont{G.~A.} \bibnamefont{Sawatzky}},
  \bibinfo{journal}{Phys. Rev. B} \textbf{\bibinfo{volume}{57}},
  \bibinfo{pages}{14584} (\bibinfo{year}{1998}).

\bibitem[{\citenamefont{Bisogni
  et~al.}(2012{\natexlab{b}})\citenamefont{Bisogni, Moretti~Sala, Bendounan,
  Brookes, Ghiringhelli, and Braicovich}}]{Bisogni2012b}
\bibinfo{author}{\bibfnamefont{V.}~\bibnamefont{Bisogni}},
  \bibinfo{author}{\bibfnamefont{M.}~\bibnamefont{Moretti~Sala}},
  \bibinfo{author}{\bibfnamefont{A.}~\bibnamefont{Bendounan}},
  \bibinfo{author}{\bibfnamefont{N.~B.} \bibnamefont{Brookes}},
  \bibinfo{author}{\bibfnamefont{G.}~\bibnamefont{Ghiringhelli}},
  \bibnamefont{and}
  \bibinfo{author}{\bibfnamefont{L.}~\bibnamefont{Braicovich}},
  \bibinfo{journal}{Phys. Rev. B} \textbf{\bibinfo{volume}{85}},
  \bibinfo{pages}{214528} (\bibinfo{year}{2012}{\natexlab{b}}).

\bibitem[{\citenamefont{Butorin et~al.}(1996)\citenamefont{Butorin, Guo,
  Magnuson, Kuiper, and Nordgren}}]{butorin1996low}
\bibinfo{author}{\bibfnamefont{S.}~\bibnamefont{Butorin}},
  \bibinfo{author}{\bibfnamefont{J.-H.} \bibnamefont{Guo}},
  \bibinfo{author}{\bibfnamefont{M.}~\bibnamefont{Magnuson}},
  \bibinfo{author}{\bibfnamefont{P.}~\bibnamefont{Kuiper}}, \bibnamefont{and}
  \bibinfo{author}{\bibfnamefont{J.}~\bibnamefont{Nordgren}},
  \bibinfo{journal}{Phys. Rev. B} \textbf{\bibinfo{volume}{54}},
  \bibinfo{pages}{4405} (\bibinfo{year}{1996}).

\bibitem[{\citenamefont{Ghiringhelli et~al.}(2004)\citenamefont{Ghiringhelli,
  Brookes, Annese, Berger, Dallera, Grioni, Perfetti, Tagliaferri, and
  Braicovich}}]{ghiringhelli2004low}
\bibinfo{author}{\bibfnamefont{G.}~\bibnamefont{Ghiringhelli}},
  \bibinfo{author}{\bibfnamefont{N.}~\bibnamefont{Brookes}},
  \bibinfo{author}{\bibfnamefont{E.}~\bibnamefont{Annese}},
  \bibinfo{author}{\bibfnamefont{H.}~\bibnamefont{Berger}},
  \bibinfo{author}{\bibfnamefont{C.}~\bibnamefont{Dallera}},
  \bibinfo{author}{\bibfnamefont{M.}~\bibnamefont{Grioni}},
  \bibinfo{author}{\bibfnamefont{L.}~\bibnamefont{Perfetti}},
  \bibinfo{author}{\bibfnamefont{A.}~\bibnamefont{Tagliaferri}},
  \bibnamefont{and}
  \bibinfo{author}{\bibfnamefont{L.}~\bibnamefont{Braicovich}},
  \bibinfo{journal}{Phys. Rev. Lett.} \textbf{\bibinfo{volume}{92}},
  \bibinfo{pages}{117406} (\bibinfo{year}{2004}).

\bibitem[{\citenamefont{Lee et~al.}(2013)\citenamefont{Lee, Johnston, Moritz,
  Lee, Yi, Zhou, Schmitt, Patthey, Strocov, Kudo et~al.}}]{lee2013role}
\bibinfo{author}{\bibfnamefont{W.}~\bibnamefont{Lee}},
  \bibinfo{author}{\bibfnamefont{S.}~\bibnamefont{Johnston}},
  \bibinfo{author}{\bibfnamefont{B.}~\bibnamefont{Moritz}},
  \bibinfo{author}{\bibfnamefont{J.}~\bibnamefont{Lee}},
  \bibinfo{author}{\bibfnamefont{M.}~\bibnamefont{Yi}},
  \bibinfo{author}{\bibfnamefont{K.}~\bibnamefont{Zhou}},
  \bibinfo{author}{\bibfnamefont{T.}~\bibnamefont{Schmitt}},
  \bibinfo{author}{\bibfnamefont{L.}~\bibnamefont{Patthey}},
  \bibinfo{author}{\bibfnamefont{V.}~\bibnamefont{Strocov}},
  \bibinfo{author}{\bibfnamefont{K.}~\bibnamefont{Kudo}}, \bibnamefont{et~al.},
  \bibinfo{journal}{Phys. Rev. Lett.} \textbf{\bibinfo{volume}{110}},
  \bibinfo{pages}{265502} (\bibinfo{year}{2013}).

\bibitem[{\citenamefont{Peng et~al.}(2015)\citenamefont{Peng, Hashimoto, Sala,
  Amorese, Brookes, Dellea, Lee, Minola, Schmitt, Yoshida
  et~al.}}]{peng2015magnetic}
\bibinfo{author}{\bibfnamefont{Y.}~\bibnamefont{Peng}},
  \bibinfo{author}{\bibfnamefont{M.}~\bibnamefont{Hashimoto}},
  \bibinfo{author}{\bibfnamefont{M.~M.} \bibnamefont{Sala}},
  \bibinfo{author}{\bibfnamefont{A.}~\bibnamefont{Amorese}},
  \bibinfo{author}{\bibfnamefont{N.}~\bibnamefont{Brookes}},
  \bibinfo{author}{\bibfnamefont{G.}~\bibnamefont{Dellea}},
  \bibinfo{author}{\bibfnamefont{W.-S.} \bibnamefont{Lee}},
  \bibinfo{author}{\bibfnamefont{M.}~\bibnamefont{Minola}},
  \bibinfo{author}{\bibfnamefont{T.}~\bibnamefont{Schmitt}},
  \bibinfo{author}{\bibfnamefont{Y.}~\bibnamefont{Yoshida}},
  \bibnamefont{et~al.}, \bibinfo{journal}{Phys. Rev. B}
  \textbf{\bibinfo{volume}{92}}, \bibinfo{pages}{064517}
  (\bibinfo{year}{2015}).

\bibitem[{\citenamefont{Balzer et~al.}(2011)\citenamefont{Balzer, Gdaniec, and
  Potthoff}}]{balzer2011krylov}
\bibinfo{author}{\bibfnamefont{M.}~\bibnamefont{Balzer}},
  \bibinfo{author}{\bibfnamefont{N.}~\bibnamefont{Gdaniec}}, \bibnamefont{and}
  \bibinfo{author}{\bibfnamefont{M.}~\bibnamefont{Potthoff}},
  \bibinfo{journal}{J. Phys. Condens. Matter} \textbf{\bibinfo{volume}{24}},
  \bibinfo{pages}{035603} (\bibinfo{year}{2011}).

\bibitem[{\citenamefont{Bon{\v{c}}a et~al.}(1999)\citenamefont{Bon{\v{c}}a,
  Trugman, and Batisti{\'c}}}]{bonvca1999holstein}
\bibinfo{author}{\bibfnamefont{J.}~\bibnamefont{Bon{\v{c}}a}},
  \bibinfo{author}{\bibfnamefont{S.}~\bibnamefont{Trugman}}, \bibnamefont{and}
  \bibinfo{author}{\bibfnamefont{I.}~\bibnamefont{Batisti{\'c}}},
  \bibinfo{journal}{Phys. Rev. B} \textbf{\bibinfo{volume}{60}},
  \bibinfo{pages}{1633} (\bibinfo{year}{1999}).

\bibitem[{\citenamefont{Schollw{\"o}ck}(2011)}]{schollwock2011density}
\bibinfo{author}{\bibfnamefont{U.}~\bibnamefont{Schollw{\"o}ck}},
  \bibinfo{journal}{Ann. Phys.} \textbf{\bibinfo{volume}{326}},
  \bibinfo{pages}{96} (\bibinfo{year}{2011}).

\bibitem[{\citenamefont{Aoki et~al.}(2014)\citenamefont{Aoki, Tsuji, Eckstein,
  Kollar, Oka, and Werner}}]{aoki2014nonequilibrium}
\bibinfo{author}{\bibfnamefont{H.}~\bibnamefont{Aoki}},
  \bibinfo{author}{\bibfnamefont{N.}~\bibnamefont{Tsuji}},
  \bibinfo{author}{\bibfnamefont{M.}~\bibnamefont{Eckstein}},
  \bibinfo{author}{\bibfnamefont{M.}~\bibnamefont{Kollar}},
  \bibinfo{author}{\bibfnamefont{T.}~\bibnamefont{Oka}}, \bibnamefont{and}
  \bibinfo{author}{\bibfnamefont{P.}~\bibnamefont{Werner}},
  \bibinfo{journal}{Rev. Mod. Phys.} \textbf{\bibinfo{volume}{86}},
  \bibinfo{pages}{779} (\bibinfo{year}{2014}).

\bibitem[{\citenamefont{Balzer and Potthoff}(2011)}]{balzer2011nonequilibrium}
\bibinfo{author}{\bibfnamefont{M.}~\bibnamefont{Balzer}} \bibnamefont{and}
  \bibinfo{author}{\bibfnamefont{M.}~\bibnamefont{Potthoff}},
  \bibinfo{journal}{Phys. Rev. B} \textbf{\bibinfo{volume}{83}},
  \bibinfo{pages}{195132} (\bibinfo{year}{2011}).

\bibitem[{\citenamefont{Spataru et~al.}(2004)\citenamefont{Spataru, Benedict,
  and Louie}}]{Spataru2004}
\bibinfo{author}{\bibfnamefont{C.~D.} \bibnamefont{Spataru}},
  \bibinfo{author}{\bibfnamefont{L.~X.} \bibnamefont{Benedict}},
  \bibnamefont{and} \bibinfo{author}{\bibfnamefont{S.~G.} \bibnamefont{Louie}},
  \bibinfo{journal}{Phys. Rev. B} \textbf{\bibinfo{volume}{69}},
  \bibinfo{pages}{205204} (\bibinfo{year}{2004}).

\bibitem[{\citenamefont{Park et~al.}(2009)\citenamefont{Park, Giustino,
  Spataru, Cohen, and Louie}}]{Park2009a}
\bibinfo{author}{\bibfnamefont{C.~H.} \bibnamefont{Park}},
  \bibinfo{author}{\bibfnamefont{F.}~\bibnamefont{Giustino}},
  \bibinfo{author}{\bibfnamefont{C.~D.} \bibnamefont{Spataru}},
  \bibinfo{author}{\bibfnamefont{M.~L.} \bibnamefont{Cohen}}, \bibnamefont{and}
  \bibinfo{author}{\bibfnamefont{S.~G.} \bibnamefont{Louie}},
  \bibinfo{journal}{Nano Lett.} \textbf{\bibinfo{volume}{9}},
  \bibinfo{pages}{4234} (\bibinfo{year}{2009}).

\bibitem[{\citenamefont{Perfetto et~al.}(2016)\citenamefont{Perfetto, Sangalli,
  Marini, and Stefanucci}}]{Perfetto2016b}
\bibinfo{author}{\bibfnamefont{E.}~\bibnamefont{Perfetto}},
  \bibinfo{author}{\bibfnamefont{D.}~\bibnamefont{Sangalli}},
  \bibinfo{author}{\bibfnamefont{A.}~\bibnamefont{Marini}}, \bibnamefont{and}
  \bibinfo{author}{\bibfnamefont{G.}~\bibnamefont{Stefanucci}},
  \bibinfo{journal}{Phys. Rev. B} \textbf{\bibinfo{volume}{94}},
  \bibinfo{pages}{245303} (\bibinfo{year}{2016}), \eprint{1607.06449}.

\bibitem[{\citenamefont{Braun et~al.}(2016)\citenamefont{Braun, Rausch,
  Potthoff, and Ebert}}]{Braun2016}
\bibinfo{author}{\bibfnamefont{J.}~\bibnamefont{Braun}},
  \bibinfo{author}{\bibfnamefont{R.}~\bibnamefont{Rausch}},
  \bibinfo{author}{\bibfnamefont{M.}~\bibnamefont{Potthoff}}, \bibnamefont{and}
  \bibinfo{author}{\bibfnamefont{H.}~\bibnamefont{Ebert}},
  \bibinfo{journal}{Phys. Rev. B} \textbf{\bibinfo{volume}{94}},
  \bibinfo{pages}{125128} (\bibinfo{year}{2016}), \eprint{1605.08596}.

\bibitem[{\citenamefont{Marques et~al.}(2012)\citenamefont{Marques, Maitra,
  Nogueira, Gross, and Rubio}}]{Marques2012a}
\bibinfo{author}{\bibfnamefont{M.~A.} \bibnamefont{Marques}},
  \bibinfo{author}{\bibfnamefont{N.~T.} \bibnamefont{Maitra}},
  \bibinfo{author}{\bibfnamefont{F.~M.} \bibnamefont{Nogueira}},
  \bibinfo{author}{\bibfnamefont{E.}~\bibnamefont{Gross}}, \bibnamefont{and}
  \bibinfo{author}{\bibfnamefont{A.}~\bibnamefont{Rubio}},
  \emph{\bibinfo{title}{{Fundamentals of Time-Dependent Density Functional
  Theory}}}, vol. \bibinfo{volume}{837} of \emph{\bibinfo{series}{Lecture Notes
  in Physics}} (\bibinfo{publisher}{Springer}, \bibinfo{year}{2012}).

\bibitem[{\citenamefont{Wopperer et~al.}(2017)\citenamefont{Wopperer, {De
  Giovannini}, and Rubio}}]{Wopperer2017}
\bibinfo{author}{\bibfnamefont{P.}~\bibnamefont{Wopperer}},
  \bibinfo{author}{\bibfnamefont{U.}~\bibnamefont{{De Giovannini}}},
  \bibnamefont{and} \bibinfo{author}{\bibfnamefont{A.}~\bibnamefont{Rubio}},
  \bibinfo{journal}{Eur. Phys. J B} \textbf{\bibinfo{volume}{90}},
  \bibinfo{pages}{51} (\bibinfo{year}{2017}).

\bibitem[{\citenamefont{Yusupov et~al.}(2010)\citenamefont{Yusupov, Mertelj,
  Kabanov, Brazovskii, Kusar, Chu, Fisher, and
  Mihailovic}}]{yusupov2010coherent}
\bibinfo{author}{\bibfnamefont{R.}~\bibnamefont{Yusupov}},
  \bibinfo{author}{\bibfnamefont{T.}~\bibnamefont{Mertelj}},
  \bibinfo{author}{\bibfnamefont{V.~V.} \bibnamefont{Kabanov}},
  \bibinfo{author}{\bibfnamefont{S.}~\bibnamefont{Brazovskii}},
  \bibinfo{author}{\bibfnamefont{P.}~\bibnamefont{Kusar}},
  \bibinfo{author}{\bibfnamefont{J.-H.} \bibnamefont{Chu}},
  \bibinfo{author}{\bibfnamefont{I.~R.} \bibnamefont{Fisher}},
  \bibnamefont{and}
  \bibinfo{author}{\bibfnamefont{D.}~\bibnamefont{Mihailovic}},
  \bibinfo{journal}{Nat. Phys.} \textbf{\bibinfo{volume}{6}},
  \bibinfo{pages}{681} (\bibinfo{year}{2010}).

\bibitem[{\citenamefont{Smallwood et~al.}(2012)\citenamefont{Smallwood, Hinton,
  Jozwiak, Zhang, Koralek, Eisaki, Lee, Orenstein, and
  Lanzara}}]{smallwood2012tracking}
\bibinfo{author}{\bibfnamefont{C.~L.} \bibnamefont{Smallwood}},
  \bibinfo{author}{\bibfnamefont{J.~P.} \bibnamefont{Hinton}},
  \bibinfo{author}{\bibfnamefont{C.}~\bibnamefont{Jozwiak}},
  \bibinfo{author}{\bibfnamefont{W.}~\bibnamefont{Zhang}},
  \bibinfo{author}{\bibfnamefont{J.~D.} \bibnamefont{Koralek}},
  \bibinfo{author}{\bibfnamefont{H.}~\bibnamefont{Eisaki}},
  \bibinfo{author}{\bibfnamefont{D.-H.} \bibnamefont{Lee}},
  \bibinfo{author}{\bibfnamefont{J.}~\bibnamefont{Orenstein}},
  \bibnamefont{and} \bibinfo{author}{\bibfnamefont{A.}~\bibnamefont{Lanzara}},
  \bibinfo{journal}{Science} \textbf{\bibinfo{volume}{336}},
  \bibinfo{pages}{1137} (\bibinfo{year}{2012}).

\bibitem[{\citenamefont{Dal~Conte et~al.}(2015)\citenamefont{Dal~Conte, Vidmar,
  Gole{\v{z}}, Mierzejewski, Soavi, Peli, Banfi, Ferrini, Comin, Ludbrook
  et~al.}}]{dal2015snapshots}
\bibinfo{author}{\bibfnamefont{S.}~\bibnamefont{Dal~Conte}},
  \bibinfo{author}{\bibfnamefont{L.}~\bibnamefont{Vidmar}},
  \bibinfo{author}{\bibfnamefont{D.}~\bibnamefont{Gole{\v{z}}}},
  \bibinfo{author}{\bibfnamefont{M.}~\bibnamefont{Mierzejewski}},
  \bibinfo{author}{\bibfnamefont{G.}~\bibnamefont{Soavi}},
  \bibinfo{author}{\bibfnamefont{S.}~\bibnamefont{Peli}},
  \bibinfo{author}{\bibfnamefont{F.}~\bibnamefont{Banfi}},
  \bibinfo{author}{\bibfnamefont{G.}~\bibnamefont{Ferrini}},
  \bibinfo{author}{\bibfnamefont{R.}~\bibnamefont{Comin}},
  \bibinfo{author}{\bibfnamefont{B.~M.} \bibnamefont{Ludbrook}},
  \bibnamefont{et~al.}, \bibinfo{journal}{Nat. Phys.}
  \textbf{\bibinfo{volume}{11}}, \bibinfo{pages}{421} (\bibinfo{year}{2015}).

\bibitem[{\citenamefont{Vishik et~al.}(2017)\citenamefont{Vishik, Mahmood,
  Alpichshev, Gedik, Higgins, and Greene}}]{vishik2017ultrafast}
\bibinfo{author}{\bibfnamefont{I.}~\bibnamefont{Vishik}},
  \bibinfo{author}{\bibfnamefont{F.}~\bibnamefont{Mahmood}},
  \bibinfo{author}{\bibfnamefont{Z.}~\bibnamefont{Alpichshev}},
  \bibinfo{author}{\bibfnamefont{N.}~\bibnamefont{Gedik}},
  \bibinfo{author}{\bibfnamefont{J.}~\bibnamefont{Higgins}}, \bibnamefont{and}
  \bibinfo{author}{\bibfnamefont{R.}~\bibnamefont{Greene}},
  \bibinfo{journal}{Phys. Rev. B} \textbf{\bibinfo{volume}{95}},
  \bibinfo{pages}{115125} (\bibinfo{year}{2017}).

\bibitem[{\citenamefont{Eckstein and
  Werner}(2011)}]{eckstein2011thermalization}
\bibinfo{author}{\bibfnamefont{M.}~\bibnamefont{Eckstein}} \bibnamefont{and}
  \bibinfo{author}{\bibfnamefont{P.}~\bibnamefont{Werner}},
  \bibinfo{journal}{Phys. Rev. B} \textbf{\bibinfo{volume}{84}},
  \bibinfo{pages}{035122} (\bibinfo{year}{2011}).

\bibitem[{\citenamefont{Sentef et~al.}(2013)\citenamefont{Sentef, Kemper,
  Moritz, Freericks, Shen, and Devereaux}}]{sentef2013examining}
\bibinfo{author}{\bibfnamefont{M.}~\bibnamefont{Sentef}},
  \bibinfo{author}{\bibfnamefont{A.~F.} \bibnamefont{Kemper}},
  \bibinfo{author}{\bibfnamefont{B.}~\bibnamefont{Moritz}},
  \bibinfo{author}{\bibfnamefont{J.~K.} \bibnamefont{Freericks}},
  \bibinfo{author}{\bibfnamefont{Z.-X.} \bibnamefont{Shen}}, \bibnamefont{and}
  \bibinfo{author}{\bibfnamefont{T.~P.} \bibnamefont{Devereaux}},
  \bibinfo{journal}{Phys. Rev. X} \textbf{\bibinfo{volume}{3}},
  \bibinfo{pages}{041033} (\bibinfo{year}{2013}).

\bibitem[{\citenamefont{Murakami et~al.}(2015)\citenamefont{Murakami, Werner,
  Tsuji, and Aoki}}]{murakami2015interaction}
\bibinfo{author}{\bibfnamefont{Y.}~\bibnamefont{Murakami}},
  \bibinfo{author}{\bibfnamefont{P.}~\bibnamefont{Werner}},
  \bibinfo{author}{\bibfnamefont{N.}~\bibnamefont{Tsuji}}, \bibnamefont{and}
  \bibinfo{author}{\bibfnamefont{H.}~\bibnamefont{Aoki}},
  \bibinfo{journal}{Phys. Rev. B} \textbf{\bibinfo{volume}{91}},
  \bibinfo{pages}{045128} (\bibinfo{year}{2015}).

\bibitem[{\citenamefont{Strand et~al.}(2017)\citenamefont{Strand, Gole{\v{z}},
  Eckstein, and Werner}}]{strand2017hund}
\bibinfo{author}{\bibfnamefont{H.~U.} \bibnamefont{Strand}},
  \bibinfo{author}{\bibfnamefont{D.}~\bibnamefont{Gole{\v{z}}}},
  \bibinfo{author}{\bibfnamefont{M.}~\bibnamefont{Eckstein}}, \bibnamefont{and}
  \bibinfo{author}{\bibfnamefont{P.}~\bibnamefont{Werner}},
  \bibinfo{journal}{Phys. Rev. B} \textbf{\bibinfo{volume}{96}},
  \bibinfo{pages}{165104} (\bibinfo{year}{2017}).

\bibitem[{\citenamefont{Gerber et~al.}(2017)\citenamefont{Gerber, Yang, Zhu,
  Soifer, Sobota, Rebec, Lee, Jia, Moritz, Jia et~al.}}]{gerber2017femtosecond}
\bibinfo{author}{\bibfnamefont{S.}~\bibnamefont{Gerber}},
  \bibinfo{author}{\bibfnamefont{S.-L.} \bibnamefont{Yang}},
  \bibinfo{author}{\bibfnamefont{D.}~\bibnamefont{Zhu}},
  \bibinfo{author}{\bibfnamefont{H.}~\bibnamefont{Soifer}},
  \bibinfo{author}{\bibfnamefont{J.}~\bibnamefont{Sobota}},
  \bibinfo{author}{\bibfnamefont{S.}~\bibnamefont{Rebec}},
  \bibinfo{author}{\bibfnamefont{J.}~\bibnamefont{Lee}},
  \bibinfo{author}{\bibfnamefont{T.}~\bibnamefont{Jia}},
  \bibinfo{author}{\bibfnamefont{B.}~\bibnamefont{Moritz}},
  \bibinfo{author}{\bibfnamefont{C.}~\bibnamefont{Jia}}, \bibnamefont{et~al.},
  \bibinfo{journal}{Science} \textbf{\bibinfo{volume}{357}},
  \bibinfo{pages}{71} (\bibinfo{year}{2017}).

\bibitem[{\citenamefont{Mandal et~al.}(2014)\citenamefont{Mandal, Cohen, and
  Haule}}]{mandal2014strong}
\bibinfo{author}{\bibfnamefont{S.}~\bibnamefont{Mandal}},
  \bibinfo{author}{\bibfnamefont{R.~E.} \bibnamefont{Cohen}}, \bibnamefont{and}
  \bibinfo{author}{\bibfnamefont{K.}~\bibnamefont{Haule}},
  \bibinfo{journal}{Phys. Rev. B} \textbf{\bibinfo{volume}{89}},
  \bibinfo{pages}{220502} (\bibinfo{year}{2014}).

\bibitem[{\citenamefont{Claassen et~al.}(2016)\citenamefont{Claassen, Jia,
  Moritz, and Devereaux}}]{claassen2016all}
\bibinfo{author}{\bibfnamefont{M.}~\bibnamefont{Claassen}},
  \bibinfo{author}{\bibfnamefont{C.}~\bibnamefont{Jia}},
  \bibinfo{author}{\bibfnamefont{B.}~\bibnamefont{Moritz}}, \bibnamefont{and}
  \bibinfo{author}{\bibfnamefont{T.~P.} \bibnamefont{Devereaux}},
  \bibinfo{journal}{Nat. Commun.} \textbf{\bibinfo{volume}{7}},
  \bibinfo{pages}{13074} (\bibinfo{year}{2016}).

\bibitem[{\citenamefont{H{\"u}bener et~al.}(2017)\citenamefont{H{\"u}bener,
  Sentef, De~Giovannini, Kemper, and Rubio}}]{hubener2017creating}
\bibinfo{author}{\bibfnamefont{H.}~\bibnamefont{H{\"u}bener}},
  \bibinfo{author}{\bibfnamefont{M.~A.} \bibnamefont{Sentef}},
  \bibinfo{author}{\bibfnamefont{U.}~\bibnamefont{De~Giovannini}},
  \bibinfo{author}{\bibfnamefont{A.~F.} \bibnamefont{Kemper}},
  \bibnamefont{and} \bibinfo{author}{\bibfnamefont{A.}~\bibnamefont{Rubio}},
  \bibinfo{journal}{Nat. Commun.} \textbf{\bibinfo{volume}{8}},
  \bibinfo{pages}{13940} (\bibinfo{year}{2017}).

\bibitem[{\citenamefont{Mentink et~al.}(2015)\citenamefont{Mentink, Balzer, and
  Eckstein}}]{mentink2015ultrafast}
\bibinfo{author}{\bibfnamefont{J.}~\bibnamefont{Mentink}},
  \bibinfo{author}{\bibfnamefont{K.}~\bibnamefont{Balzer}}, \bibnamefont{and}
  \bibinfo{author}{\bibfnamefont{M.}~\bibnamefont{Eckstein}},
  \bibinfo{journal}{Nat. Commun.} \textbf{\bibinfo{volume}{6}},
  \bibinfo{pages}{6708} (\bibinfo{year}{2015}).

\bibitem[{\citenamefont{Sobota et~al.}(2012)\citenamefont{Sobota, Yang,
  Analytis, Chen, Fisher, Kirchmann, and Shen}}]{sobota2012ultrafast}
\bibinfo{author}{\bibfnamefont{J.~A.} \bibnamefont{Sobota}},
  \bibinfo{author}{\bibfnamefont{S.}~\bibnamefont{Yang}},
  \bibinfo{author}{\bibfnamefont{J.~G.} \bibnamefont{Analytis}},
  \bibinfo{author}{\bibfnamefont{Y.}~\bibnamefont{Chen}},
  \bibinfo{author}{\bibfnamefont{I.~R.} \bibnamefont{Fisher}},
  \bibinfo{author}{\bibfnamefont{P.~S.} \bibnamefont{Kirchmann}},
  \bibnamefont{and} \bibinfo{author}{\bibfnamefont{Z.-X.} \bibnamefont{Shen}},
  \bibinfo{journal}{Phys. Rev. Lett.} \textbf{\bibinfo{volume}{108}},
  \bibinfo{pages}{117403} (\bibinfo{year}{2012}).

\bibitem[{\citenamefont{Niesner et~al.}(2012)\citenamefont{Niesner, Fauster,
  Eremeev, Menshchikova, Koroteev, Protogenov, Chulkov, Tereshchenko, Kokh,
  Alekperov et~al.}}]{niesner2012unoccupied}
\bibinfo{author}{\bibfnamefont{D.}~\bibnamefont{Niesner}},
  \bibinfo{author}{\bibfnamefont{T.}~\bibnamefont{Fauster}},
  \bibinfo{author}{\bibfnamefont{S.~V.} \bibnamefont{Eremeev}},
  \bibinfo{author}{\bibfnamefont{T.~V.} \bibnamefont{Menshchikova}},
  \bibinfo{author}{\bibfnamefont{Y.~M.} \bibnamefont{Koroteev}},
  \bibinfo{author}{\bibfnamefont{A.~P.} \bibnamefont{Protogenov}},
  \bibinfo{author}{\bibfnamefont{E.~V.} \bibnamefont{Chulkov}},
  \bibinfo{author}{\bibfnamefont{O.~E.} \bibnamefont{Tereshchenko}},
  \bibinfo{author}{\bibfnamefont{K.~A.} \bibnamefont{Kokh}},
  \bibinfo{author}{\bibfnamefont{O.}~\bibnamefont{Alekperov}},
  \bibnamefont{et~al.}, \bibinfo{journal}{Phys. Rev. B}
  \textbf{\bibinfo{volume}{86}}, \bibinfo{pages}{205403}
  (\bibinfo{year}{2012}).

\bibitem[{\citenamefont{Sobota et~al.}(2013)\citenamefont{Sobota, Yang, Kemper,
  Lee, Schmitt, Li, Moore, Analytis, Fisher, Kirchmann
  et~al.}}]{sobota2013direct}
\bibinfo{author}{\bibfnamefont{J.~A.} \bibnamefont{Sobota}},
  \bibinfo{author}{\bibfnamefont{S.-L.} \bibnamefont{Yang}},
  \bibinfo{author}{\bibfnamefont{A.~F.} \bibnamefont{Kemper}},
  \bibinfo{author}{\bibfnamefont{J.}~\bibnamefont{Lee}},
  \bibinfo{author}{\bibfnamefont{F.~T.} \bibnamefont{Schmitt}},
  \bibinfo{author}{\bibfnamefont{W.}~\bibnamefont{Li}},
  \bibinfo{author}{\bibfnamefont{R.~G.} \bibnamefont{Moore}},
  \bibinfo{author}{\bibfnamefont{J.~G.} \bibnamefont{Analytis}},
  \bibinfo{author}{\bibfnamefont{I.~R.} \bibnamefont{Fisher}},
  \bibinfo{author}{\bibfnamefont{P.~S.} \bibnamefont{Kirchmann}},
  \bibnamefont{et~al.}, \bibinfo{journal}{Phys. Rev. Lett.}
  \textbf{\bibinfo{volume}{111}}, \bibinfo{pages}{136802}
  (\bibinfo{year}{2013}).

\bibitem[{\citenamefont{Sobota et~al.}(2014{\natexlab{a}})\citenamefont{Sobota,
  Yang, Leuenberger, Kemper, Analytis, Fisher, Kirchmann, Devereaux, and
  Shen}}]{sobota2014distinguishing}
\bibinfo{author}{\bibfnamefont{J.~A.} \bibnamefont{Sobota}},
  \bibinfo{author}{\bibfnamefont{S.-L.} \bibnamefont{Yang}},
  \bibinfo{author}{\bibfnamefont{D.}~\bibnamefont{Leuenberger}},
  \bibinfo{author}{\bibfnamefont{A.~F.} \bibnamefont{Kemper}},
  \bibinfo{author}{\bibfnamefont{J.~G.} \bibnamefont{Analytis}},
  \bibinfo{author}{\bibfnamefont{I.~R.} \bibnamefont{Fisher}},
  \bibinfo{author}{\bibfnamefont{P.~S.} \bibnamefont{Kirchmann}},
  \bibinfo{author}{\bibfnamefont{T.~P.} \bibnamefont{Devereaux}},
  \bibnamefont{and} \bibinfo{author}{\bibfnamefont{Z.-X.} \bibnamefont{Shen}},
  \bibinfo{journal}{Phys. Rev. Lett.} \textbf{\bibinfo{volume}{113}},
  \bibinfo{pages}{157401} (\bibinfo{year}{2014}{\natexlab{a}}).

\bibitem[{\citenamefont{Sobota et~al.}(2014{\natexlab{b}})\citenamefont{Sobota,
  Yang, Leuenberger, Kemper, Analytis, Fisher, Kirchmann, Devereaux, and
  Shen}}]{sobota2014ultrafast}
\bibinfo{author}{\bibfnamefont{J.~A.} \bibnamefont{Sobota}},
  \bibinfo{author}{\bibfnamefont{S.-L.} \bibnamefont{Yang}},
  \bibinfo{author}{\bibfnamefont{D.}~\bibnamefont{Leuenberger}},
  \bibinfo{author}{\bibfnamefont{A.~F.} \bibnamefont{Kemper}},
  \bibinfo{author}{\bibfnamefont{J.~G.} \bibnamefont{Analytis}},
  \bibinfo{author}{\bibfnamefont{I.~R.} \bibnamefont{Fisher}},
  \bibinfo{author}{\bibfnamefont{P.~S.} \bibnamefont{Kirchmann}},
  \bibinfo{author}{\bibfnamefont{T.~P.} \bibnamefont{Devereaux}},
  \bibnamefont{and} \bibinfo{author}{\bibfnamefont{Z.-X.} \bibnamefont{Shen}},
  \bibinfo{journal}{J Electron Spectrosc.} \textbf{\bibinfo{volume}{195}},
  \bibinfo{pages}{249} (\bibinfo{year}{2014}{\natexlab{b}}).

\bibitem[{\citenamefont{Jozwiak et~al.}(2016)\citenamefont{Jozwiak, Sobota,
  Gotlieb, Kemper, Rotundu, Birgeneau, Hussain, Lee, Shen, and
  Lanzara}}]{jozwiak2016spin}
\bibinfo{author}{\bibfnamefont{C.}~\bibnamefont{Jozwiak}},
  \bibinfo{author}{\bibfnamefont{J.~A.} \bibnamefont{Sobota}},
  \bibinfo{author}{\bibfnamefont{K.}~\bibnamefont{Gotlieb}},
  \bibinfo{author}{\bibfnamefont{A.~F.} \bibnamefont{Kemper}},
  \bibinfo{author}{\bibfnamefont{C.~R.} \bibnamefont{Rotundu}},
  \bibinfo{author}{\bibfnamefont{R.~J.} \bibnamefont{Birgeneau}},
  \bibinfo{author}{\bibfnamefont{Z.}~\bibnamefont{Hussain}},
  \bibinfo{author}{\bibfnamefont{D.-H.} \bibnamefont{Lee}},
  \bibinfo{author}{\bibfnamefont{Z.-X.} \bibnamefont{Shen}}, \bibnamefont{and}
  \bibinfo{author}{\bibfnamefont{A.}~\bibnamefont{Lanzara}},
  \bibinfo{journal}{Nat. Commun.} \textbf{\bibinfo{volume}{7}},
  \bibinfo{pages}{13143} (\bibinfo{year}{2016}).

\bibitem[{\citenamefont{Sonoda and Munakata}(2004)}]{sonoda2004unoccupied}
\bibinfo{author}{\bibfnamefont{Y.}~\bibnamefont{Sonoda}} \bibnamefont{and}
  \bibinfo{author}{\bibfnamefont{T.}~\bibnamefont{Munakata}},
  \bibinfo{journal}{Phys. Rev. B} \textbf{\bibinfo{volume}{70}},
  \bibinfo{pages}{134517} (\bibinfo{year}{2004}).

\bibitem[{\citenamefont{Yang et~al.}(2017)\citenamefont{Yang, Sobota, He, Wang,
  Leuenberger, Soifer, Hashimoto, Lu, Eisaki, Moritz
  et~al.}}]{yang2017revealing}
\bibinfo{author}{\bibfnamefont{S.-L.} \bibnamefont{Yang}},
  \bibinfo{author}{\bibfnamefont{J.}~\bibnamefont{Sobota}},
  \bibinfo{author}{\bibfnamefont{Y.}~\bibnamefont{He}},
  \bibinfo{author}{\bibfnamefont{Y.}~\bibnamefont{Wang}},
  \bibinfo{author}{\bibfnamefont{D.}~\bibnamefont{Leuenberger}},
  \bibinfo{author}{\bibfnamefont{H.}~\bibnamefont{Soifer}},
  \bibinfo{author}{\bibfnamefont{M.}~\bibnamefont{Hashimoto}},
  \bibinfo{author}{\bibfnamefont{D.}~\bibnamefont{Lu}},
  \bibinfo{author}{\bibfnamefont{H.}~\bibnamefont{Eisaki}},
  \bibinfo{author}{\bibfnamefont{B.}~\bibnamefont{Moritz}},
  \bibnamefont{et~al.}, \bibinfo{journal}{Phys. Rev. B}
  \textbf{\bibinfo{volume}{96}}, \bibinfo{pages}{245112}
  (\bibinfo{year}{2017}).

\bibitem[{\citenamefont{Hellmann et~al.}(2012)\citenamefont{Hellmann, Rohwer,
  Kall{\"a}ne, Hanff, Sohrt, Stange, Carr, Murnane, Kapteyn, Kipp
  et~al.}}]{hellmann2012time}
\bibinfo{author}{\bibfnamefont{S.}~\bibnamefont{Hellmann}},
  \bibinfo{author}{\bibfnamefont{T.}~\bibnamefont{Rohwer}},
  \bibinfo{author}{\bibfnamefont{M.}~\bibnamefont{Kall{\"a}ne}},
  \bibinfo{author}{\bibfnamefont{K.}~\bibnamefont{Hanff}},
  \bibinfo{author}{\bibfnamefont{C.}~\bibnamefont{Sohrt}},
  \bibinfo{author}{\bibfnamefont{A.}~\bibnamefont{Stange}},
  \bibinfo{author}{\bibfnamefont{A.}~\bibnamefont{Carr}},
  \bibinfo{author}{\bibfnamefont{M.~M.} \bibnamefont{Murnane}},
  \bibinfo{author}{\bibfnamefont{H.~C.} \bibnamefont{Kapteyn}},
  \bibinfo{author}{\bibfnamefont{L.}~\bibnamefont{Kipp}}, \bibnamefont{et~al.},
  \bibinfo{journal}{Nat. Commun.} \textbf{\bibinfo{volume}{3}},
  \bibinfo{pages}{1069} (\bibinfo{year}{2012}).

\bibitem[{\citenamefont{Conte et~al.}(2012)\citenamefont{Conte, Giannetti,
  Coslovich, Cilento, Bossini, Abebaw, Banfi, Ferrini, Eisaki, Greven
  et~al.}}]{conte2012disentangling}
\bibinfo{author}{\bibfnamefont{S.~D.} \bibnamefont{Conte}},
  \bibinfo{author}{\bibfnamefont{C.}~\bibnamefont{Giannetti}},
  \bibinfo{author}{\bibfnamefont{G.}~\bibnamefont{Coslovich}},
  \bibinfo{author}{\bibfnamefont{F.}~\bibnamefont{Cilento}},
  \bibinfo{author}{\bibfnamefont{D.}~\bibnamefont{Bossini}},
  \bibinfo{author}{\bibfnamefont{T.}~\bibnamefont{Abebaw}},
  \bibinfo{author}{\bibfnamefont{F.}~\bibnamefont{Banfi}},
  \bibinfo{author}{\bibfnamefont{G.}~\bibnamefont{Ferrini}},
  \bibinfo{author}{\bibfnamefont{H.}~\bibnamefont{Eisaki}},
  \bibinfo{author}{\bibfnamefont{M.}~\bibnamefont{Greven}},
  \bibnamefont{et~al.}, \bibinfo{journal}{Science}
  \textbf{\bibinfo{volume}{335}}, \bibinfo{pages}{1600} (\bibinfo{year}{2012}).

\bibitem[{\citenamefont{Yang et~al.}(2015)\citenamefont{Yang, Sobota,
  Leuenberger, Kemper, Lee, Schmitt, Li, Moore, Kirchmann, and
  Shen}}]{yang2015thickness}
\bibinfo{author}{\bibfnamefont{S.}~\bibnamefont{Yang}},
  \bibinfo{author}{\bibfnamefont{J.~A.} \bibnamefont{Sobota}},
  \bibinfo{author}{\bibfnamefont{D.}~\bibnamefont{Leuenberger}},
  \bibinfo{author}{\bibfnamefont{A.~F.} \bibnamefont{Kemper}},
  \bibinfo{author}{\bibfnamefont{J.~J.} \bibnamefont{Lee}},
  \bibinfo{author}{\bibfnamefont{F.~T.} \bibnamefont{Schmitt}},
  \bibinfo{author}{\bibfnamefont{W.}~\bibnamefont{Li}},
  \bibinfo{author}{\bibfnamefont{R.~G.} \bibnamefont{Moore}},
  \bibinfo{author}{\bibfnamefont{P.~S.} \bibnamefont{Kirchmann}},
  \bibnamefont{and} \bibinfo{author}{\bibfnamefont{Z.-X.} \bibnamefont{Shen}},
  \bibinfo{journal}{Nano Lett.} \textbf{\bibinfo{volume}{15}},
  \bibinfo{pages}{4150} (\bibinfo{year}{2015}).

\bibitem[{\citenamefont{Matsunaga et~al.}(2013)\citenamefont{Matsunaga, Hamada,
  Makise, Uzawa, Terai, Wang, and Shimano}}]{matsunaga2013higgs}
\bibinfo{author}{\bibfnamefont{R.}~\bibnamefont{Matsunaga}},
  \bibinfo{author}{\bibfnamefont{Y.~I.} \bibnamefont{Hamada}},
  \bibinfo{author}{\bibfnamefont{K.}~\bibnamefont{Makise}},
  \bibinfo{author}{\bibfnamefont{Y.}~\bibnamefont{Uzawa}},
  \bibinfo{author}{\bibfnamefont{H.}~\bibnamefont{Terai}},
  \bibinfo{author}{\bibfnamefont{Z.}~\bibnamefont{Wang}}, \bibnamefont{and}
  \bibinfo{author}{\bibfnamefont{R.}~\bibnamefont{Shimano}},
  \bibinfo{journal}{Phys. Rev. Lett.} \textbf{\bibinfo{volume}{111}},
  \bibinfo{pages}{057002} (\bibinfo{year}{2013}).

\bibitem[{\citenamefont{Matsunaga et~al.}(2014)\citenamefont{Matsunaga, Tsuji,
  Fujita, Sugioka, Makise, Uzawa, Terai, Wang, Aoki, and
  Shimano}}]{matsunaga2014light}
\bibinfo{author}{\bibfnamefont{R.}~\bibnamefont{Matsunaga}},
  \bibinfo{author}{\bibfnamefont{N.}~\bibnamefont{Tsuji}},
  \bibinfo{author}{\bibfnamefont{H.}~\bibnamefont{Fujita}},
  \bibinfo{author}{\bibfnamefont{A.}~\bibnamefont{Sugioka}},
  \bibinfo{author}{\bibfnamefont{K.}~\bibnamefont{Makise}},
  \bibinfo{author}{\bibfnamefont{Y.}~\bibnamefont{Uzawa}},
  \bibinfo{author}{\bibfnamefont{H.}~\bibnamefont{Terai}},
  \bibinfo{author}{\bibfnamefont{Z.}~\bibnamefont{Wang}},
  \bibinfo{author}{\bibfnamefont{H.}~\bibnamefont{Aoki}}, \bibnamefont{and}
  \bibinfo{author}{\bibfnamefont{R.}~\bibnamefont{Shimano}},
  \bibinfo{journal}{Science} \textbf{\bibinfo{volume}{345}},
  \bibinfo{pages}{1145} (\bibinfo{year}{2014}).

\bibitem[{\citenamefont{Sherman et~al.}(2015)\citenamefont{Sherman, Pracht,
  Gorshunov, Poran, Jesudasan, Chand, Raychaudhuri, Swanson, Trivedi, Auerbach
  et~al.}}]{sherman2015higgs}
\bibinfo{author}{\bibfnamefont{D.}~\bibnamefont{Sherman}},
  \bibinfo{author}{\bibfnamefont{U.~S.} \bibnamefont{Pracht}},
  \bibinfo{author}{\bibfnamefont{B.}~\bibnamefont{Gorshunov}},
  \bibinfo{author}{\bibfnamefont{S.}~\bibnamefont{Poran}},
  \bibinfo{author}{\bibfnamefont{J.}~\bibnamefont{Jesudasan}},
  \bibinfo{author}{\bibfnamefont{M.}~\bibnamefont{Chand}},
  \bibinfo{author}{\bibfnamefont{P.}~\bibnamefont{Raychaudhuri}},
  \bibinfo{author}{\bibfnamefont{M.}~\bibnamefont{Swanson}},
  \bibinfo{author}{\bibfnamefont{N.}~\bibnamefont{Trivedi}},
  \bibinfo{author}{\bibfnamefont{A.}~\bibnamefont{Auerbach}},
  \bibnamefont{et~al.}, \bibinfo{journal}{Nat. Phys.}
  \textbf{\bibinfo{volume}{11}}, \bibinfo{pages}{188} (\bibinfo{year}{2015}).

\bibitem[{\citenamefont{Bittner et~al.}(2015)\citenamefont{Bittner, Einzel,
  Klam, and Manske}}]{bittner2015leggett}
\bibinfo{author}{\bibfnamefont{N.}~\bibnamefont{Bittner}},
  \bibinfo{author}{\bibfnamefont{D.}~\bibnamefont{Einzel}},
  \bibinfo{author}{\bibfnamefont{L.}~\bibnamefont{Klam}}, \bibnamefont{and}
  \bibinfo{author}{\bibfnamefont{D.}~\bibnamefont{Manske}},
  \bibinfo{journal}{Phys. Rev. Lett.} \textbf{\bibinfo{volume}{115}},
  \bibinfo{pages}{227002} (\bibinfo{year}{2015}).

\bibitem[{\citenamefont{Krull et~al.}(2016)\citenamefont{Krull, Bittner, Uhrig,
  Manske, and Schnyder}}]{krull2016coupling}
\bibinfo{author}{\bibfnamefont{H.}~\bibnamefont{Krull}},
  \bibinfo{author}{\bibfnamefont{N.}~\bibnamefont{Bittner}},
  \bibinfo{author}{\bibfnamefont{G.}~\bibnamefont{Uhrig}},
  \bibinfo{author}{\bibfnamefont{D.}~\bibnamefont{Manske}}, \bibnamefont{and}
  \bibinfo{author}{\bibfnamefont{A.}~\bibnamefont{Schnyder}},
  \bibinfo{journal}{Nat. Commun.} \textbf{\bibinfo{volume}{7}},
  \bibinfo{pages}{11921} (\bibinfo{year}{2016}).

\bibitem[{\citenamefont{Nosarzewski et~al.}(2017)\citenamefont{Nosarzewski,
  Moritz, Freericks, Kemper, and Devereaux}}]{nosarzewski2017amplitude}
\bibinfo{author}{\bibfnamefont{B.}~\bibnamefont{Nosarzewski}},
  \bibinfo{author}{\bibfnamefont{B.}~\bibnamefont{Moritz}},
  \bibinfo{author}{\bibfnamefont{J.}~\bibnamefont{Freericks}},
  \bibinfo{author}{\bibfnamefont{A.}~\bibnamefont{Kemper}}, \bibnamefont{and}
  \bibinfo{author}{\bibfnamefont{T.}~\bibnamefont{Devereaux}},
  \bibinfo{journal}{Phys. Rev. B} \textbf{\bibinfo{volume}{96}},
  \bibinfo{pages}{184518} (\bibinfo{year}{2017}).

\bibitem[{\citenamefont{Kemper et~al.}(2015)\citenamefont{Kemper, Sentef,
  Moritz, Freericks, and Devereaux}}]{kemper2015direct}
\bibinfo{author}{\bibfnamefont{A.}~\bibnamefont{Kemper}},
  \bibinfo{author}{\bibfnamefont{M.}~\bibnamefont{Sentef}},
  \bibinfo{author}{\bibfnamefont{B.}~\bibnamefont{Moritz}},
  \bibinfo{author}{\bibfnamefont{J.}~\bibnamefont{Freericks}},
  \bibnamefont{and}
  \bibinfo{author}{\bibfnamefont{T.}~\bibnamefont{Devereaux}},
  \bibinfo{journal}{Phys. Rev. B} \textbf{\bibinfo{volume}{92}},
  \bibinfo{pages}{224517} (\bibinfo{year}{2015}).

\bibitem[{\citenamefont{Wang et~al.}(2016)\citenamefont{Wang, Moritz, Chen,
  Jia, van Veenendaal, and Devereaux}}]{wang2016using}
\bibinfo{author}{\bibfnamefont{Y.}~\bibnamefont{Wang}},
  \bibinfo{author}{\bibfnamefont{B.}~\bibnamefont{Moritz}},
  \bibinfo{author}{\bibfnamefont{C.-C.} \bibnamefont{Chen}},
  \bibinfo{author}{\bibfnamefont{C.}~\bibnamefont{Jia}},
  \bibinfo{author}{\bibfnamefont{M.}~\bibnamefont{van Veenendaal}},
  \bibnamefont{and} \bibinfo{author}{\bibfnamefont{T.~P.}
  \bibnamefont{Devereaux}}, \bibinfo{journal}{Phys. Rev. Lett.}
  \textbf{\bibinfo{volume}{116}}, \bibinfo{pages}{086401}
  (\bibinfo{year}{2016}).

\bibitem[{\citenamefont{Wang et~al.}(2013)\citenamefont{Wang, Steinberg,
  Jarillo-Herrero, and Gedik}}]{wang2013observation}
\bibinfo{author}{\bibfnamefont{Y.}~\bibnamefont{Wang}},
  \bibinfo{author}{\bibfnamefont{H.}~\bibnamefont{Steinberg}},
  \bibinfo{author}{\bibfnamefont{P.}~\bibnamefont{Jarillo-Herrero}},
  \bibnamefont{and} \bibinfo{author}{\bibfnamefont{N.}~\bibnamefont{Gedik}},
  \bibinfo{journal}{Science} \textbf{\bibinfo{volume}{342}},
  \bibinfo{pages}{453} (\bibinfo{year}{2013}).

\bibitem[{\citenamefont{Mahmood et~al.}(2016)\citenamefont{Mahmood, Chan,
  Alpichshev, and Gardner}}]{mahmood2016selective}
\bibinfo{author}{\bibfnamefont{F.}~\bibnamefont{Mahmood}},
  \bibinfo{author}{\bibfnamefont{C.~K.} \bibnamefont{Chan}},
  \bibinfo{author}{\bibfnamefont{Z.}~\bibnamefont{Alpichshev}},
  \bibnamefont{and} \bibinfo{author}{\bibfnamefont{D.}~\bibnamefont{Gardner}},
  \bibinfo{journal}{Nat. Phys.} \textbf{\bibinfo{volume}{12}},
  \bibinfo{pages}{306} (\bibinfo{year}{2016}).

\bibitem[{\citenamefont{Oka and Aoki}(2009)}]{oka2009photovoltaic}
\bibinfo{author}{\bibfnamefont{T.}~\bibnamefont{Oka}} \bibnamefont{and}
  \bibinfo{author}{\bibfnamefont{H.}~\bibnamefont{Aoki}},
  \bibinfo{journal}{Phys. Rev. B} \textbf{\bibinfo{volume}{79}},
  \bibinfo{pages}{081406} (\bibinfo{year}{2009}).

\bibitem[{\citenamefont{Inoue and Tanaka}(2010)}]{inoue2010photoinduced}
\bibinfo{author}{\bibfnamefont{J.-i.} \bibnamefont{Inoue}} \bibnamefont{and}
  \bibinfo{author}{\bibfnamefont{A.}~\bibnamefont{Tanaka}},
  \bibinfo{journal}{Phys. Rev. Lett.} \textbf{\bibinfo{volume}{105}},
  \bibinfo{pages}{017401} (\bibinfo{year}{2010}).

\bibitem[{\citenamefont{Kitagawa et~al.}(2011)\citenamefont{Kitagawa, Oka,
  Brataas, Fu, and Demler}}]{kitagawa2011transport}
\bibinfo{author}{\bibfnamefont{T.}~\bibnamefont{Kitagawa}},
  \bibinfo{author}{\bibfnamefont{T.}~\bibnamefont{Oka}},
  \bibinfo{author}{\bibfnamefont{A.}~\bibnamefont{Brataas}},
  \bibinfo{author}{\bibfnamefont{L.}~\bibnamefont{Fu}}, \bibnamefont{and}
  \bibinfo{author}{\bibfnamefont{E.}~\bibnamefont{Demler}},
  \bibinfo{journal}{Phys. Rev. B} \textbf{\bibinfo{volume}{84}},
  \bibinfo{pages}{235108} (\bibinfo{year}{2011}).

\bibitem[{\citenamefont{Lindner et~al.}(2011)\citenamefont{Lindner, Refael, and
  Galitski}}]{lindner2011floquet}
\bibinfo{author}{\bibfnamefont{N.~H.} \bibnamefont{Lindner}},
  \bibinfo{author}{\bibfnamefont{G.}~\bibnamefont{Refael}}, \bibnamefont{and}
  \bibinfo{author}{\bibfnamefont{V.}~\bibnamefont{Galitski}},
  \bibinfo{journal}{Nat. Phys.} \textbf{\bibinfo{volume}{7}},
  \bibinfo{pages}{490} (\bibinfo{year}{2011}).

\bibitem[{\citenamefont{Sentef et~al.}(2015)\citenamefont{Sentef, Claassen,
  Kemper, Moritz, Oka, Freericks, and Devereaux}}]{sentef2015theory}
\bibinfo{author}{\bibfnamefont{M.}~\bibnamefont{Sentef}},
  \bibinfo{author}{\bibfnamefont{M.}~\bibnamefont{Claassen}},
  \bibinfo{author}{\bibfnamefont{A.}~\bibnamefont{Kemper}},
  \bibinfo{author}{\bibfnamefont{B.}~\bibnamefont{Moritz}},
  \bibinfo{author}{\bibfnamefont{T.}~\bibnamefont{Oka}},
  \bibinfo{author}{\bibfnamefont{J.}~\bibnamefont{Freericks}},
  \bibnamefont{and}
  \bibinfo{author}{\bibfnamefont{T.}~\bibnamefont{Devereaux}},
  \bibinfo{journal}{Nat. Commun.} \textbf{\bibinfo{volume}{6}},
  \bibinfo{pages}{7047} (\bibinfo{year}{2015}).

\bibitem[{\citenamefont{Balzer et~al.}(2015)\citenamefont{Balzer, Wolf,
  McCulloch, Werner, and Eckstein}}]{balzer2015nonthermal}
\bibinfo{author}{\bibfnamefont{K.}~\bibnamefont{Balzer}},
  \bibinfo{author}{\bibfnamefont{F.~A.} \bibnamefont{Wolf}},
  \bibinfo{author}{\bibfnamefont{I.~P.} \bibnamefont{McCulloch}},
  \bibinfo{author}{\bibfnamefont{P.}~\bibnamefont{Werner}}, \bibnamefont{and}
  \bibinfo{author}{\bibfnamefont{M.}~\bibnamefont{Eckstein}},
  \bibinfo{journal}{Phys. Rev. X} \textbf{\bibinfo{volume}{5}},
  \bibinfo{pages}{031039} (\bibinfo{year}{2015}).

\bibitem[{\citenamefont{Knap et~al.}(2016)\citenamefont{Knap, Babadi, Refael,
  Martin, and Demler}}]{knap2016dynamical}
\bibinfo{author}{\bibfnamefont{M.}~\bibnamefont{Knap}},
  \bibinfo{author}{\bibfnamefont{M.}~\bibnamefont{Babadi}},
  \bibinfo{author}{\bibfnamefont{G.}~\bibnamefont{Refael}},
  \bibinfo{author}{\bibfnamefont{I.}~\bibnamefont{Martin}}, \bibnamefont{and}
  \bibinfo{author}{\bibfnamefont{E.}~\bibnamefont{Demler}},
  \bibinfo{journal}{Phys. Rev. B} \textbf{\bibinfo{volume}{94}},
  \bibinfo{pages}{214504} (\bibinfo{year}{2016}).

\bibitem[{\citenamefont{Wang et~al.}(2018)\citenamefont{Wang, Chen, Moritz, and
  Devereaux}}]{wang2017light}
\bibinfo{author}{\bibfnamefont{Y.}~\bibnamefont{Wang}},
  \bibinfo{author}{\bibfnamefont{C.-C.} \bibnamefont{Chen}},
  \bibinfo{author}{\bibfnamefont{B.}~\bibnamefont{Moritz}}, \bibnamefont{and}
  \bibinfo{author}{\bibfnamefont{T.}~\bibnamefont{Devereaux}},
  \bibinfo{journal}{Phys. Rev. Lett.} \textbf{\bibinfo{volume}{120}},
  \bibinfo{pages}{246402} (\bibinfo{year}{2018}).

\bibitem[{\citenamefont{D'Alessio and Rigol}(2014)}]{DAlessioPRX2014}
\bibinfo{author}{\bibfnamefont{L.}~\bibnamefont{D'Alessio}} \bibnamefont{and}
  \bibinfo{author}{\bibfnamefont{M.}~\bibnamefont{Rigol}},
  \bibinfo{journal}{Phys. Rev. X} \textbf{\bibinfo{volume}{4}},
  \bibinfo{pages}{041048} (\bibinfo{year}{2014}).

\bibitem[{\citenamefont{Lazarides et~al.}(2014)\citenamefont{Lazarides, Das,
  and Moessner}}]{LazaridesPRE2014}
\bibinfo{author}{\bibfnamefont{A.}~\bibnamefont{Lazarides}},
  \bibinfo{author}{\bibfnamefont{A.}~\bibnamefont{Das}}, \bibnamefont{and}
  \bibinfo{author}{\bibfnamefont{R.}~\bibnamefont{Moessner}},
  \bibinfo{journal}{Phys. Rev. E} \textbf{\bibinfo{volume}{90}},
  \bibinfo{pages}{012110} (\bibinfo{year}{2014}).

\bibitem[{\citenamefont{D'Alessio and Polkovnikov}(2013)}]{DAlessioAnnPhys2013}
\bibinfo{author}{\bibfnamefont{L.}~\bibnamefont{D'Alessio}} \bibnamefont{and}
  \bibinfo{author}{\bibfnamefont{A.}~\bibnamefont{Polkovnikov}},
  \bibinfo{journal}{Ann. Phys.} \textbf{\bibinfo{volume}{333}},
  \bibinfo{pages}{19} (\bibinfo{year}{2013}).

\bibitem[{\citenamefont{Ponte et~al.}(2015)\citenamefont{Ponte, Papi\'c,
  Huveneers, and Abanin}}]{PontePRL2015}
\bibinfo{author}{\bibfnamefont{P.}~\bibnamefont{Ponte}},
  \bibinfo{author}{\bibfnamefont{Z.}~\bibnamefont{Papi\'c}},
  \bibinfo{author}{\bibfnamefont{F.}~\bibnamefont{Huveneers}},
  \bibnamefont{and} \bibinfo{author}{\bibfnamefont{D.~A.}
  \bibnamefont{Abanin}}, \bibinfo{journal}{Phys. Rev. Lett.}
  \textbf{\bibinfo{volume}{114}}, \bibinfo{pages}{140401}
  (\bibinfo{year}{2015}).

\bibitem[{\citenamefont{Lazarides et~al.}(2015)\citenamefont{Lazarides, Das,
  and Moessner}}]{LazaridesPRL2015}
\bibinfo{author}{\bibfnamefont{A.}~\bibnamefont{Lazarides}},
  \bibinfo{author}{\bibfnamefont{A.}~\bibnamefont{Das}}, \bibnamefont{and}
  \bibinfo{author}{\bibfnamefont{R.}~\bibnamefont{Moessner}},
  \bibinfo{journal}{Phys. Rev. Lett.} \textbf{\bibinfo{volume}{115}},
  \bibinfo{pages}{030402} (\bibinfo{year}{2015}).

\bibitem[{\citenamefont{Canovi et~al.}(2016)\citenamefont{Canovi, Kollar, and
  Eckstein}}]{CanoviPRE2016}
\bibinfo{author}{\bibfnamefont{E.}~\bibnamefont{Canovi}},
  \bibinfo{author}{\bibfnamefont{M.}~\bibnamefont{Kollar}}, \bibnamefont{and}
  \bibinfo{author}{\bibfnamefont{M.}~\bibnamefont{Eckstein}},
  \bibinfo{journal}{Phys. Rev. E} \textbf{\bibinfo{volume}{93}},
  \bibinfo{pages}{012130} (\bibinfo{year}{2016}).

\bibitem[{\citenamefont{Kuwahara et~al.}(2016)\citenamefont{Kuwahara, Mori, and
  Saito}}]{kuwahara2016floquet}
\bibinfo{author}{\bibfnamefont{T.}~\bibnamefont{Kuwahara}},
  \bibinfo{author}{\bibfnamefont{T.}~\bibnamefont{Mori}}, \bibnamefont{and}
  \bibinfo{author}{\bibfnamefont{K.}~\bibnamefont{Saito}},
  \bibinfo{journal}{Ann. Phys.} \textbf{\bibinfo{volume}{367}},
  \bibinfo{pages}{96} (\bibinfo{year}{2016}).

\bibitem[{\citenamefont{Mori et~al.}(2016)\citenamefont{Mori, Kuwahara, and
  Saito}}]{MoriPRL2016}
\bibinfo{author}{\bibfnamefont{T.}~\bibnamefont{Mori}},
  \bibinfo{author}{\bibfnamefont{T.}~\bibnamefont{Kuwahara}}, \bibnamefont{and}
  \bibinfo{author}{\bibfnamefont{K.}~\bibnamefont{Saito}},
  \bibinfo{journal}{Phys. Rev. Lett.} \textbf{\bibinfo{volume}{116}},
  \bibinfo{pages}{120401} (\bibinfo{year}{2016}).

\bibitem[{\citenamefont{Bukov et~al.}(2015)\citenamefont{Bukov, Gopalakrishnan,
  Knap, and Demler}}]{bukov2015prethermal}
\bibinfo{author}{\bibfnamefont{M.}~\bibnamefont{Bukov}},
  \bibinfo{author}{\bibfnamefont{S.}~\bibnamefont{Gopalakrishnan}},
  \bibinfo{author}{\bibfnamefont{M.}~\bibnamefont{Knap}}, \bibnamefont{and}
  \bibinfo{author}{\bibfnamefont{E.}~\bibnamefont{Demler}},
  \bibinfo{journal}{Phys. Rev. Lett.} \textbf{\bibinfo{volume}{115}},
  \bibinfo{pages}{205301} (\bibinfo{year}{2015}).

\bibitem[{\citenamefont{Abanin et~al.}(2015)\citenamefont{Abanin, Roeck, and
  Huveneers}}]{AbaninPRL2015}
\bibinfo{author}{\bibfnamefont{D.~A.} \bibnamefont{Abanin}},
  \bibinfo{author}{\bibfnamefont{W.~D.} \bibnamefont{Roeck}}, \bibnamefont{and}
  \bibinfo{author}{\bibfnamefont{F.}~\bibnamefont{Huveneers}},
  \bibinfo{journal}{Phys. Rev. Lett.} \textbf{\bibinfo{volume}{115}},
  \bibinfo{pages}{256803} (\bibinfo{year}{2015}).

\bibitem[{\citenamefont{Ho and Abanin}(2016)}]{HoARXIV2016}
\bibinfo{author}{\bibfnamefont{W.~W.} \bibnamefont{Ho}} \bibnamefont{and}
  \bibinfo{author}{\bibfnamefont{D.~A.} \bibnamefont{Abanin}},
  \bibinfo{journal}{arXiv:1611.05024}  (\bibinfo{year}{2016}).

\bibitem[{\citenamefont{Itin and Katsnelson}(2015)}]{ItinPRL2015}
\bibinfo{author}{\bibfnamefont{A.~P.} \bibnamefont{Itin}} \bibnamefont{and}
  \bibinfo{author}{\bibfnamefont{M.~I.} \bibnamefont{Katsnelson}},
  \bibinfo{journal}{Phys. Rev. Lett} \textbf{\bibinfo{volume}{115}},
  \bibinfo{pages}{075301} (\bibinfo{year}{2015}).

\bibitem[{\citenamefont{Claassen et~al.}(2017)\citenamefont{Claassen, Jiang,
  Moritz, and Devereaux}}]{claassen2017dynamical}
\bibinfo{author}{\bibfnamefont{M.}~\bibnamefont{Claassen}},
  \bibinfo{author}{\bibfnamefont{H.-C.} \bibnamefont{Jiang}},
  \bibinfo{author}{\bibfnamefont{B.}~\bibnamefont{Moritz}}, \bibnamefont{and}
  \bibinfo{author}{\bibfnamefont{T.~P.} \bibnamefont{Devereaux}},
  \bibinfo{journal}{Nat. Commun.} \textbf{\bibinfo{volume}{8}},
  \bibinfo{pages}{1192} (\bibinfo{year}{2017}).

\bibitem[{\citenamefont{Bukov et~al.}(2016{\natexlab{a}})\citenamefont{Bukov,
  Kolodrubetz, and Polkovnikov}}]{bukov2016schrieffer}
\bibinfo{author}{\bibfnamefont{M.}~\bibnamefont{Bukov}},
  \bibinfo{author}{\bibfnamefont{M.}~\bibnamefont{Kolodrubetz}},
  \bibnamefont{and}
  \bibinfo{author}{\bibfnamefont{A.}~\bibnamefont{Polkovnikov}},
  \bibinfo{journal}{Phys. Rev. Lett.} \textbf{\bibinfo{volume}{116}},
  \bibinfo{pages}{125301} (\bibinfo{year}{2016}{\natexlab{a}}).

\bibitem[{\citenamefont{Eckardt}(2017)}]{eckardt2017colloquium}
\bibinfo{author}{\bibfnamefont{A.}~\bibnamefont{Eckardt}},
  \bibinfo{journal}{Rev. Mod. Phys.} \textbf{\bibinfo{volume}{89}},
  \bibinfo{pages}{011004} (\bibinfo{year}{2017}).

\bibitem[{\citenamefont{Coulthard et~al.}(2016)\citenamefont{Coulthard, Clark,
  Al-Assam, Cavalleri, and Jaksch}}]{coulthard2016enhancement}
\bibinfo{author}{\bibfnamefont{J.}~\bibnamefont{Coulthard}},
  \bibinfo{author}{\bibfnamefont{S.~R.} \bibnamefont{Clark}},
  \bibinfo{author}{\bibfnamefont{S.}~\bibnamefont{Al-Assam}},
  \bibinfo{author}{\bibfnamefont{A.}~\bibnamefont{Cavalleri}},
  \bibnamefont{and} \bibinfo{author}{\bibfnamefont{D.}~\bibnamefont{Jaksch}},
  \bibinfo{journal}{arXiv:1608.03964}  (\bibinfo{year}{2016}).

\bibitem[{\citenamefont{Bukov et~al.}(2016{\natexlab{b}})\citenamefont{Bukov,
  Heyl, Huse, and Polkovnikov}}]{bukov2016heating}
\bibinfo{author}{\bibfnamefont{M.}~\bibnamefont{Bukov}},
  \bibinfo{author}{\bibfnamefont{M.}~\bibnamefont{Heyl}},
  \bibinfo{author}{\bibfnamefont{D.~A.} \bibnamefont{Huse}}, \bibnamefont{and}
  \bibinfo{author}{\bibfnamefont{A.}~\bibnamefont{Polkovnikov}},
  \bibinfo{journal}{Phys. Rev. B} \textbf{\bibinfo{volume}{93}},
  \bibinfo{pages}{155132} (\bibinfo{year}{2016}{\natexlab{b}}).

\bibitem[{\citenamefont{Wang et~al.}(2017)\citenamefont{Wang, Claassen, Moritz,
  and Devereaux}}]{wang2017producing}
\bibinfo{author}{\bibfnamefont{Y.}~\bibnamefont{Wang}},
  \bibinfo{author}{\bibfnamefont{M.}~\bibnamefont{Claassen}},
  \bibinfo{author}{\bibfnamefont{B.}~\bibnamefont{Moritz}}, \bibnamefont{and}
  \bibinfo{author}{\bibfnamefont{T.}~\bibnamefont{Devereaux}},
  \bibinfo{journal}{Phys. Rev. B} \textbf{\bibinfo{volume}{96}},
  \bibinfo{pages}{235142} (\bibinfo{year}{2017}).

\bibitem[{\citenamefont{Perfetti et~al.}(2008)\citenamefont{Perfetti, Loukakos,
  Lisowski, Bovensiepen, Wolf, Berger, Biermann, and
  Georges}}]{perfetti2008femtosecond}
\bibinfo{author}{\bibfnamefont{L.}~\bibnamefont{Perfetti}},
  \bibinfo{author}{\bibfnamefont{P.~A.} \bibnamefont{Loukakos}},
  \bibinfo{author}{\bibfnamefont{M.}~\bibnamefont{Lisowski}},
  \bibinfo{author}{\bibfnamefont{U.}~\bibnamefont{Bovensiepen}},
  \bibinfo{author}{\bibfnamefont{M.}~\bibnamefont{Wolf}},
  \bibinfo{author}{\bibfnamefont{H.}~\bibnamefont{Berger}},
  \bibinfo{author}{\bibfnamefont{S.}~\bibnamefont{Biermann}}, \bibnamefont{and}
  \bibinfo{author}{\bibfnamefont{A.}~\bibnamefont{Georges}},
  \bibinfo{journal}{New J Phys.} \textbf{\bibinfo{volume}{10}},
  \bibinfo{pages}{053019} (\bibinfo{year}{2008}).

\bibitem[{\citenamefont{Liu et~al.}(2012)\citenamefont{Liu, Hwang, Tao,
  Strikwerda, and Fan}}]{liu2012terahertz}
\bibinfo{author}{\bibfnamefont{M.}~\bibnamefont{Liu}},
  \bibinfo{author}{\bibfnamefont{H.~Y.} \bibnamefont{Hwang}},
  \bibinfo{author}{\bibfnamefont{H.}~\bibnamefont{Tao}},
  \bibinfo{author}{\bibfnamefont{A.~C.} \bibnamefont{Strikwerda}},
  \bibnamefont{and} \bibinfo{author}{\bibfnamefont{K.}~\bibnamefont{Fan}},
  \bibinfo{journal}{Nature}  (\bibinfo{year}{2012}).

\bibitem[{\citenamefont{Schmitt et~al.}(2008)\citenamefont{Schmitt, Kirchmann,
  Bovensiepen, Moore, Rettig, Krenz, Chu, Ru, Perfetti, Lu
  et~al.}}]{schmitt2008transient}
\bibinfo{author}{\bibfnamefont{F.}~\bibnamefont{Schmitt}},
  \bibinfo{author}{\bibfnamefont{P.~S.} \bibnamefont{Kirchmann}},
  \bibinfo{author}{\bibfnamefont{U.}~\bibnamefont{Bovensiepen}},
  \bibinfo{author}{\bibfnamefont{R.~G.} \bibnamefont{Moore}},
  \bibinfo{author}{\bibfnamefont{L.}~\bibnamefont{Rettig}},
  \bibinfo{author}{\bibfnamefont{M.}~\bibnamefont{Krenz}},
  \bibinfo{author}{\bibfnamefont{J.~H.} \bibnamefont{Chu}},
  \bibinfo{author}{\bibfnamefont{N.}~\bibnamefont{Ru}},
  \bibinfo{author}{\bibfnamefont{L.}~\bibnamefont{Perfetti}},
  \bibinfo{author}{\bibfnamefont{D.~H.} \bibnamefont{Lu}},
  \bibnamefont{et~al.}, \bibinfo{journal}{Science}
  \textbf{\bibinfo{volume}{321}}, \bibinfo{pages}{1649} (\bibinfo{year}{2008}).

\bibitem[{\citenamefont{Stojchevska et~al.}(2014)\citenamefont{Stojchevska,
  Vaskivskyi, Mertelj, Kusar, Svetin, Brazovskii, and
  Mihailovic}}]{stojchevska2014ultrafast}
\bibinfo{author}{\bibfnamefont{L.}~\bibnamefont{Stojchevska}},
  \bibinfo{author}{\bibfnamefont{I.}~\bibnamefont{Vaskivskyi}},
  \bibinfo{author}{\bibfnamefont{T.}~\bibnamefont{Mertelj}},
  \bibinfo{author}{\bibfnamefont{P.}~\bibnamefont{Kusar}},
  \bibinfo{author}{\bibfnamefont{D.}~\bibnamefont{Svetin}},
  \bibinfo{author}{\bibfnamefont{S.}~\bibnamefont{Brazovskii}},
  \bibnamefont{and}
  \bibinfo{author}{\bibfnamefont{D.}~\bibnamefont{Mihailovic}},
  \bibinfo{journal}{Science} \textbf{\bibinfo{volume}{344}},
  \bibinfo{pages}{177} (\bibinfo{year}{2014}).

\bibitem[{\citenamefont{Zhang et~al.}(2016)\citenamefont{Zhang, Tan, Liu,
  Teitelbaum, Post, Jin, Nelson, Basov, Wu, and
  Averitt}}]{zhang2016cooperative}
\bibinfo{author}{\bibfnamefont{J.}~\bibnamefont{Zhang}},
  \bibinfo{author}{\bibfnamefont{X.}~\bibnamefont{Tan}},
  \bibinfo{author}{\bibfnamefont{M.}~\bibnamefont{Liu}},
  \bibinfo{author}{\bibfnamefont{S.~W.} \bibnamefont{Teitelbaum}},
  \bibinfo{author}{\bibfnamefont{K.~W.} \bibnamefont{Post}},
  \bibinfo{author}{\bibfnamefont{F.}~\bibnamefont{Jin}},
  \bibinfo{author}{\bibfnamefont{K.~A.} \bibnamefont{Nelson}},
  \bibinfo{author}{\bibfnamefont{D.}~\bibnamefont{Basov}},
  \bibinfo{author}{\bibfnamefont{W.}~\bibnamefont{Wu}}, \bibnamefont{and}
  \bibinfo{author}{\bibfnamefont{R.~D.} \bibnamefont{Averitt}},
  \bibinfo{journal}{Nat. Mater.} \textbf{\bibinfo{volume}{15}},
  \bibinfo{pages}{956} (\bibinfo{year}{2016}).

\bibitem[{\citenamefont{Wegkamp et~al.}(2014)\citenamefont{Wegkamp, Herzog,
  Xian, Gatti, Cudazzo, McGahan, Marvel, Haglund~Jr, Rubio, Wolf
  et~al.}}]{wegkamp2014instantaneous}
\bibinfo{author}{\bibfnamefont{D.}~\bibnamefont{Wegkamp}},
  \bibinfo{author}{\bibfnamefont{M.}~\bibnamefont{Herzog}},
  \bibinfo{author}{\bibfnamefont{L.}~\bibnamefont{Xian}},
  \bibinfo{author}{\bibfnamefont{M.}~\bibnamefont{Gatti}},
  \bibinfo{author}{\bibfnamefont{P.}~\bibnamefont{Cudazzo}},
  \bibinfo{author}{\bibfnamefont{C.~L.} \bibnamefont{McGahan}},
  \bibinfo{author}{\bibfnamefont{R.~E.} \bibnamefont{Marvel}},
  \bibinfo{author}{\bibfnamefont{R.~F.} \bibnamefont{Haglund~Jr}},
  \bibinfo{author}{\bibfnamefont{A.}~\bibnamefont{Rubio}},
  \bibinfo{author}{\bibfnamefont{M.}~\bibnamefont{Wolf}}, \bibnamefont{et~al.},
  \bibinfo{journal}{Phys. Rev. Lett.} \textbf{\bibinfo{volume}{113}},
  \bibinfo{pages}{216401} (\bibinfo{year}{2014}).

\bibitem[{\citenamefont{Werner et~al.}(2012)\citenamefont{Werner, Tsuji, and
  Eckstein}}]{werner2012nonthermal}
\bibinfo{author}{\bibfnamefont{P.}~\bibnamefont{Werner}},
  \bibinfo{author}{\bibfnamefont{N.}~\bibnamefont{Tsuji}}, \bibnamefont{and}
  \bibinfo{author}{\bibfnamefont{M.}~\bibnamefont{Eckstein}},
  \bibinfo{journal}{Phys. Rev. B} \textbf{\bibinfo{volume}{86}},
  \bibinfo{pages}{205101} (\bibinfo{year}{2012}).

\bibitem[{\citenamefont{Moeckel and Kehrein}(2008)}]{moeckel2008interaction}
\bibinfo{author}{\bibfnamefont{M.}~\bibnamefont{Moeckel}} \bibnamefont{and}
  \bibinfo{author}{\bibfnamefont{S.}~\bibnamefont{Kehrein}},
  \bibinfo{journal}{Phys. Rev. Lett.} \textbf{\bibinfo{volume}{100}},
  \bibinfo{pages}{175702} (\bibinfo{year}{2008}).

\bibitem[{\citenamefont{Mankowsky et~al.}(2014)\citenamefont{Mankowsky, Subedi,
  F{\"o}rst, and Mariager}}]{mankowsky2014nonlinear}
\bibinfo{author}{\bibfnamefont{R.}~\bibnamefont{Mankowsky}},
  \bibinfo{author}{\bibfnamefont{A.}~\bibnamefont{Subedi}},
  \bibinfo{author}{\bibfnamefont{M.}~\bibnamefont{F{\"o}rst}},
  \bibnamefont{and} \bibinfo{author}{\bibfnamefont{S.~O.}
  \bibnamefont{Mariager}}, \bibinfo{journal}{Nature}  (\bibinfo{year}{2014}).

\bibitem[{\citenamefont{Denny et~al.}(2015)\citenamefont{Denny, Clark, Laplace,
  Cavalleri, and Jaksch}}]{denny2015proposed}
\bibinfo{author}{\bibfnamefont{S.}~\bibnamefont{Denny}},
  \bibinfo{author}{\bibfnamefont{S.}~\bibnamefont{Clark}},
  \bibinfo{author}{\bibfnamefont{Y.}~\bibnamefont{Laplace}},
  \bibinfo{author}{\bibfnamefont{A.}~\bibnamefont{Cavalleri}},
  \bibnamefont{and} \bibinfo{author}{\bibfnamefont{D.}~\bibnamefont{Jaksch}},
  \bibinfo{journal}{Phys. Rev. Lett.} \textbf{\bibinfo{volume}{114}},
  \bibinfo{pages}{137001} (\bibinfo{year}{2015}).

\bibitem[{\citenamefont{Hoeppner et~al.}(2015)\citenamefont{Hoeppner, Zhu,
  Rexin, Cavalleri, and Mathey}}]{hoeppner2015redistribution}
\bibinfo{author}{\bibfnamefont{R.}~\bibnamefont{Hoeppner}},
  \bibinfo{author}{\bibfnamefont{B.}~\bibnamefont{Zhu}},
  \bibinfo{author}{\bibfnamefont{T.}~\bibnamefont{Rexin}},
  \bibinfo{author}{\bibfnamefont{A.}~\bibnamefont{Cavalleri}},
  \bibnamefont{and} \bibinfo{author}{\bibfnamefont{L.}~\bibnamefont{Mathey}},
  \bibinfo{journal}{Phys. Rev. B} \textbf{\bibinfo{volume}{91}},
  \bibinfo{pages}{104507} (\bibinfo{year}{2015}).

\bibitem[{\citenamefont{Nava et~al.}(2018)\citenamefont{Nava, Giannetti,
  Georges, Tosatti, and Fabrizio}}]{nava2018cooling}
\bibinfo{author}{\bibfnamefont{A.}~\bibnamefont{Nava}},
  \bibinfo{author}{\bibfnamefont{C.}~\bibnamefont{Giannetti}},
  \bibinfo{author}{\bibfnamefont{A.}~\bibnamefont{Georges}},
  \bibinfo{author}{\bibfnamefont{E.}~\bibnamefont{Tosatti}}, \bibnamefont{and}
  \bibinfo{author}{\bibfnamefont{M.}~\bibnamefont{Fabrizio}},
  \bibinfo{journal}{Nat. Phys.} \textbf{\bibinfo{volume}{14}},
  \bibinfo{pages}{154} (\bibinfo{year}{2018}).

\bibitem[{\citenamefont{Werner and Eckstein}(2015)}]{werner2015field}
\bibinfo{author}{\bibfnamefont{P.}~\bibnamefont{Werner}} \bibnamefont{and}
  \bibinfo{author}{\bibfnamefont{M.}~\bibnamefont{Eckstein}},
  \bibinfo{journal}{Europhys. Lett.} \textbf{\bibinfo{volume}{109}},
  \bibinfo{pages}{37002} (\bibinfo{year}{2015}).

\bibitem[{\citenamefont{Babadi et~al.}(2017)\citenamefont{Babadi, Knap, Martin,
  Refael, and Demler}}]{babadi2017theory}
\bibinfo{author}{\bibfnamefont{M.}~\bibnamefont{Babadi}},
  \bibinfo{author}{\bibfnamefont{M.}~\bibnamefont{Knap}},
  \bibinfo{author}{\bibfnamefont{I.}~\bibnamefont{Martin}},
  \bibinfo{author}{\bibfnamefont{G.}~\bibnamefont{Refael}}, \bibnamefont{and}
  \bibinfo{author}{\bibfnamefont{E.}~\bibnamefont{Demler}},
  \bibinfo{journal}{Phys. Rev. B} \textbf{\bibinfo{volume}{96}},
  \bibinfo{pages}{014512} (\bibinfo{year}{2017}).

\bibitem[{\citenamefont{Kennes et~al.}(2017)\citenamefont{Kennes, Wilner,
  Reichman, and Millis}}]{kennes2017transient}
\bibinfo{author}{\bibfnamefont{D.~M.} \bibnamefont{Kennes}},
  \bibinfo{author}{\bibfnamefont{E.~Y.} \bibnamefont{Wilner}},
  \bibinfo{author}{\bibfnamefont{D.~R.} \bibnamefont{Reichman}},
  \bibnamefont{and} \bibinfo{author}{\bibfnamefont{A.~J.}
  \bibnamefont{Millis}}, \bibinfo{journal}{Nat. Phys.}
  \textbf{\bibinfo{volume}{13}}, \bibinfo{pages}{479} (\bibinfo{year}{2017}).

\bibitem[{\citenamefont{Murakami
  et~al.}(2017{\natexlab{a}})\citenamefont{Murakami, Tsuji, Eckstein, and
  Werner}}]{murakami2017nonequilibrium}
\bibinfo{author}{\bibfnamefont{Y.}~\bibnamefont{Murakami}},
  \bibinfo{author}{\bibfnamefont{N.}~\bibnamefont{Tsuji}},
  \bibinfo{author}{\bibfnamefont{M.}~\bibnamefont{Eckstein}}, \bibnamefont{and}
  \bibinfo{author}{\bibfnamefont{P.}~\bibnamefont{Werner}},
  \bibinfo{journal}{Phys. Rev. B} \textbf{\bibinfo{volume}{96}},
  \bibinfo{pages}{045125} (\bibinfo{year}{2017}{\natexlab{a}}).

\bibitem[{\citenamefont{Sentef}(2017)}]{sentef2017light}
\bibinfo{author}{\bibfnamefont{M.}~\bibnamefont{Sentef}},
  \bibinfo{journal}{Phys. Rev. B} \textbf{\bibinfo{volume}{95}},
  \bibinfo{pages}{205111} (\bibinfo{year}{2017}).

\bibitem[{\citenamefont{Komnik and Thorwart}(2016)}]{komnik2016bcs}
\bibinfo{author}{\bibfnamefont{A.}~\bibnamefont{Komnik}} \bibnamefont{and}
  \bibinfo{author}{\bibfnamefont{M.}~\bibnamefont{Thorwart}},
  \bibinfo{journal}{Euro. Phys. J B} \textbf{\bibinfo{volume}{89}},
  \bibinfo{pages}{244} (\bibinfo{year}{2016}).

\bibitem[{\citenamefont{Sentef et~al.}(2016)\citenamefont{Sentef, Kemper,
  Georges, and Kollath}}]{sentef2016theory}
\bibinfo{author}{\bibfnamefont{M.~A.} \bibnamefont{Sentef}},
  \bibinfo{author}{\bibfnamefont{A.~F.} \bibnamefont{Kemper}},
  \bibinfo{author}{\bibfnamefont{A.}~\bibnamefont{Georges}}, \bibnamefont{and}
  \bibinfo{author}{\bibfnamefont{C.}~\bibnamefont{Kollath}},
  \bibinfo{journal}{Phys. Rev. B} \textbf{\bibinfo{volume}{93}},
  \bibinfo{pages}{144506} (\bibinfo{year}{2016}).

\bibitem[{\citenamefont{Kim et~al.}(2016)\citenamefont{Kim, Nomura, Ferrero,
  Seth, Parcollet, and Georges}}]{kim2016enhancing}
\bibinfo{author}{\bibfnamefont{M.}~\bibnamefont{Kim}},
  \bibinfo{author}{\bibfnamefont{Y.}~\bibnamefont{Nomura}},
  \bibinfo{author}{\bibfnamefont{M.}~\bibnamefont{Ferrero}},
  \bibinfo{author}{\bibfnamefont{P.}~\bibnamefont{Seth}},
  \bibinfo{author}{\bibfnamefont{O.}~\bibnamefont{Parcollet}},
  \bibnamefont{and} \bibinfo{author}{\bibfnamefont{A.}~\bibnamefont{Georges}},
  \bibinfo{journal}{Phys. Rev. B} \textbf{\bibinfo{volume}{94}},
  \bibinfo{pages}{155152} (\bibinfo{year}{2016}).

\bibitem[{\citenamefont{Mazza and Georges}(2017)}]{mazza2017nonequilibrium}
\bibinfo{author}{\bibfnamefont{G.}~\bibnamefont{Mazza}} \bibnamefont{and}
  \bibinfo{author}{\bibfnamefont{A.}~\bibnamefont{Georges}},
  \bibinfo{journal}{Phys. Rev. B} \textbf{\bibinfo{volume}{96}},
  \bibinfo{pages}{064515} (\bibinfo{year}{2017}).

\bibitem[{\citenamefont{Fausti et~al.}(2011)\citenamefont{Fausti, Tobey, Dean,
  Kaiser, Dienst, Hoffmann, Pyon, Takayama, Takagi, and
  Cavalleri}}]{fausti2011light}
\bibinfo{author}{\bibfnamefont{D.}~\bibnamefont{Fausti}},
  \bibinfo{author}{\bibfnamefont{R.~I.} \bibnamefont{Tobey}},
  \bibinfo{author}{\bibfnamefont{N.}~\bibnamefont{Dean}},
  \bibinfo{author}{\bibfnamefont{S.}~\bibnamefont{Kaiser}},
  \bibinfo{author}{\bibfnamefont{A.}~\bibnamefont{Dienst}},
  \bibinfo{author}{\bibfnamefont{M.~C.} \bibnamefont{Hoffmann}},
  \bibinfo{author}{\bibfnamefont{S.}~\bibnamefont{Pyon}},
  \bibinfo{author}{\bibfnamefont{T.}~\bibnamefont{Takayama}},
  \bibinfo{author}{\bibfnamefont{H.}~\bibnamefont{Takagi}}, \bibnamefont{and}
  \bibinfo{author}{\bibfnamefont{A.}~\bibnamefont{Cavalleri}},
  \bibinfo{journal}{Science} \textbf{\bibinfo{volume}{331}},
  \bibinfo{pages}{189} (\bibinfo{year}{2011}).

\bibitem[{\citenamefont{Raines et~al.}(2015)\citenamefont{Raines, Stanev, and
  Galitski}}]{raines2015enhancement}
\bibinfo{author}{\bibfnamefont{Z.~M.} \bibnamefont{Raines}},
  \bibinfo{author}{\bibfnamefont{V.}~\bibnamefont{Stanev}}, \bibnamefont{and}
  \bibinfo{author}{\bibfnamefont{V.~M.} \bibnamefont{Galitski}},
  \bibinfo{journal}{Phys. Rev. B} \textbf{\bibinfo{volume}{91}},
  \bibinfo{pages}{184506} (\bibinfo{year}{2015}).

\bibitem[{\citenamefont{Patel and Eberlein}(2016)}]{patel2016light}
\bibinfo{author}{\bibfnamefont{A.~A.} \bibnamefont{Patel}} \bibnamefont{and}
  \bibinfo{author}{\bibfnamefont{A.}~\bibnamefont{Eberlein}},
  \bibinfo{journal}{Phys. Rev. B} \textbf{\bibinfo{volume}{93}},
  \bibinfo{pages}{195139} (\bibinfo{year}{2016}).

\bibitem[{\citenamefont{Sentef et~al.}(2017)\citenamefont{Sentef, Tokuno,
  Georges, and Kollath}}]{sentef2017theory}
\bibinfo{author}{\bibfnamefont{M.~A.} \bibnamefont{Sentef}},
  \bibinfo{author}{\bibfnamefont{A.}~\bibnamefont{Tokuno}},
  \bibinfo{author}{\bibfnamefont{A.}~\bibnamefont{Georges}}, \bibnamefont{and}
  \bibinfo{author}{\bibfnamefont{C.}~\bibnamefont{Kollath}},
  \bibinfo{journal}{Phys. Rev. Lett.} \textbf{\bibinfo{volume}{118}},
  \bibinfo{pages}{087002} (\bibinfo{year}{2017}).

\bibitem[{\citenamefont{Ido et~al.}(2017)\citenamefont{Ido, Ohgoe, and
  Imada}}]{ido2017correlation}
\bibinfo{author}{\bibfnamefont{K.}~\bibnamefont{Ido}},
  \bibinfo{author}{\bibfnamefont{T.}~\bibnamefont{Ohgoe}}, \bibnamefont{and}
  \bibinfo{author}{\bibfnamefont{M.}~\bibnamefont{Imada}},
  \bibinfo{journal}{Sci. Adv.} \textbf{\bibinfo{volume}{3}},
  \bibinfo{pages}{e1700718} (\bibinfo{year}{2017}).

\bibitem[{\citenamefont{Mor et~al.}(2017)\citenamefont{Mor, Herzog,
  Gole{\v{z}}, Werner, Eckstein, Katayama, Nohara, Takagi, Mizokawa, Monney
  et~al.}}]{mor2017ultrafast}
\bibinfo{author}{\bibfnamefont{S.}~\bibnamefont{Mor}},
  \bibinfo{author}{\bibfnamefont{M.}~\bibnamefont{Herzog}},
  \bibinfo{author}{\bibfnamefont{D.}~\bibnamefont{Gole{\v{z}}}},
  \bibinfo{author}{\bibfnamefont{P.}~\bibnamefont{Werner}},
  \bibinfo{author}{\bibfnamefont{M.}~\bibnamefont{Eckstein}},
  \bibinfo{author}{\bibfnamefont{N.}~\bibnamefont{Katayama}},
  \bibinfo{author}{\bibfnamefont{M.}~\bibnamefont{Nohara}},
  \bibinfo{author}{\bibfnamefont{H.}~\bibnamefont{Takagi}},
  \bibinfo{author}{\bibfnamefont{T.}~\bibnamefont{Mizokawa}},
  \bibinfo{author}{\bibfnamefont{C.}~\bibnamefont{Monney}},
  \bibnamefont{et~al.}, \bibinfo{journal}{Phys. Rev. Lett.}
  \textbf{\bibinfo{volume}{119}}, \bibinfo{pages}{086401}
  (\bibinfo{year}{2017}).

\bibitem[{\citenamefont{Murakami
  et~al.}(2017{\natexlab{b}})\citenamefont{Murakami, Gole{\v{z}}, Eckstein, and
  Werner}}]{murakami2017photoinduced}
\bibinfo{author}{\bibfnamefont{Y.}~\bibnamefont{Murakami}},
  \bibinfo{author}{\bibfnamefont{D.}~\bibnamefont{Gole{\v{z}}}},
  \bibinfo{author}{\bibfnamefont{M.}~\bibnamefont{Eckstein}}, \bibnamefont{and}
  \bibinfo{author}{\bibfnamefont{P.}~\bibnamefont{Werner}},
  \bibinfo{journal}{Phys. Rev. Lett.} \textbf{\bibinfo{volume}{119}},
  \bibinfo{pages}{247601} (\bibinfo{year}{2017}{\natexlab{b}}).

\end{thebibliography}

\end{document}